%% file: main.tex
\documentclass[trackchanges,twocolumn]{aastex7}
\usepackage{xcolor}
\newcommand{\package}[1]{\textsl{#1}}

\newcommand{\var}[1]{%
 \ensuremath{\texttt{\MakeUppercase{#1}}}\xspace
}

\makeatletter
\renewcommand\frontmatter@affiliationfont{\footnotesize}
\makeatother

\usepackage{amsmath}

\usepackage{lineno}

\input{shorthand.tex}

\shorttitle{DELVE-MC Overview}
\shorttitle{Nidever et al.}

\shortauthors{Nidever et al.}

\usepackage{hyperref}
\begin{document}

\title{DECam Local Volume Exploration Magellanic Clouds (DELVE-MC) Survey Overview}
\input{authors.tex}

\vspace{-25pt}
\begin{abstract}
The Small and Large Magellanic Clouds, the largest satellites of the Milky Way, provide a nearby laboratory for low-mass galaxies. Building on the multi-band DECam photometry of the SMASH survey, the DECam Local Volume Exploration Magellanic Clouds (DELVE-MC) survey has mapped the Magellanic System contiguously with deep broadband imaging over 2170~deg$^2$, reaching $\sim$24.5--25.5 mag in the $griz$ bands using new and archival data. The primary goals are to (i) map stellar populations across the MCs, particularly in their outskirts, (ii) derive spatially resolved star formation histories, (iii) search for new MC satellites, and (iv) study the 3-D structure of the MCs using variable stars. We present an overview of the survey, observing strategy, data reduction, and the first public data release as part of DELVE DR3. DELVE-MC images are processed with the DECam Community Pipeline, and photometry is produced using the custom \delvered\ software. PSF photometry is generated with DAOPHOT/ALLSTAR and calibrated with Gaia DR2 astrometry using a photometric strategy similar to that of the NOIRLab Source Catalog (NSC). Forced photometry is performed on 0.25\degr$\times$0.25\degr\ bricks using master source lists from multi-band S/N-weighted coadds, followed by DAOPHOT/ALLFRAME to determine exposure-level fluxes. The final data products include a ``measurement'' table of individual measurements and an ``object'' table of mean source properties. DELVE-MC provides $\sim$19 billion measurements of $\sim$1.27 billion unique objects enabling light-curve studies and investigations of the MCs. Data are available through the NOIRLab Astro Data Lab.
\end{abstract}

\vspace{-25pt}

\section{Introduction}
\label{sec:intro}
\setcounter{footnote}{0}

The Magellanic Clouds (MCs) have likely been observed by humans for tens of thousands of years \citep{Dennefeld2020}, but detailed scientific studies only began in the 20th century.
The MCs offer one of the best opportunities to study stellar populations, star formation, stellar dynamics, galaxy interactions, galactic tidal effects, and many other astrophysical phenomena in great detail from an external perspective.  In addition, they occupy an important rung on the extragalactic distance ladder \citep[e.g.,][]{Freedman2010}.
The extended Magellanic System, including tidal debris, stellar halos, and gaseous features, spans thousands of square degrees across the southern sky \citep[e.g.,][]{Belokurov2019}. Capturing this full extent in a coherent and deep photometric dataset is essential for tracing the past interactions of the MCs, understanding their disruption history, and identifying faint stellar substructures such as streams, shells, and satellite companions. While previous surveys have provided coverage of the main bodies and selected regions of the outskirts of the Clouds, much of the low-density periphery has remained unexplored at the depth and resolution needed to resolve ancient stellar populations and low-surface-brightness features.

\subsection{Imaging survey history}

The outskirts of the MCs preserve important signatures of their dynamical and interaction history. 
Extended stellar envelopes, tidal arms, diffuse substructures, and satellite companions provide evidence of past interactions between the LMC and SMC as well as the hierarchical growth of low-mass galaxies. 
Recent studies have identified stellar structures extending far beyond the traditional optical boundaries of the Clouds, including tidal features and stellar debris likely associated with repeated interactions between the two galaxies \citep[e.g.,][]{Mackey2016,Choi2018b,Mackey2018,Belokurov2019,Cullinane2022a,Oden2025}.
In addition, several candidate and confirmed dwarf satellite galaxies associated with the MC system have now been discovered \citep{Kallivayalil2018, Patel2020, Erkal2020, Pace2022, Pace2025}, suggesting that the Clouds themselves may host a rich satellite population.
Similar satellite populations around LMC analogs in the Local Volume have recently been identified by the DELVE-DEEP survey \citep{Fielder2025}, consistent with recent theoretical and simulation studies of hierarchical growth in low-mass galaxies \citep[e.g.,][]{Deason2022,KadoFong2022}.

Deep photometric imaging is particularly powerful for studying these low-surface-brightness structures because it enables the use of the numerous old main-sequence and main-sequence turnoff (oMSTO) stars as tracers of the extended stellar populations of the MCs. 
Reaching below the oldest MSTO also enables robust derivation of spatially resolved star-formation histories (SFHs) and helps break degeneracies between age and metallicity in color--magnitude diagram analyses. 
Wide-field imaging surveys over the past two decades have progressively expanded the depth, spatial coverage, and temporal sampling of the Magellanic System, opening new windows into its stellar populations, three-dimensional structure, variability, and interaction history.

\begin{figure*}[t]
\centering
\includegraphics[width=0.48\textwidth,angle=0]{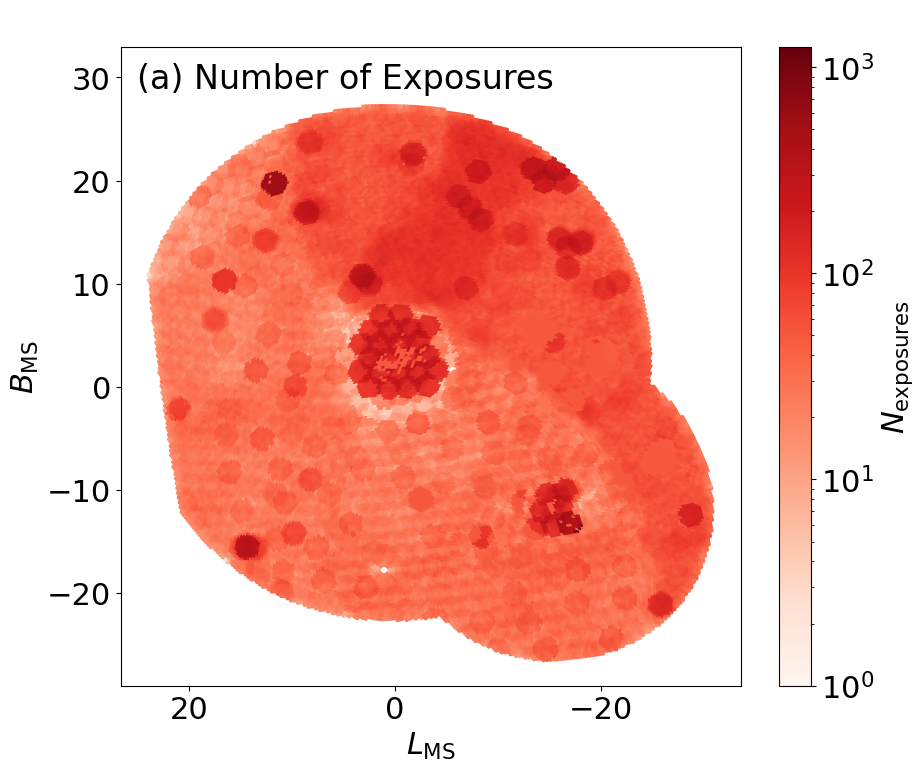}
\includegraphics[width=0.48\textwidth,angle=0]{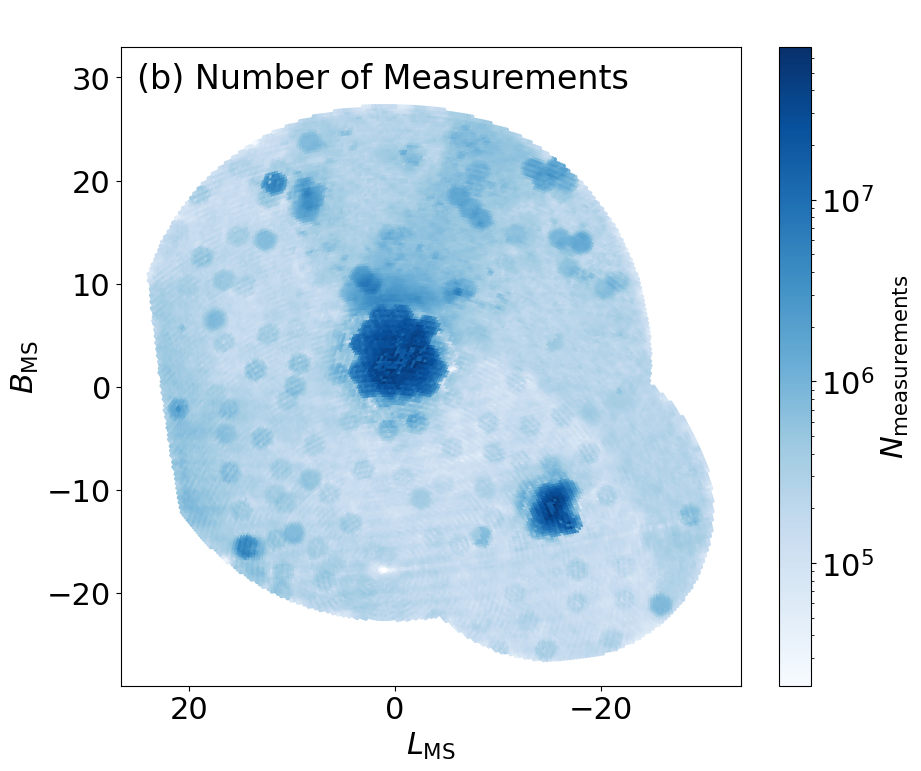} \\
\includegraphics[width=0.48\textwidth,angle=0]{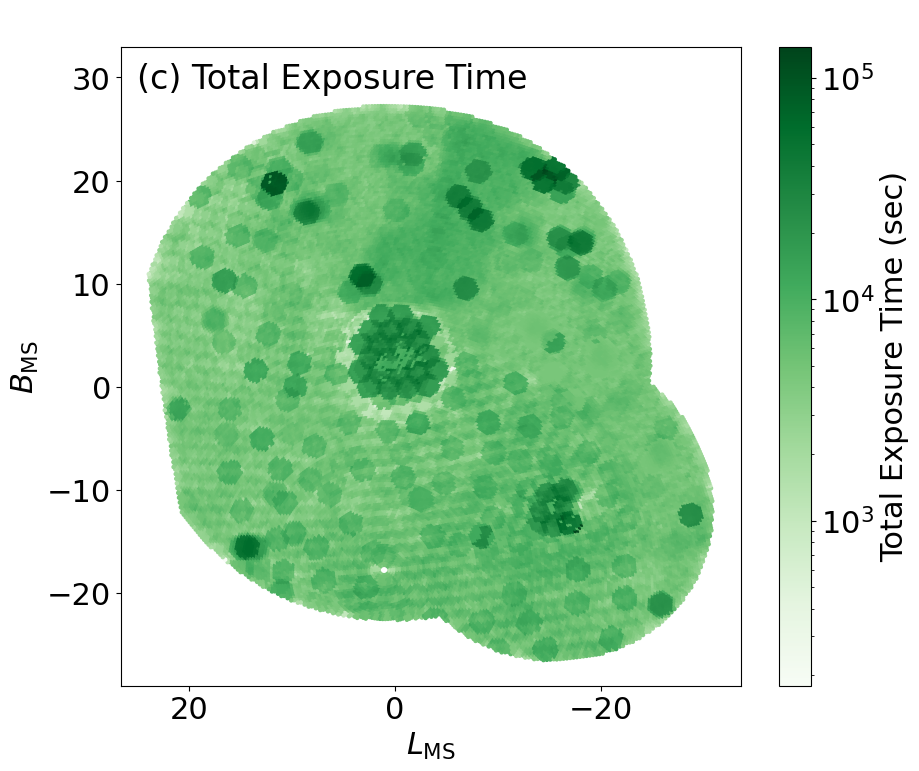}
\includegraphics[width=0.48\textwidth,angle=0]{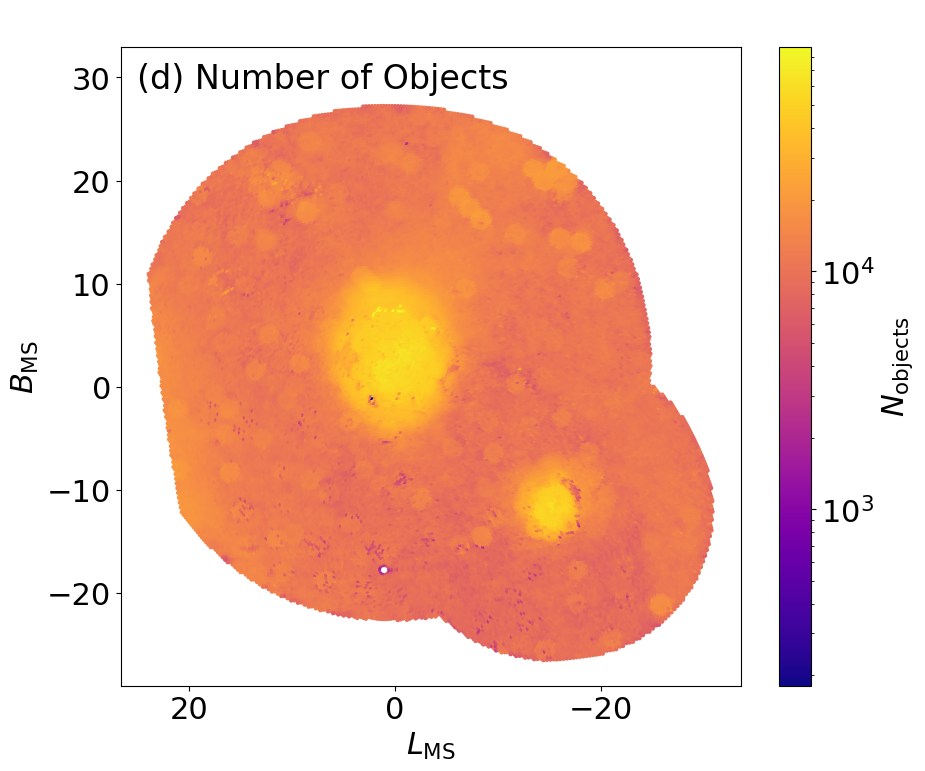}
\caption{The DELVE-MC region in Magellanic Stream coordinates \citep{Nidever2008}.  Statistics per 0.25\degr $\times$ 0.25\dgr brick.
({\em a}) Number of individual exposures, ({\em b}) number of measurements,
({\em c}) total exposure time, and ({\em d}) number of unique objects.
Some features are clearly visible: (1) the deep SMASH coverage in the central regions of the Clouds, (2) the SMASH outlier fields, (3) the DES boundary in the upper-right portion of the footprint, (4) the enhanced, contiguous coverage from the DECam Science Verification (SV) survey immediately inside that boundary (most apparent in panel $b$), and (5) the Galactic-plane cutoff ($|b|\geq10$\degr).
}
\label{fig:map}
\end{figure*}

Photographic studies of the MCs date back to the late 19th and early 20th centuries, particularly through the Harvard Observatory variable-star surveys \citep[e.g.,][]{Leavitt1908,Leavitt1912,PayneGaposchkin1971}.
Later wide-field photographic surveys
\citep[e.g.,][]{deVaucouleurs1955a,deVaucouleurs1955b,Irwin1991,Gardiner1992} laid the groundwork for modern studies of the MCs
by revealing large-scale structural features and
stellar density variations.
From the 1970s through the 1990s, photometric surveys of the MCs advanced dramatically, shifting from broad photographic plate mapping to precise CCD-based studies.
Smaller programs in the 1970s produced some of the first color–magnitude diagrams (CMDs) including Westerlund’s foundational work on stellar populations \citep{Westerlund1970}.
These studies cleared the path for understanding the stellar content and SFHs of the Clouds, but were limited in spatial extent to typically less than 0.5 square degrees in the main bodies. Subsequent wide-field CCD programs expanded this coverage. In particular, the Magellanic Clouds Photometric Survey \cite[MCPS;][]{Zaritsky2002, Zaritsky2004} provided uniform (but shallow) $UBVI$ photometry across the central regions of both Clouds (64~deg$^2$ for the LMC and 18~deg$^2$ for the SMC). These foundational surveys revealed complex stellar structures, age gradients, spatial variation of the dust extinction, and evidence for past interactions, setting the stage for the deeper and more expansive digital surveys of the 2000s and beyond.

Near-infrared surveys such as the Two Micron All Sky Survey \citep[2MASS;][]{Skrutskie2006} also played an important role in advancing studies of the Magellanic System by enabling wide-area mapping of evolved stellar populations. Using 2MASS data, \citet{vdM2001} and \citet{vdMarelCioni2001} constructed large-scale stellar density maps of the LMC and derived constraints on its disk geometry, inclination, and intrinsic elongation using RGB and AGB stars. These studies revealed that the LMC possesses a significant ellipticity and a complex outer structure that is not fully aligned with its disk kinematics, suggesting the influence of tidal interactions.

During the 2000s, deep imaging studies began reaching below the oldest main-sequence turnoff (oMSTO) in selected regions of the MCs, enabling detailed spatially resolved star-formation histories for the first time \citep[e.g.,][]{Noel2007,Noel2009,Meschin2014}. Photometry reaching below the oMSTO is critical for deriving robust star-formation histories and for breaking the age--metallicity degeneracy in color--magnitude diagram analyses. However, these studies were generally limited to relatively small pencil-beam fields, as achieving both oMSTO depth and wide contiguous spatial coverage over the full Magellanic System remained observationally challenging.

One of the first surveys to combine substantial spatial coverage with photometry reaching the oMSTO was the VISTA Survey of the Magellanic Clouds system \citep[VMC;][]{Cioni2011}, a deep, multi-epoch near-infrared imaging campaign covering $\sim$170 deg$^2$ of the LMC, SMC, Bridge, and Stream regions in the infrared $YJK_s$ bands. The survey has produced significant insights into the structure, stellar populations, and SFH of the Magellanic System. The VMC has enabled precise 3-D mapping of the Clouds by utilizing variable stars such as classical Cepheids and RR Lyrae, and derivation of spatially-resolved SFHs using stellar populations down to the oldest main-sequence turnoff \citep{Rubele2012,Rubele2015,Rubele2018,Mazzi2021}. The survey has revealed age gradients and morphological differences between young and old populations \citep{ElYoussoufi2019}, as well as large-scale tidal distortions, especially in the SMC. Analysis of red giant branch slopes has uncovered a shallow metallicity gradient across the SMC \citep{Choudhury2020}. Additionally, studies of young stellar distributions have shown scale-free, fractal clustering indicative of turbulence-driven star formation \citep{Miller2022}. VMC has also enabled the age-dating and structural analysis of hundreds of star clusters, offering new constraints on the cluster formation history of the Clouds \citep{Piatti2015}. Finally, their long survey time baseline has enabled them to obtain accurate proper motions for a large number of stars, giving us the largest catalog of MC stellar proper motions \citep{Niederhofer2021, Niederhofer2022, Vijayasree2025}. Together, these results have greatly enhanced our understanding of the evolution and interaction history of the Magellanic System.

In addition, the optical YMCA survey \citep[Yes, Magellanic Clouds Again;][]{Gatto2024} is an optical imaging program using the VLT Survey Telescope (VST) to map the outer regions of the MCs in multiple bands.
Targeting approximately 110 square degrees across the outskirts of both the LMC and SMC --- including a continuous strip between them --- YMCA reaches to $g,i$ $\sim$ $23.5$ 
and uncovers predominantly intermediate-age and old stellar populations. 
The YMCA has already identified candidate clusters formed during a close Magellanic encounter $\sim$2.5-3 Gyr ago, challenging the classical ``age gap'' paradigm: the pronounced scarcity of populous clusters with ages between approximately 3 and 10 Gyr, despite continued field-star formation during that interval \citep[e.g.,][]{DaCosta1991,Geisler1997}.
Furthermore, the fourth phase of the Optical Gravitational Lensing Experiment \citep[OGLE-IV;][]{Udalski2015} has been contiguously covering the main bodies of the MCs and the Bridge region with relatively shallow $VI$ ($\sim$21) 
photometry, but with many repeat observations that make it useful for studying the distribution of Magellanic variable stars \citep{Skowron2014,Soszynski2016,J-D2016,J-D2017}.

The advent of the Dark Energy Camera \citep[DECam;][]{Flaugher2015} on the Blanco 4-m telescope enabled a new generation of deep, wide-field optical surveys of the Magellanic periphery. The Dark Energy Survey \citep[DES;][]{DES} provided contiguous multi-band imaging over much of the southern sky, including regions surrounding the MCs, and contributed to the discovery of numerous ultra-faint dwarf galaxy candidates \citep{Bechtol2015,DrlicaWagner2015,Koposov2015}. Building on this capability, the Magellanic Satellites Survey \citep[MagLiteS;][]{DrlicaWagner2016} used DECam imaging around the southern periphery of the MCs to search for faint satellites and stellar substructures, leading to the discovery of candidate Magellanic satellite galaxies \citep{DrlicaWagner2016,Torrealba2018}. Wide-field DECam imaging by \citet{Mackey2016,Mackey2018} revealed extended tidal features around the LMC, including the northern stellar arm and southern hooks, providing strong evidence for ongoing tidal interactions between the MCs and with the Milky Way. Together, these surveys demonstrated the power of deep, wide-field imaging for uncovering the low-surface-brightness structure of the Magellanic periphery and laid the foundation for subsequent contiguous surveys of the full Magellanic System.

The \textit{Gaia} data releases \citep{GaiaDR2,GaiaDR3}
have opened a new era in the study of the MCs by providing high-precision astrometry and photometry for millions of stars across the system. These data have enabled the construction of detailed kinematic maps, the identification of tidal distortions, and the analysis of internal rotation and proper motion gradients in both the LMC and SMC \citep[e.g.,][]{Helmi18,Zivick2019,Choi2022,Schmidt2022,Jimenez-Arranz2023}.
\citet{Luri2021} presented a comprehensive overview of the astrometric properties of the Clouds using \textit{Gaia} EDR3, revealing coherent rotation in the LMC, complex proper motion patterns in the SMC, and evidence for tidal features consistent with recent interactions between the two galaxies.
In addition, \citet{Massana2024} and \citet{Omkumar2025} used Gaia XP stellar parameter information to generate independent metallicity maps across the entire Magellanic System.

The Survey of the MAgellanic Stellar History \citep[SMASH;][]{Nidever2017,Nidever2021a} conducted a deep, multi-band imaging survey exploring the extended structure and stellar content of the Magellanic System. Utilizing the Dark Energy Camera \citep[DECam;][]{Flaugher2015} on the Blanco 4-meter telescope at Cerro Tololo Inter-American Observatory (CTIO), SMASH targeted $\sim$480 deg$^2$ of carefully selected fields across a broader $\sim$2400 deg$^2$ region around the LMC and SMC. Its primary goals included identifying low-surface-brightness features, constraining spatially-resolved SFHs, and searching for a stellar counterpart to the Magellanic Stream. SMASH provided some of the deepest wide-area photometry of the Clouds to date, leading to discoveries such as new stellar substructures \citep{Choi2018b}, insights into disk warping \citep{Choi2018a}, discovery of MC satellites \citep{Martin2015,Martin2016}, resolved SFHs \citep{Ruiz-Lara:2020,Massana2022,Sakowska2026}, and evidence for extended stellar peripheries \citep{Nidever2019a,Massana:2020}. The dataset established a crucial foundation for subsequent surveys aiming to build a more complete view of the Magellanic periphery.

\subsection{DELVE-MC survey}
The DECam Local Volume Exploration \citep[DELVE;][]{Drlica-Wagner:2021} Survey is a $\sim$150-night program using NSF's 4-m Blanco Telescope in Chile to perform an unprecedented observational census of the faintest satellite galaxies \citep[e.g.,][]{Mau2020, Cerny2023, Tan2025}. One component, ``DELVE-MC'', comprises deeper, contiguous coverage of $2{,}170 \deg^2$ in $g,r,i$ centered on the MCs (see \autoref{fig:map}). DELVE-MC builds on the SMASH survey by contiguously mapping the MC periphery via $\sim$3,200 new exposures that makes the deepest, wide-area survey of the Clouds to date. The DELVE-MC program combines new observations with existing public data to maximize depth and available color information within the allocated DELVE observing time. Based on currently processed data, DELVE-MC reaches $5\sigma$ depths of $g/r/i/z$=25.7/25.3/24.8/24.3 mag.

Many previous wide-area studies of the diffuse Magellanic outskirts have relied primarily on evolved giant stars, which are bright enough for all-sky surveys but provide relatively sparse tracers. By reaching well below the oldest main-sequence turnoff across the full Magellanic System, DELVE-MC greatly increases the density of individually resolved tracers available for mapping faint stellar substructures and extended halos.

With its combination of contiguous, deep, multi-band imaging across the full Magellanic System, DELVE-MC enables a comprehensive view into the formation and evolution of low-mass galaxies. The survey’s sensitivity to ancient main-sequence populations and very low-surface-brightness structures makes it possible to obtain a comprehensive census of stellar substructures in the Magellanic periphery, including tidal streams, diffuse envelopes, and possible remnants of past accretion events. These maps will illuminate the dynamical history of the LMC–SMC interaction, trace the build-up of their stellar halos, and provide crucial context for the Magellanic dwarf satellite population. In addition, the spatially resolved photometry from DELVE-MC will support detailed star-formation histories across the Clouds and will help clarify how their chemical enrichment and structural evolution depend on environment and interaction history.

The dataset will also serve as a foundation for precise three-dimensional mapping using variable-star distance indicators. In combination with the high-cadence time-domain observations from the Vera C. Rubin Observatory Legacy Survey of Space and Time \citep[LSST;][]{Ivezic2019}, DELVE-MC will provide deep spatial coverage and long temporal baselines that extend variability studies to much fainter magnitudes and lower surface-brightness regions of the Magellanic System. When combined with forthcoming spectroscopic surveys like the Sloan Digital Sky Survey-V \citep[SDSS-V;][]{Kollmeier2026} Magellanic Genesis Survey \citep{Nidever2026}, DELVE-MC will provide a transformative, system-wide view of the Magellanic Clouds that reveals the fossil record of their growth and repeated interactions, establishing the Clouds as a benchmark laboratory for understanding the hierarchical evolution of dwarf galaxies.

In this overview paper, we present the DELVE-MC survey, including its scientific goals, observing strategy, data processing, and overall performance. We also highlight early science results that illustrate the survey’s potential to advance our understanding of the structure and evolution of the Magellanic System.

\begin{deluxetable}{lc}
\centering 
\tablecaption{DELVE-MC Survey \label{table:program}}
\tablecolumns{5}
\tablewidth{500pt}
\tablehead{
  \colhead{Property} & \colhead{Value}
}
\startdata
Depth (95\%)  &  $g/r/i/z$=25.7/25.3/24.8/24.3 mag  \\
$\langle$Completeness$\rangle$ (50\%) &  $g/r/i/z$=25.2/25.0/24.5/23.9 mag \\
Coverage     &  2{,}170 deg$^2$  \\
Exposures (new/total)     &  $\sim$3{,}200/33{,}968  \\
Exp. Time (new/total) &  254h / 1365h \\
Astrometric RMS  &  10.5 mas \\
Photometric RMS & 11 mmag \\
$\langle N_{\rm epochs} \rangle$ (all bands) &  $\sim$14  \\
$\langle \Delta T \rangle$  &  $\sim$1700d
\enddata
\end{deluxetable}

\section{Data set}
\label{sec:data}

To fully map the main stellar components of the Magellanic System including the periphery, we delineate a region extending to a radius of 25\dgr from the LMC and 15\dgr from the SMC but requiring a Galactic latitude less than $-$10\dgr to limit the contamination from the Milky Way (MW) disk.  This amounts to an area of 2170 deg$^2$.  \autoref{fig:map} shows this region and highlights the
(a) number of exposures, (b) number of measurements, (c) total exposure time, and (d) number of objects in each patch of the sky (0.25\degr $\times$ 0.25\dgr ``brick''). 
\autoref{fig:nexp_bricks} shows the distribution of the number of exposures per brick, highlighting the substantial multi-epoch observations across much of the survey.
The data set consists of 35{,}146 DECam exposures from 02 November 2012 up until 27 January 2023.
This is a subset of all of the available data that passed our quality cuts on seeing (FWHM $<$ 2\arcsec) and photometric zero-points ($<$0.5 mag from the expected airmass-corrected zero-point), and are in the $ugrizY$ bands.
Of those exposures, 32{,}714 pass the additional exposure time $\geq~$90s selection criteria to be included in the forced photometry analysis.
Most of the exposures are public, community data from the NOIRLab Astro Data Archive\footnote{\url{https://astroarchive.noirlab.edu/}}
with $\sim$3,200 exposures taken by DELVE itself.



The DELVE-MC exposures mapped 1{,}075 deg$^2$ in regions not previously covered sufficiently by SMASH, DES, or MagLiteS \citep{DrlicaWagner2016}, and had total integration times of 800s (3$\times$267s) for $g$,$r$ and 1000s (3$\times$333s) for $i$.  
For DELVE-MC, we followed the DELVE-WIDE tiling approach, using a sequence of three dithered exposures patterned after the DECaLS \citep{Dey2019} icosahedral layout but with larger dither steps spanning approximately three DECam CCDs. This strategy improves the uniformity and completeness of the contiguous survey footprint by ensuring that most sky positions are observed on multiple detectors and by filling the CCD gaps.
The new DELVE-MC observations were taken preferentially in good seeing (PSF Full Width at Half Maximum [FWHM] $<$1.1\arcsec~as estimated for an $i$-band exposure at zenith) during DELVE's observing campaign.
Table \ref{table:program} summarizes the main characteristics of the DELVE-MC survey.






\section{Reduction and Photometry}
\label{sec:phot}

\subsection{Reduction}
\label{subsec:reduction}

The reduction and analysis tools are similar to those used by SMASH as well as some routines used for the NOIRLab Source Catalog \citep[NSC;][]{Nidever2018,Nidever2021b} Data Release 2 (DR2).  The DELVE-MC uses images that are processed by the NOAO Community Pipelines (CP) for instrumental signature removal  \citep{Valdes2014}\footnote{\url{https://www.noao.edu/noao/staff/fvaldes/CPDocPrelim}}.  See \S 3.1 of the NSC DR1 paper \citep{Nidever2018} for a description of the main CP processing steps.
For photometry, we use the 
\daophot software \citep{Stetson1987,Stetson1990,Stetson1994}
to obtain Point Spread Function (PSF) measurements for each exposure
(\allstar)
and then
perform forced photometry (\allframe) on the set of exposures for each unique region of the sky to obtain the deepest and most accurate photometric measurements.



\subsection{delvered}
\label{subsec:delvered}

We use the custom \delvered package\footnote{\url{https://github.com/dnidever/delvered}} \citep{delvered:2025} to perform our photometric processing.  
The \delvered package relies on the \photred package\footnote{\url{https://github.com/dnidever/PHOTRED}} \citep{Nidever2018,photred:2025}, which 
was also used for the MAgellanic Periphery Survey \citep[MAPS;][]{Majewski2009,Nidever2011,Nidever2013} and SMASH data processing and is a wrapper around the \daophot suite of software.  See \citet{Nidever2017} for a detailed description of \photred and the SMASH data processing.

There are two main differences between SMASH and DELVE-MC processing: (1) photometric calibration, and (2) the forced photometry reference frame.  For SMASH, the photometric calibration was performed in a more ``classical'' mode whereby zero-points, extinction terms, and chip-to-chip color terms are determined on a regular basis (nightly or per run) using standard star observations, for photometric nights.  The SMASH data were calibrated to the SDSS system using color-terms and are therefore on a ``quasi-SDSS'' system because
the DECam filters were used and these
have some non-negligible differences from the standard SDSS filters (e.g., in the $u$-band; see Appendix in \citealt{Abbott2021}).  However, with the larger number of exposures and nights for DELVE-MC, this classical approach became impractical.  Therefore, we adopted the NOIRLab Source Catalog \citep[NSC;][]{Nidever2018,Nidever2021b} calibration scheme, which uses the ``native'' DECam photometric system (no color terms) and uses ``model magnitudes'' --- magnitudes derived from a linear combination of photometry from large surveys such as \Gaia, 2MASS \citep{Skrutskie2006}, and SkyMapper \citep{Wolf2018,Onken2024} --- to determine a global zero-point for each unique exposure. A new \zeropoint stage was added to the \photred software to enable the measurement of exposure-level zero-points for DELVE-MC calibration.

The deep SMASH exposures ($\sim$300s) used only small dithers ($\sim$ 
a few arcseconds) so that the \allframe forced photometry could easily be performed on a chip-by-chip basis (i.e., all \ccdnum=1 images together, all \ccdnum=2 images together, etc.) and using one chip image as the reference frame. Shorter exposures with larger offsets were then used to tie the chips together photometrically on the same system.  With the large archival component of the DELVE-MC dataset, this approach was also impractical. Instead of using a single chip as the basis for the reference frame, we adopted a tile or brick scheme that defined a unique region in the sky (i.e., essentially a tangent plane World Coordinate System or WCS) as the reference frame. Any chip image that overlapped this brick was included in the \allframe forced photometry processing. Using this scheme required some modifications to the \photred software to use an arbitrary reference frame.

For SMASH, the main processing was performed by \photred while some extra SMASH-specific steps (such as ``uber-calibration'') were performed by a second custom software package \smashred\footnote{\url{https://github.com/dnidever/SMASHRED}}.
The \delvered processing is logically split into two main steps: (1) exposure-level processing, and (2) forced-photometry (brick) processing. Every exposure in the DELVE-MC region is processed through the exposure level, while quality checks are performed on exposures before they are processed at the brick level. The two steps are described in more detail below.


\begin{center}
\begin{figure}[t]
\includegraphics[width=1.0\hsize,angle=0]{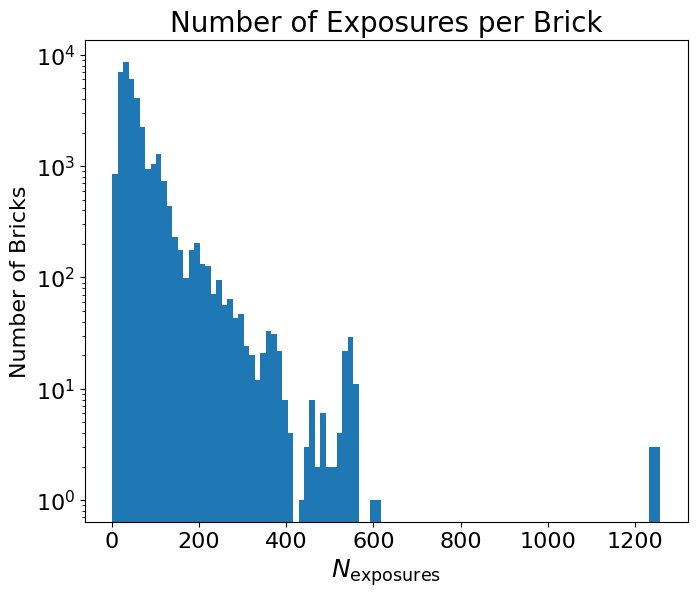}
\caption{The distribution of the number of exposures per brick in the DELVE-MC survey.}
\label{fig:nexp_bricks}
\end{figure}
\end{center}

\begin{center}
\begin{figure*}[t]
\includegraphics[width=1.0\hsize,angle=0]{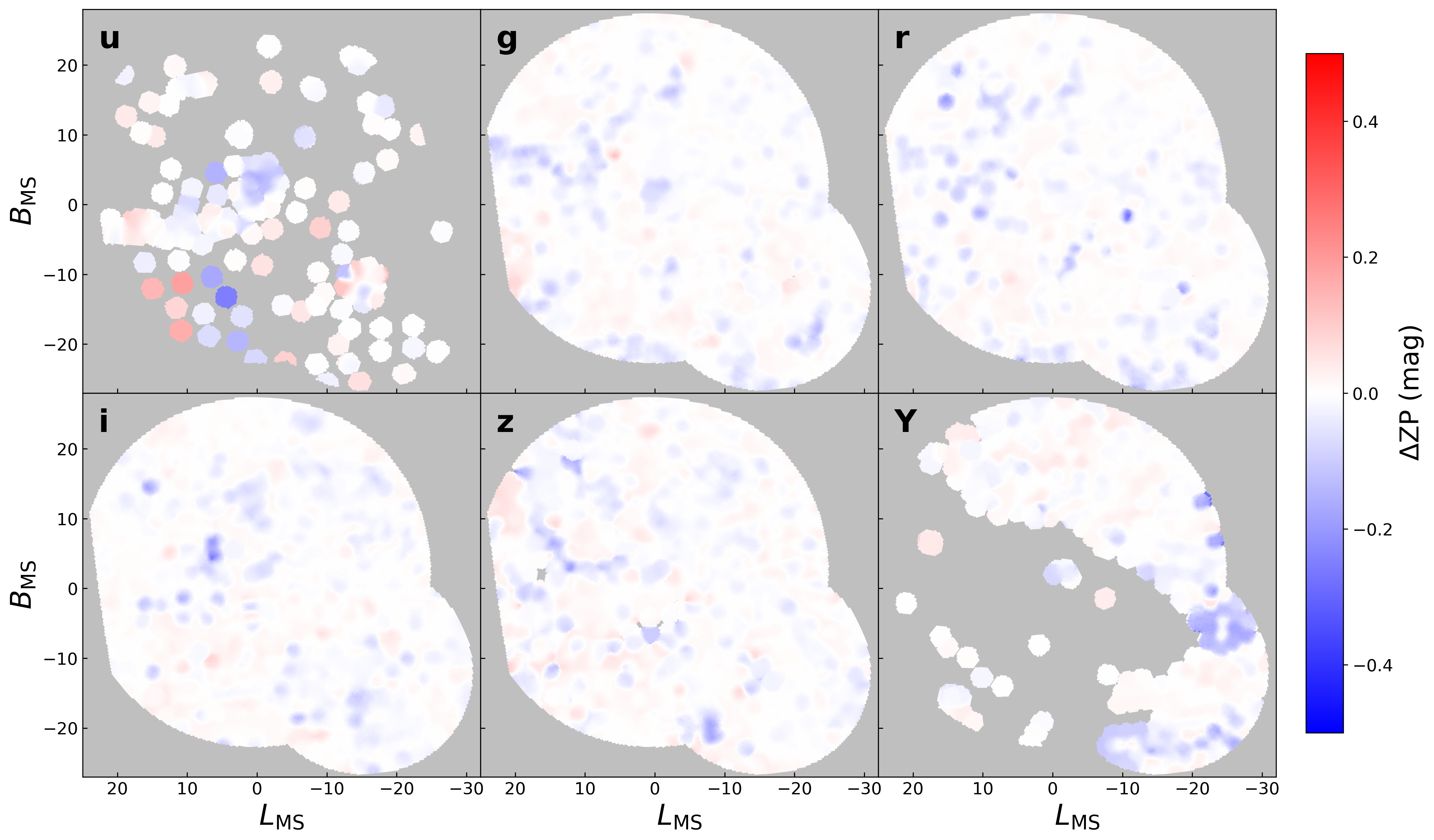}
\caption{Mean zero-point residuals in the $ugrizY$ bands. After correcting the zero points to unit airmass and removing weekly trends, we compute the mean zero point in each survey brick and subtract the survey-wide median of the brick means for that band. The larger fluctuations in the $u$ band may reflect the smaller number of exposures per brick, which reduces the averaging of exposure-to-exposure variations in atmospheric transparency. Red and blue indicate positive and negative residuals, respectively, and white indicates $\Delta\mathrm{ZP}\approx 0$. Gray marks regions with no usable measurements in the corresponding band.}

\label{fig:zeropoints}
\end{figure*}
\end{center}

\subsubsection{Exposure-Level Processing}
\label{subsubsec:exposurelevel}

The exposure-level processing is run on a night-by-night basis and parallelization is performed on the multiple data units (chip, exposure, or field) for each \photred processing ``stage''.  Each night is ``prepared'' by taking the exposures in the DELVE-MC region for that night and grouping them into unique ``fields'' (any exposures with central coordinates within 0.1\dgr of each other), as well as generating \Gaia DR2 \citep{Gaia2016,GaiaDR2} reference catalogs for each chip image.  This also generates the initial ``inlist'' for the first \photred stage \wcs.


The {\tt delvered\_exposures} routine is run on a range or list of nights that
will loop over those nights, processing them one at a time.  For each night, ten \photred stages are run.  These are explained fully in \S 5 of \citet{Nidever2017}, which describes \photred in detail. We briefly review the stages here (the type of data that is grouped for each stage is given in the trailing square brackets, e.g., [chip]):

\begin{enumerate}
\item \wcs: Improves the WCS of each chip image using the \Gaia DR2 reference astrometry that is transformed (including proper motion information) to the exposure epoch [chip].
\item \daophot: Determines the PSF and runs \daophot \allstar PSF photometry on each chip image. This produces a \psf file and a source catalog [chip].
\item \match: Groups images of the same chip in each field, picks a reference image, matches sources and calculates linear coordinate transformations between each image and the reference image [field-chip].
\item \apcor: Calculates the aperture correction (going from PSF to total flux) for each chip image using \daophot~\daogrow \citep{Stetson1990} [chip].
\item \astrom: Computes astrometric coordinates ($\alpha$, $\delta$) for all sources in a field-chip source catalog using the WCS of the reference image [field-chip].
\item \zeropoint: Calculates the photometric zeropoint for each exposure with model magnitudes using the same methods as in NSC DR2 \citep[see \S 3.2.2 of][]{Nidever2021b} [exposure].
\item \calib: Applies photometric calibration to each source catalog using the zeropoint, exposure time and aperture corrections [field-chip].
\item \combine: Combines chip-level source catalogs into a single catalog for each field [field-chip].
\item \deredden: Adds \citet{Schlegel1998} reddening information to each source in a catalog [field].
\item \save: Renames the final catalog and produces a summary table for each field [field].
\end{enumerate}

At the end of the night-level processing, a summary table is created for the entire night that contains summary information for all chips, exposures and fields.


The spatial distribution of the mean $griz$ zero-point residuals across the DELVE-MC footprint is shown in \autoref{fig:zeropoints}. For each band, the mean corrected zero point in each brick is expressed relative to the survey-wide median zero point. After correcting for airmass and removing weekly zero-point trends, the remaining large-scale residuals are generally small, indicating relatively uniform photometric calibration across the survey.

The main information from the exposure-level processing that is needed for the forced photometry processing are the chip images, image-level \psf files, source catalogs, aperture corrections, and the exposure photometric zeropoints.
In addition, the exposure-level processing generates photometry that can itself be used for time-series analysis independent of the deep, forced-photometry processing.

\subsubsection{Forced-Photometry Processing}
\label{subsubsec:forced}

The forced photometry is run on a brick level for all exposures that overlap the brick, where a brick is defined as a 0\dotdeg25 $\times$ 0\dotdeg25 patch of the sky.  We use a WCS with a tangent plane projection, 0.263\arcsec~pixel size (the average DECam pixel size), and 3600 $\times$ 3600 pixels, which provides an 88 pixel buffer around the edge of the brick. 
These bricks are laid out on the sky in such a way that there is a small amount of overlap between neighboring bricks, therefore, caution must be used when combining brick-level catalogs to only include sources in the ``unique'' areas of the bricks (\texttt{brickuniq}=1 in the object table).
We use the DESI Legacy Imaging Surveys \citep{Dey2019} tiling scheme of the sky, which contains 662,174 bricks for the full-sky.  
In our selected DELVE-MC region ($R_{\rm LMC} < 25$\degr, $R_{\rm SMC} < 15$\dgr and $|b| > 10$\degr), we select 35,052 bricks that wholly cover our footprint (see \autoref{fig:map}).

The forced photometry is performed on a brick-by-brick basis with parallelization over bricks, such that each brick has a single computing process. Generally, 10 bricks are run simultaneously on a single server.
At the beginning of the brick-level processing, an {\tt sqlite3} database of all of the nightly exposure-level processing summary files is created with chip-level and exposure-level information.

When processing a brick (size 15\arcmin~$\times$ 15\arcmin), the initial step is to find the overlapping chips (size 9\arcmin~$\times$ 18\arcmin) by querying the summary database. An initial rough search is performed for chip images that have centers within 30\arcmin~of the brick center. A second, more rigorous, overlap check is performed using the chip vertices (or corners). Only chips that overlap the brick area by more than 2 pixels ($\sim$0.5\arcsec) are kept for further analysis. 

Next, quality cuts are applied to the exposures.  Only exposures with seeing FWHM$~\leq~$2\arcsec, $\ge~$90s exposure time, $grizY$ bands, and an additive (the negative of those found in the exposure-level processing) airmass-corrected zeropoint $\ge~$$-$0.5 mag are included.  Known bad exposures compiled from the DELVE WIDE processing \citep{Drlica-Wagner:2021} are also excluded.

\begin{center}
\begin{figure}[t]
\centering
\includegraphics[width=0.95\hsize,angle=0]{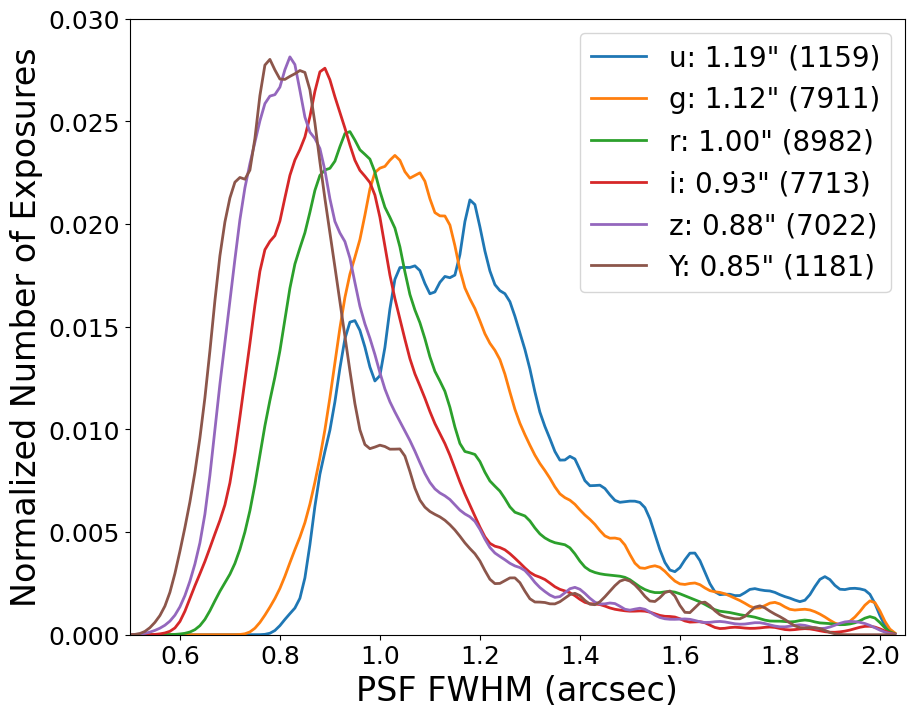}
\caption{Distribution of seeing FWHM values for each band. The median seeing and number of exposures per band is given in the legend.}
\label{fig:fwhmhist}
\end{figure}
\end{center}

The resulting distribution of image quality for the accepted exposures is shown in \autoref{fig:fwhmhist}. The median seeing ranges from approximately 0.85$^{\prime\prime}$ in the $Y$-band to 1.2$^{\prime\prime}$ in the $u$-band, with most $griz$ exposures having sub-arcsecond image quality.

The goal of the forced-photometry processing is to measure fluxes for a consistent set of sources across all overlapping exposures of a brick. Following the general approach implemented in the \photred \allframe stage, a deep coadded image is first constructed and used to define the master source catalog. Crowded-field photometry is then performed simultaneously on all of the individual exposures at the positions of the sources in this master catalog using \allframe. This approach improves source detection and photometric precision, particularly for faint objects that may not be detectable in individual exposures.


The main forced-photometry processing steps are as follows:

\begin{enumerate}
    \item {\tt Match}: This performs the same function as the exposure-level \match step, but the ``reference image'' is now the brick WCS frame.
    \item {\tt Coadd}: Creates a deep, multi-band coadded image that serves as the detection image for the brick. Each chip image is resampled onto the brick WCS tangent plane, bad pixels are masked, the sky is subtracted, and the image is photometrically scaled. The images are then averaged using S/N weighting (using IRAF IMCOMBINE; \citealt{Tody1986,Tody1993,Fitzpatrick2024}). A final weighted and scaled sky image is added back at the end.
    \item {\tt PSF}: Generates a \daophot PSF model for the coadded image.
    \item {\tt Master}: Detects sources in the coadded image and generates a master source list for the brick. Initial source detection is performed with Source Extractor \citep{Bertin1996}, after which 
    \allframe is run iteratively to refine source centroids and reject spurious detections. The final master catalog defines the positions at which photometry is measured in all overlapping individual exposures.
    \item \allframe: Runs \daophot \allframe on all individual chip images using the exposure-level processing \psf models and the master source catalog.
    Fluxes are simultaneously fit for each source in every overlapping exposure, including sources too faint to be independently detected in single images. Source positions are allowed to vary slightly, but are constrained to remain consistent across all images (i.e., assuming negligible proper motion over the survey baseline), while image-to-reference-frame transformations are refined simultaneously.
    \item {\tt Astrometry}: Calculates astrometric coordinates ($\alpha$,$\delta$) for the sources using the values determined in the \allframe step and the brick WCS reference frame.
    \item {\tt Calibrate}: Calibrates the chip-level photometry using the exposure zeropoint, exposure time and aperture corrections. Then flux-weighted average photometry and scatter are calculated per band per object (included in the {\tt object} table).
\end{enumerate}

This procedure produces consistent multi-epoch photometry tied to a common astrometric reference frame while maximizing sensitivity to faint sources across the full set of overlapping exposures.

Three final forced photometry files are created for each brick: {\tt object} (object-level information), {\tt meas} (epoch or exposure-level measurements), and {\tt meta} (metadata) files.

While most bricks only required an hour to process through the pipeline, some in the central regions required as long as $\sim$24 hours.  Fewer bricks were run in parallel when crowded regions were being processed.

\begin{center}
\begin{figure*}[t]
\includegraphics[width=1.0\hsize,angle=0]{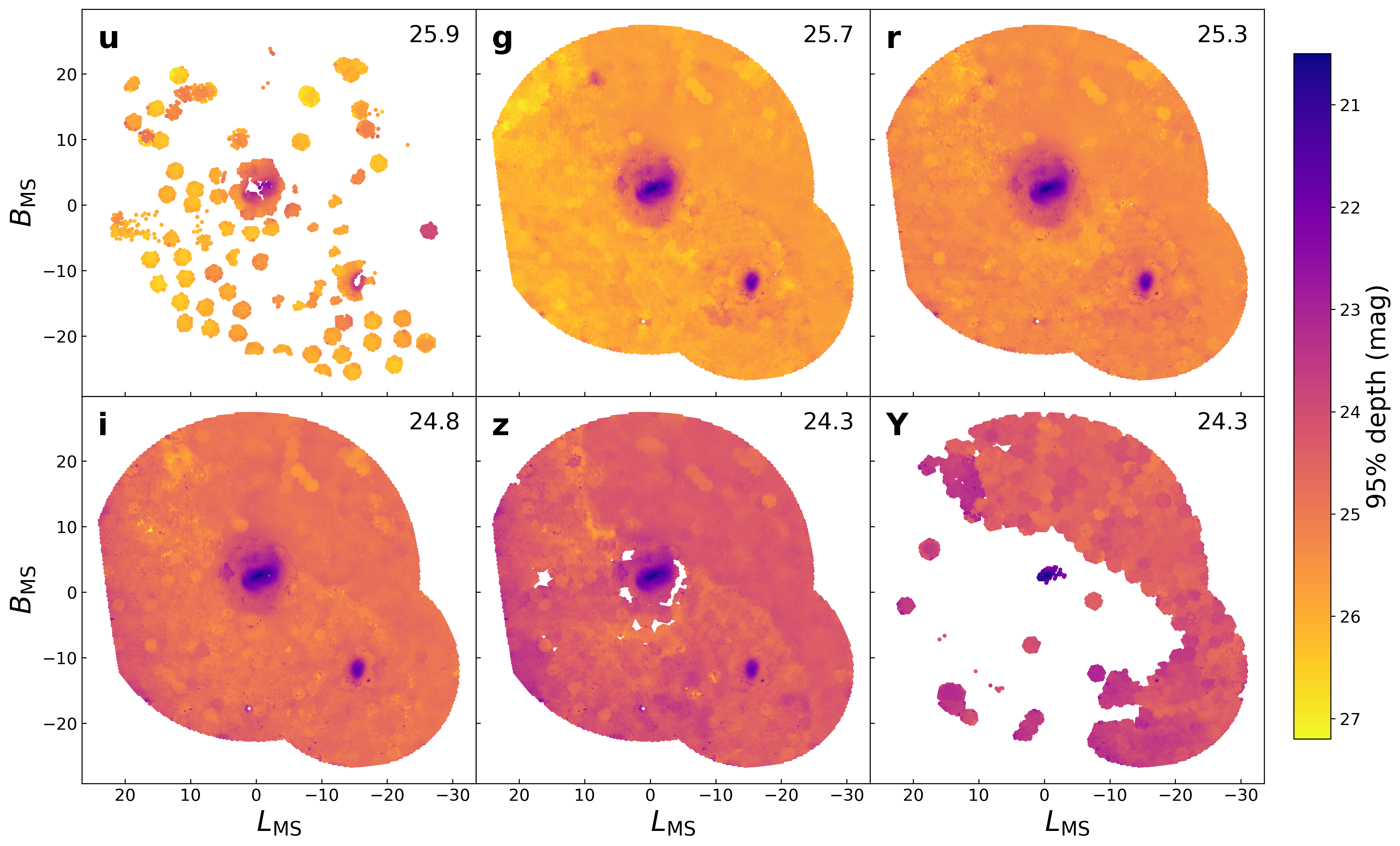}
\caption{The 95th percentile depth per band in each brick in the different passbands. The median depth across all bricks is given in the upper right-hand corner of each panel.}
\label{fig:depth}
\end{figure*}
\end{center}

\subsubsection{Picking the Best Exposures}
\label{subsubsec:bestexposure}

Bricks in the crowded central regions of the LMC and SMC can have hundreds to thousands of overlapping exposures, causing the memory required by the simultaneous ALLFRAME processing to become prohibitive. We therefore adopted a maximum of 50 exposures as a practical compromise between cumulative depth, multi-band sampling, and tractable memory and processing time. This limit applies to every brick with more than 50 otherwise acceptable exposures, including the small number of highly sampled outlying fields.



For any brick with more than 50 otherwise acceptable exposures, we use an algorithm to select the subset included in the ALLFRAME processing. The goal is to retain the highest-quality exposures overall while also providing strong sampling in each represented band. Because the usefulness of an exposure depends not only on its depth but also on the fraction of the brick that it covers, the selection algorithm considers both estimated depth and fractional overlap.

For each chip, we calculate an ``effective exposure time'', \teff~\citep{Neilsen2015}, which accounts for variations in transparency, seeing, and background levels.

\begin{equation}
{\rm t}_{\rm eff} \equiv \eta^2 \left( \frac{\rm FWHM}{0.9''} \right)^{-2} \left( \frac{b}{b_{dark}} \right)^{-1} \times {\rm EXPTIME}
\end{equation}

\noindent
where $\eta$ is the atmospheric transmission (i.e., $\exp{(-\Delta{\rm ZP})}$, where $\Delta$ZP$=$ZP$-$$\langle {\rm ZP} \rangle_{\rm band}$), FWHM is the seeing in arcsec, and $b$ is the background level.  \teff~is very useful as the signal-to-noise-ratio S/N of a point source in the sky background dominated regime varies as $\sqrt{{\rm t}_{\rm eff}}$.

\begin{center}
\begin{figure*}[t]
\includegraphics[width=0.47\hsize,angle=0]{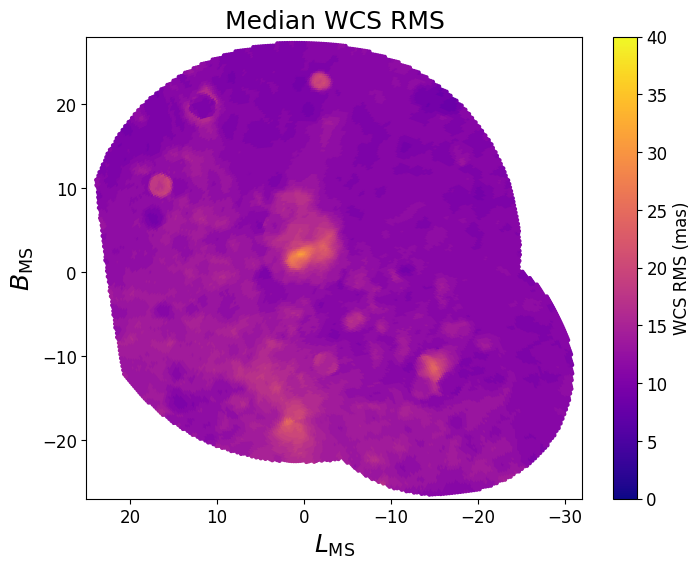}
\includegraphics[width=0.50\hsize,angle=0]{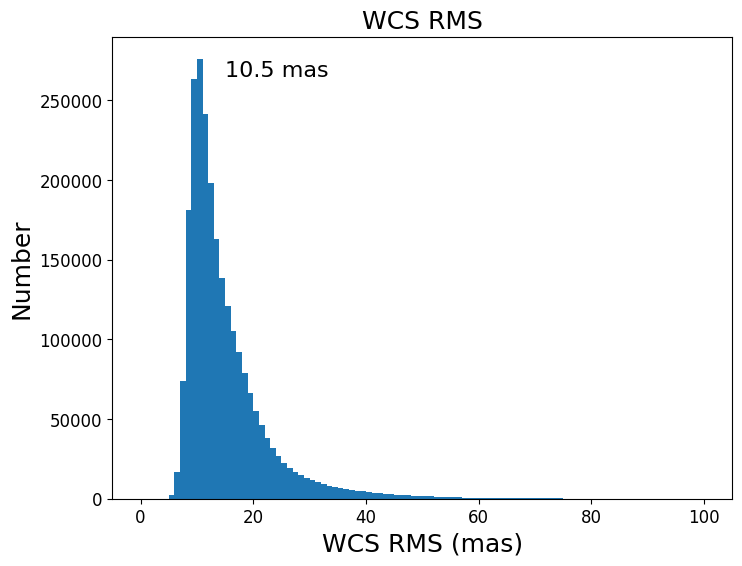}
\caption{
Astrometric RMS of bright stars ($G<$20 and measured proper motions below 200 mas yr$^{-1}$) in milliarcseconds (mas). (Left) Median RMS for each brick, and (right) histogram of chips RMS with a peak at 10.5 mas.}
\label{fig:wcsrms}
\end{figure*}
\end{center}

To handle the spatial dimension of the problem, we evaluate the sum of \teff, or sum of (S/N)$^2$, from multiple images on a 3600 $\times$ 3600 pixel grid within the brick. For a given exposure, the pixels that overlap the brick are increased by the chip's \teff. Since S/N adds in quadrature or linearly in (S/N)$^2$, we can simply continue adding the \teff~values from the chips to the image and finally take the square root to obtain the final S/N image. Our final metric is the sum of the S/N values across all pixels in the brick, $\Sigma$(S/N). Ultimately, we want to compare exposures against each other.  Therefore, we evaluated the difference of $\Sigma$(S/N) with and without the contribution of an exposure, $\Delta\Sigma({\rm S/N}) = \Sigma({\rm S/N_{\rm with}}) - \Sigma({\rm S/N_{\rm without}}$). If an exposure covers ``new'' area (with no previous exposure coverage), then the S/N will add linearly. However, coverage in an area that already has some exposure coverage will add in quadrature and will result in a smaller increase in S/N. Consequently, this scheme favors adding new area over incremental improvements in S/N via multiple overlapping exposures.

\begin{center}
\begin{figure}[b]
\includegraphics[width=0.98\hsize,angle=0]{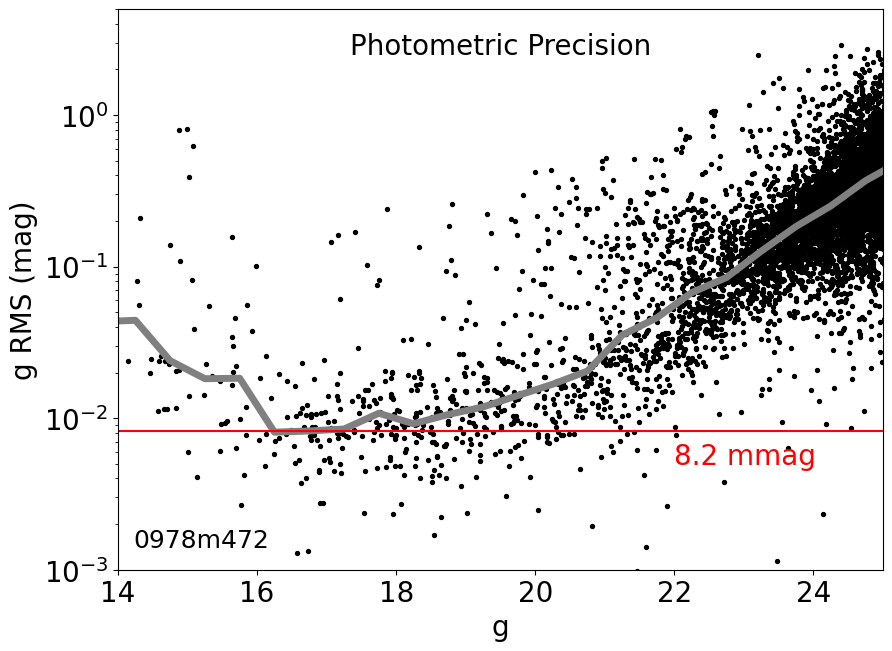}
\caption{Example of how the photometric precision is calculated using the $g$ RMS scatter of objects in the 0978m472 brick. The gray line indicates median values in 0.5 mag bins, and the derived precision of 8.2 mmag is shown by the red horizontal line.}
\label{fig:photrms_example}
\end{figure}
\end{center}

Initially, we added the best ten exposures (across all bands) using the ten highest $\Sigma({\rm S/N})$ values, essentially just the \teff $\times$covering fraction. While \teff~is a relative quantity comparing values with band-specific transparencies and backgrounds, the reference FWHM is constant. As the characteristic FWHM changes with band ($griz$$\approx$1.07/0.97/0.91/0.86\arcsec~in our data), the redder bands are slightly favored in this selection. 
In the next step, we added the best remaining exposure per band, using the top $\Delta\Sigma({\rm S/N})$ for the remaining exposures in the band. Iteration continued until the 50 exposure limit has been reached.

\begin{center}
\begin{figure*}[ht!]
\includegraphics[width=0.33\hsize,angle=0]{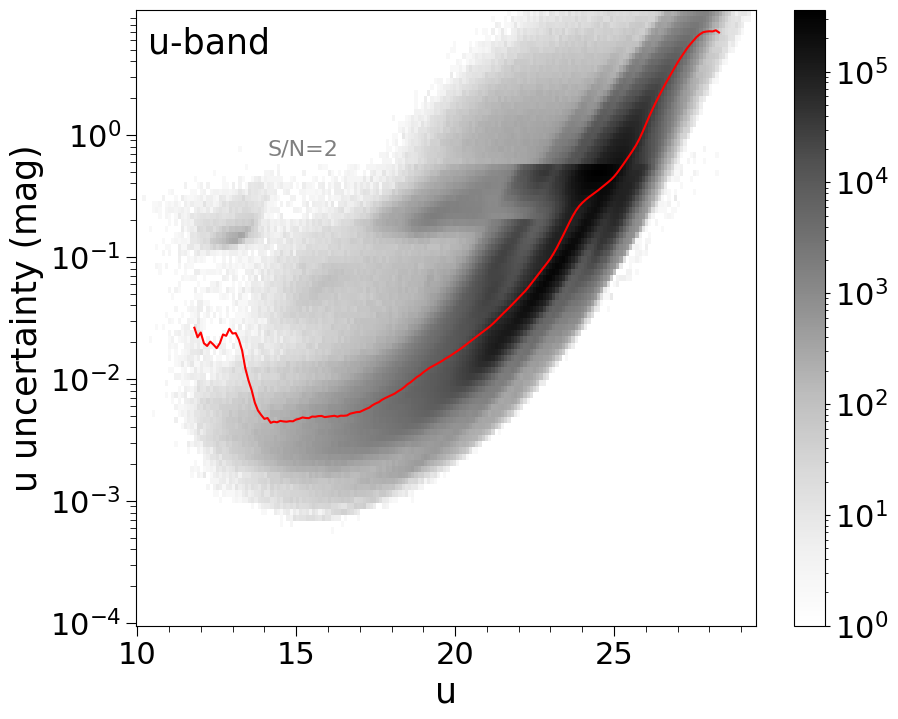}
\includegraphics[width=0.33\hsize,angle=0]{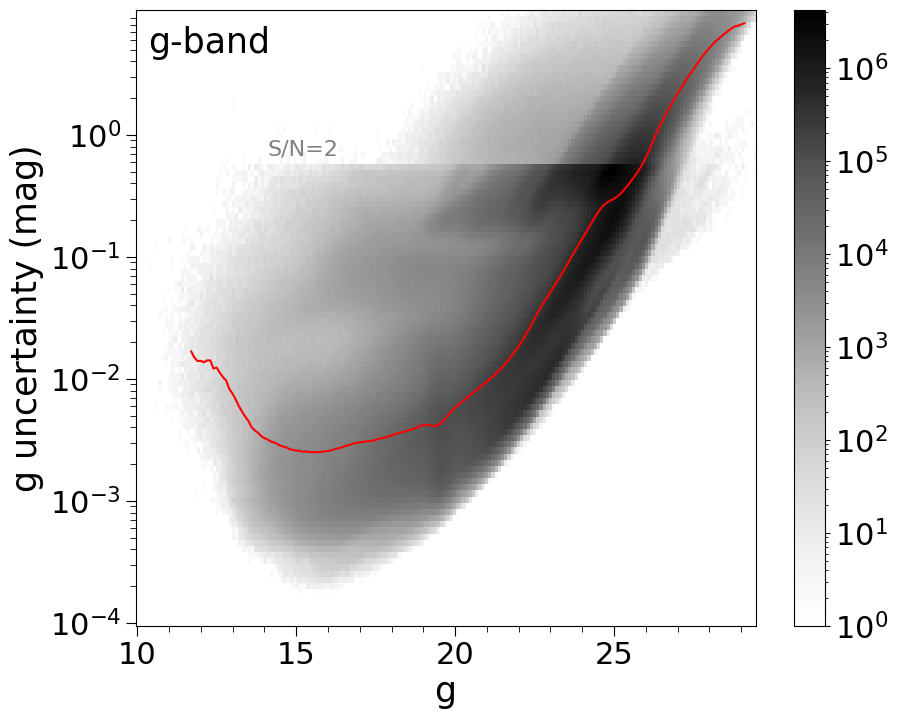}
\includegraphics[width=0.33\hsize,angle=0]{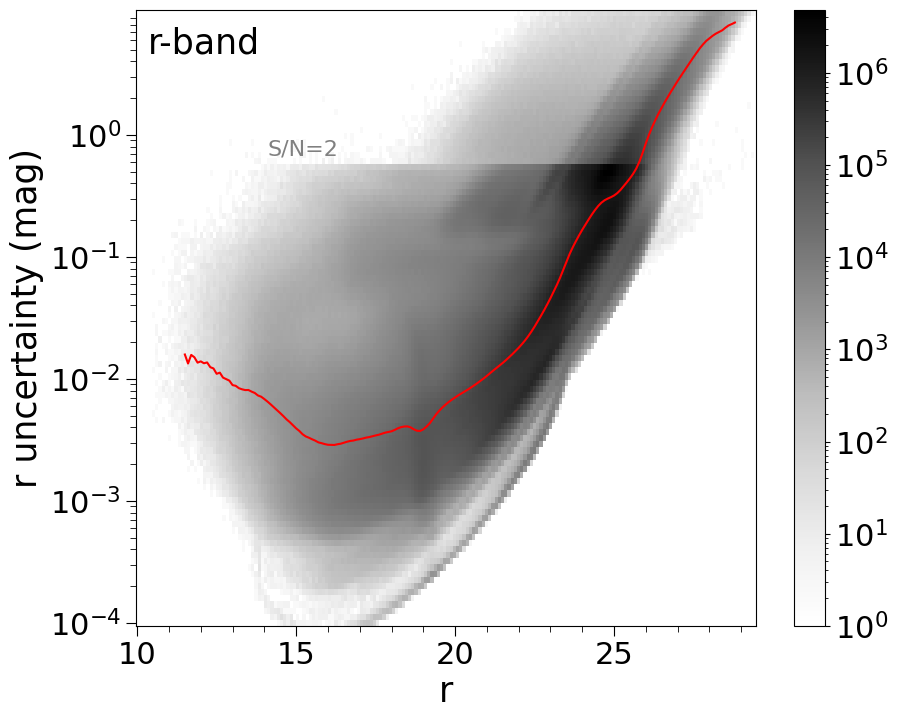} \\
\includegraphics[width=0.33\hsize,angle=0]{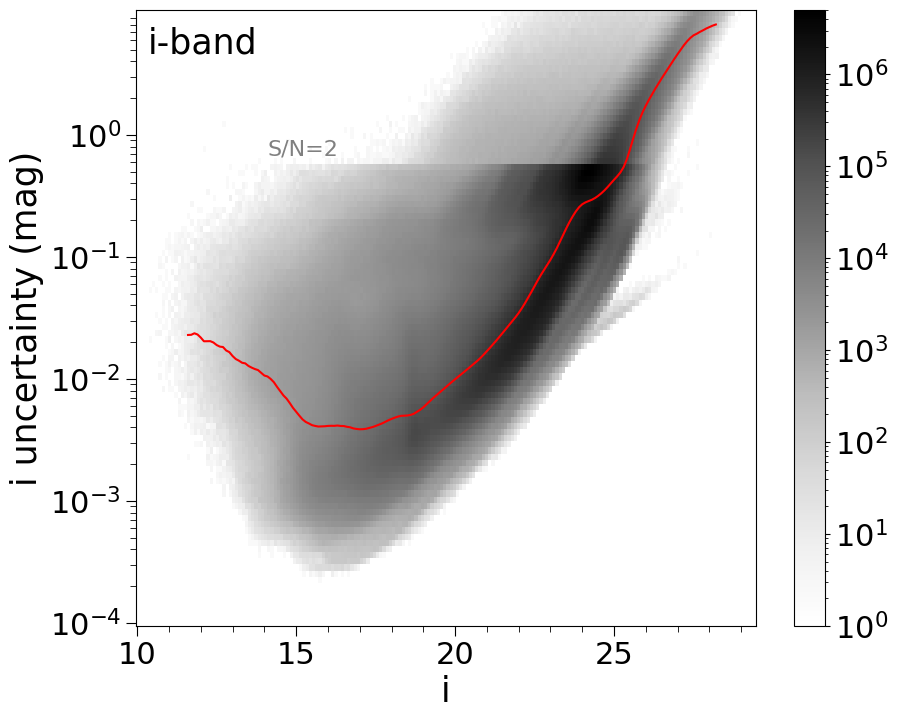}
\includegraphics[width=0.33\hsize,angle=0]{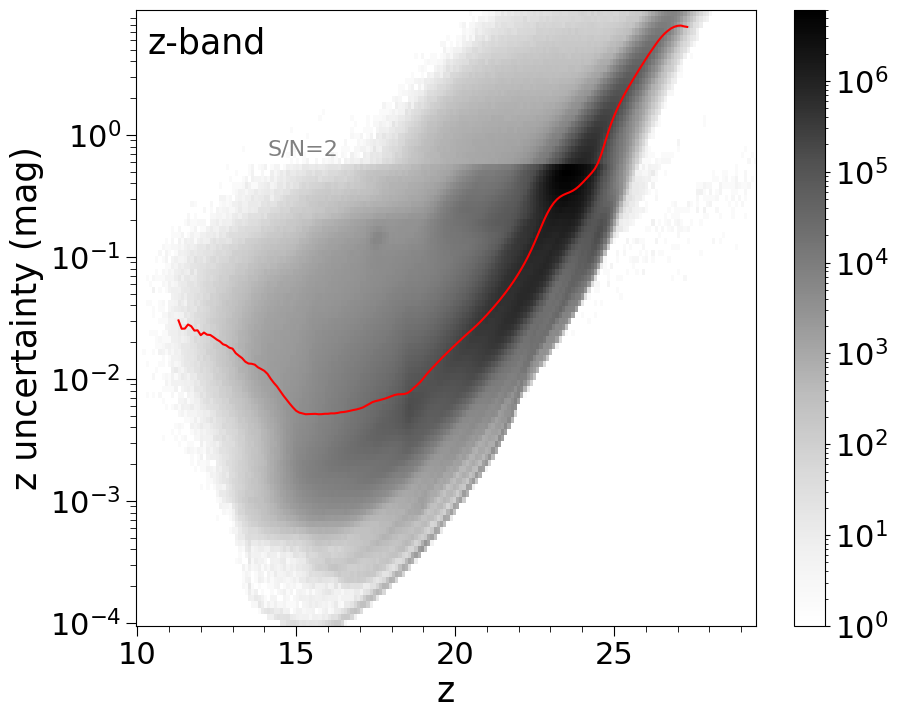}
\includegraphics[width=0.33\hsize,angle=0]{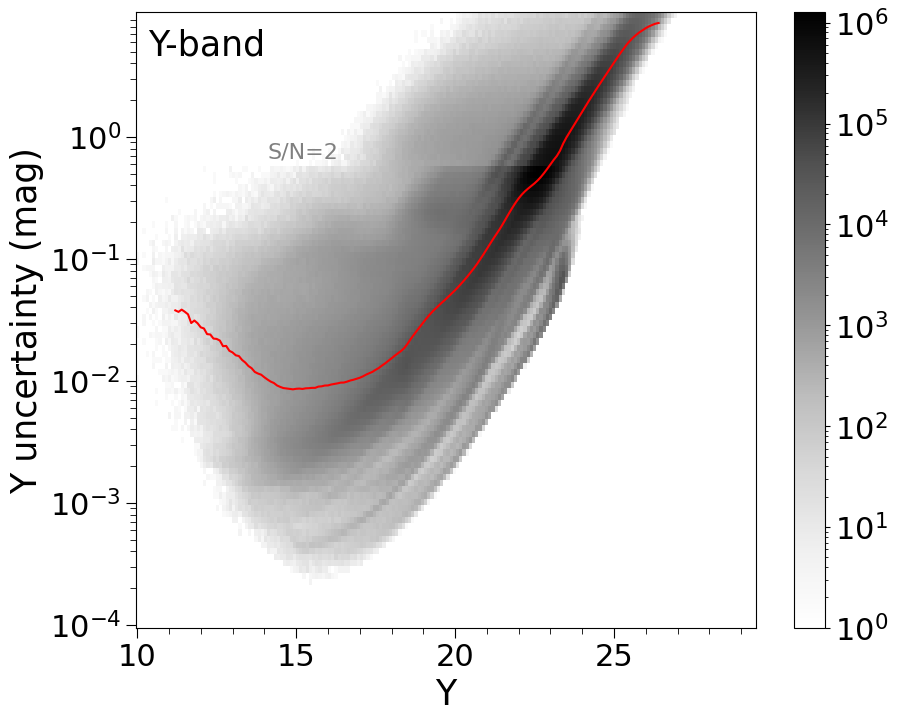} \\
\caption{Distribution of mean photometric uncertainties for all objects in the $ugrizY$ bands. The red line shows the mean uncertainty as a function of magnitude. The ``break'' at an uncertainty of 0.543 corresponds to S/N=2 and is the limit for single-exposure ALLSTAR measurements. All values above this line come from \allframe forced measurements.
}
\label{fig:photerr}
\end{figure*}
\end{center}

When $\Delta\Sigma({\rm S/N})$ values are calculated for a given band, the order in which the exposures are added matters. Therefore, we developed an iterative sub-algorithm that attempts to maximize $\Sigma({\rm S/N})$ with the fewest number of exposures. In the first iteration, the exposures are sorted by \teff. $\Delta\Sigma({\rm S/N})$ is evaluated for each exposure in turn, and the summed \teff~map is updated.  As a result, the $\Delta\Sigma({\rm S/N})$ of a given exposure is relative to all of the (presumably ``better'') exposures that came before it. On the second and subsequent iterations, the exposures are sorted by $\Delta\Sigma({\rm S/N})$ from the previous iteration. The procedure is repeated until the order stops changing or ten iterations have been reached.
The selection is evaluated separately within each available band after the initial ten-exposure selection, so that every represented band continues to receive high-value exposures. Nevertheless, the current scheme does not explicitly optimize or require matched spatial coverage between different bands; users requiring uniform multi-band samples should therefore inspect the per-band coverage information.

\subsubsection{Joint Catalogs}
\label{subsubsec:jointcats}

Not all exposures are included in the forced photometry analysis, especially for central bricks, and some bright stars are saturated in the deeper exposures used for the forced photometry.
Therefore, we create a final product called ``joint'' catalogs that combines the forced photometry with single-epoch photometry of exposures that were not included in the forced analysis.  We performed tests to confirm that there are no systematic offsets between the \allstar and \allframe photometric measurements for the same image. With no offset assured, these two sets of photometric values can be safely merged together.

To merge the two sets of photometry, a final set of unique objects is required. Objects not included in the forced photometry catalog can come from areas not covered by the chips in the forced analysis but covered by the rest of the chips, or from objects that are so bright that they are saturated in the forced photometry images (or coadd image). 
A special check is performed for possible ``missing measurements'' in chips that are included in the forced analysis, because the exposure-level \allstar catalog can contain reliable photometric measurements for some sources even if they are not contained in the forced photometry master list (e.g., bright stars).
During the merging process, we start with the forced photometry object list and catalog of measurements. We then loop through all of the ``extra'' exposure-level chips, cross-match sources using $\alpha$ and $\delta$ (using 1\arcsec~radius) to the current object list, and add any matches to the measurements table. Any remaining sources without a match are added as new rows to the object list.  

At the end of the cross-matching and merging process, a flux-weighted average magnitude is calculated for each object and band separately (\texttt{XMAG}; e.g., \texttt{GMAG} and \texttt{ZMAG}). In calculating these values, the shallower exposure-level measurements are downweighted relative to the generally deeper forced-photometry measurements. This prevents low-precision measurements from shallow exposures from degrading otherwise reliable forced photometry while still allowing useful exposure-level measurements to contribute to the final average.

\begin{deluxetable*}{lccccccc}
\centering 
\tablecaption{DELVE-MC Survey Performance Statistics\label{table:stats}}
\tablecolumns{8}
\tablewidth{500pt}
\tablehead{
  \colhead{Property}  & \colhead{all} & \colhead{$u$}  & \colhead{$g$}  & \colhead{$r$}  & \colhead{$i$}  & \colhead{$z$} & \colhead{$Y$}
}
\startdata
Exposures  &  33968 & 1159 & 7911 & 8982 & 7713 & 7022 & 1881  \\
Seeing (\arcsec)  & 0.99 & 1.19 & 1.12 & 0.98 & 0.93 & 0.88 & 0.85  \\
Airmass  & 1.34 & 1.33 & 1.36 & 1.36 & 1.34 & 1.33 & 1.27  \\
Depth (95\%)  & 25 & 25.9 & 25.7 & 25.3 & 24.8 & 24.3 & 24.3  \\
Depth (10$\sigma$)  & 23.5 & 23.4 & 24.3 & 24.2 & 23.7 & 22.7 & 21.5  \\
Completeness (50\%) & 25 &  24.2 & 25.2 & 25.0 &24.5 & 23.9 & 23.9 \\
Coverage (deg$^2$) &  2{,}170 & 349 & 2{,}167 & 2{,}168 & 2{,}164 & 2{,}123 & 1{,}059 \\
Astrometric precision (mas) &  11 &  24 &  17 &  11 &  10 &  10 &  10\\
Photometric precision (mmag) & 11 & 14 & 10.7 & 9.8 & 11.3 & 11.6 & 10.0
\enddata
\end{deluxetable*}

The released catalog contains measurements in the six broadband filters included in the DELVE-MC processing ($ugrizY$), although their spatial coverage is highly unequal. The new DELVE-MC observations were obtained in $gri$, while the $u$, $z$, and $Y$ measurements come from archival observations. As quantified above, $u$ and $Y$ cover only 349 and 1,059~deg$^2$, respectively, compared with more than 2,100~deg$^2$ in each of $griz$. We therefore show $griz$ in the principal footprint-wide calibration and depth maps, while presenting all six included filters in figures for which a direct global distribution remains informative (Figures~\ref{fig:photerr}, \ref{fig:lumfunction}, and \ref{fig:completeness}).

\begin{center}
\begin{figure*}[t]
\includegraphics[width=0.48\hsize,angle=0]{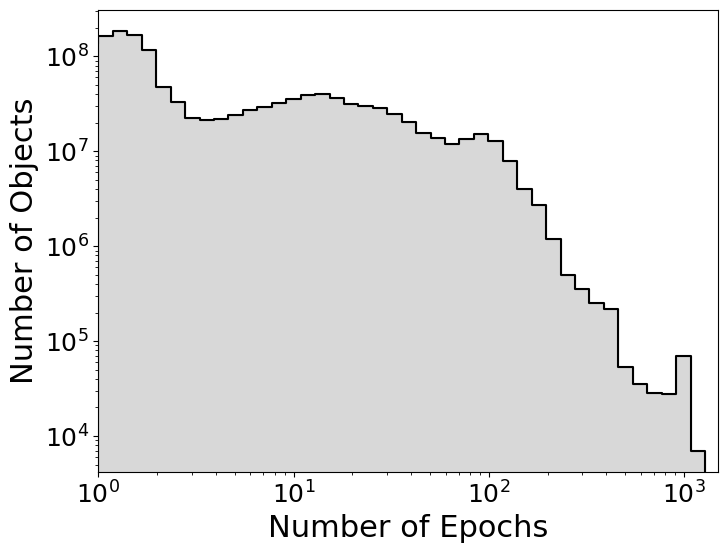}
\includegraphics[width=0.48\hsize,angle=0]{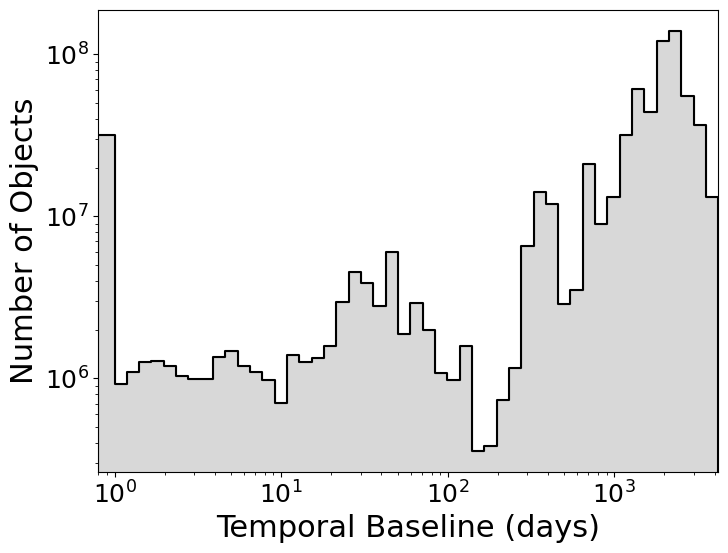}
\caption{Object statistics. ({\em Left}) Number of detections and ({\em Right}) temporal baseline in days per object.}
\label{fig:nepochs}
\end{figure*}
\end{center}

The \texttt{depthflag} column in the \texttt{object} table can be used to ascertain if exposure-level photometry (\texttt{depthflag}=1), forced photometry (\texttt{depthflag}=2), or both are used in the photometry of an object (\texttt{depthflag}=3).  In the \texttt{meas} table, the \texttt{forced} binary column indicates if the photometric measurement was from the exposure-level (\texttt{forced}=0) or forced analysis (\texttt{forced}=1).

Three final joint files are created for each brick similar to the forced-photometry files: {\tt object} (object-level information), {\tt meas} (epoch or exposure-level measurements), and {\tt meta} (metadata) files. The final joint {\tt object} catalogs were cross-matched with \Gaia\ DR3 \citep{GaiaDR3}, and selected \Gaia\ DR3 astrometric and photometric quantities are included in the {\tt gaia\_*} columns. This catalog-level cross-match is distinct from the use of \Gaia\ DR2 as the astrometric reference for the exposure-level calibration described above.




\section{Description and Achieved Performance of the Final Catalog}
\label{sec:performance}

The DELVE-MC catalogs presented in this paper are publicly released as part of the third DELVE data release (DELVE DR3). Earlier DELVE releases focused primarily on the DELVE-WIDE and DELVE-DEEP survey components and dwarf-galaxy searches \citep{Drlica-Wagner:2021,DrlicaWagner2022}, while DR3 includes the first public release of the DELVE-MC forced-photometry catalogs and associated multi-epoch measurements. All data products are available through the NOIRLab Astro Data Lab.

The DELVE-MC catalog covers more than 2170 square degrees of the sky and catalogs over 1.2 billion unique objects (\autoref{fig:map}).  It includes more than 19 billion individual measurements --- 4.6 times the number in SMASH DR2 --- from 33,120 exposures spanning over 10 years. Over 560 million objects have data in three or more bands and can be used to construct color-color diagrams.
A summary of performance statistics is given in \autoref{table:stats}.



Maps of the 95th percentile depths are shown in \autoref{fig:depth}. The median depths are 25.7, 25.3, 24.8,and 24.3 mag in the $g, r, i$, and $z$ bands.
For each chip image, the astrometric RMS is calculated using {\em Gaia} DR2 reference stars ($G<$20 and measured proper motions below 200 mas yr$^{-1}$) whose positions are propagated using proper motions to the epoch of the exposure and compared to the WCS image solution. {\em Gaia} DR2 was retained because the DELVE-MC \delvered processing had already begun before {\em Gaia} EDR3 became available; retaining the same reference system also ensured uniform processing of the full archival data set. The 10.5 mas value reported here is therefore the residual scatter of individual DECam chip solutions relative to the {\em Gaia} reference frame, rather than the intrinsic {\em Gaia} positional uncertainty. The improved {\em Gaia} EDR3 proper motions would change the propagated reference positions by only a few milliarcseconds for most calibration stars over the DELVE-MC time baseline, smaller than the measured single-exposure residual, although future reductions can benefit from the newer reference catalog.
The left panel of \autoref{fig:wcsrms} shows median WCS RMS for each brick, while the histogram for all chips is in the right panel with a median of 10.5 mas.
The photometric precision is estimated by calculating the median RMS scatter of stars in 0.5 magnitude bins and finding the minimum threshold value, which is normally around 16--18th mag (\autoref{fig:photrms_example}), although it varies from band to band.
The photometric precision in each brick has a median of $\sim$11~mmag independent of the band, with some variation as a function of position (and generally the errors are somewhat larger in the more crowded
central regions than the sparser periphery; see \autoref{fig:photrms} in Appendix \ref{sec:appendix}). 
The $z$-band, which was not included in the DELVE-MC observing campaign, is ``patchy'' because there are not enough exposures to calculate the RMS scatter in certain regions.
Although an effort has been made to improve the photometric calibration in crowded and dusty regions by using more accurate extinction corrections \citep[e.g., the RJCE method;][]{Majewski2011} and newer reference catalogs in the southern sky (e.g., Skymapper DR1, \citealt{Wolf2018} and ATLAS-Refcat2, \citealt{Tonry2018}), some issues remain. We advise some caution when using the photometry in the very crowded and high extinction regions (e.g., the LMC bar).

\begin{center}
\begin{figure*}[t]
\includegraphics[width=0.49\hsize,angle=0]{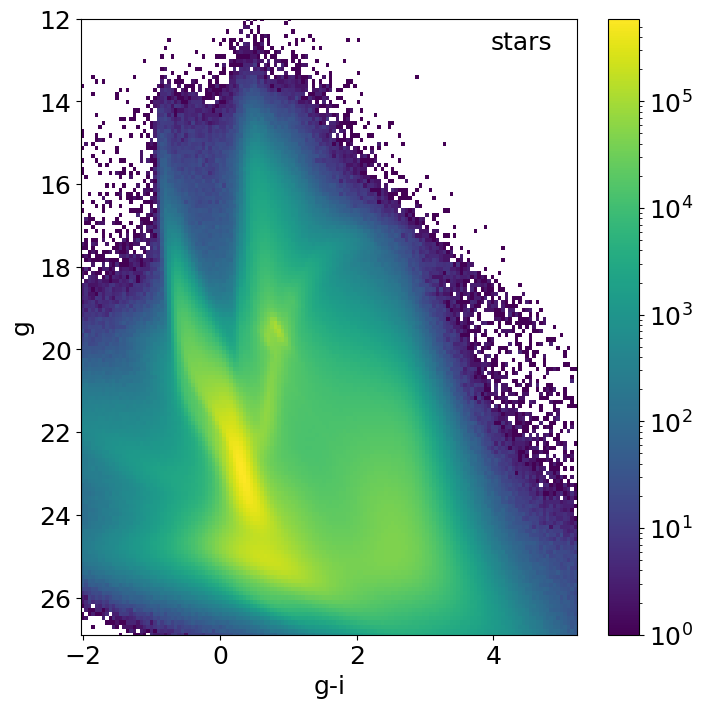}
\includegraphics[width=0.49\hsize,angle=0]{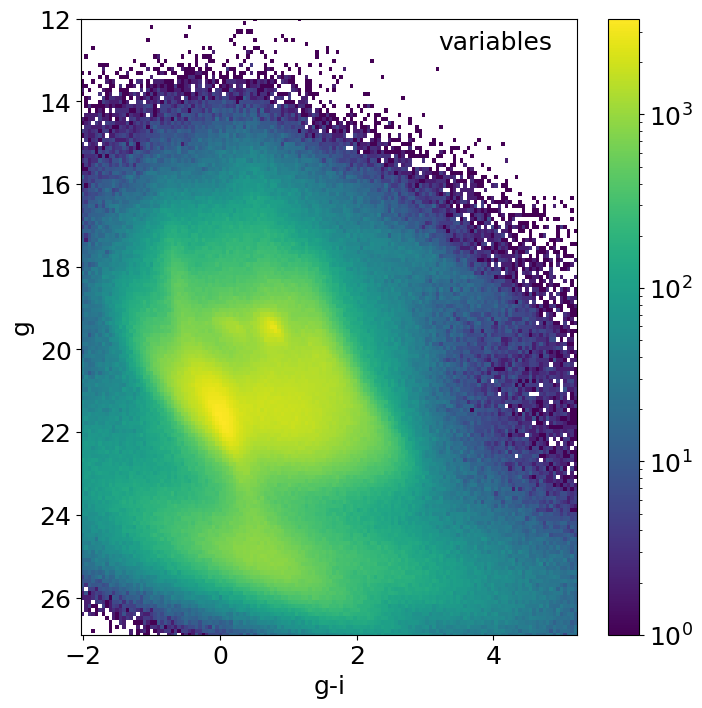}
\caption{Color Magnitude Diagrams of all ({\em left}) stars (226M), and ({\em right}) photometrically variable objects (4.5M) at the 10$\sigma$ level.}
\label{fig:cmds}
\end{figure*}
\end{center}

\begin{center}
\begin{figure}[t]
\includegraphics[width=0.99\hsize,angle=0]{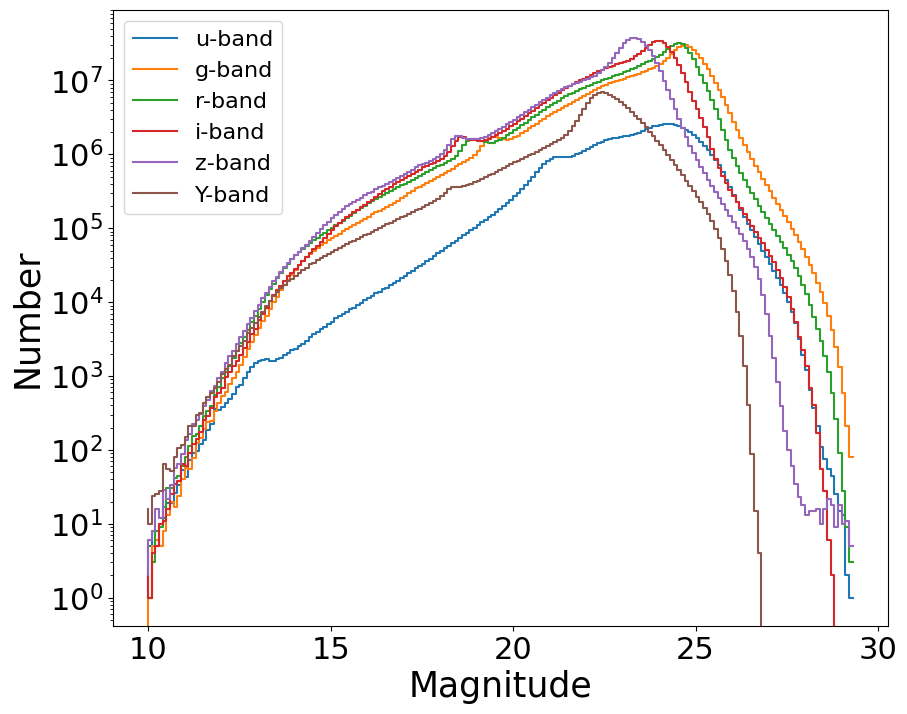}
\caption{Luminosity function of stars in all six $ugrizY$ bands. The bump around $\sim$19--20 mag is due to Magellanic red clump stars.}
\label{fig:lumfunction}
\end{figure}
\end{center}

\autoref{fig:photerr} shows the distribution of photometric uncertainties as a function of magnitude for all $ugrizY$ bands. The characteristic break near an uncertainty of 0.54 mag corresponds to the S/N=2 limit of the single-exposure ALLSTAR measurements, while fainter measurements are recovered through the deeper \allframe forced photometry. 
A weaker secondary locus near an uncertainty of $\sim$0.2 mag is visible in several bands and is most prominent in the $u$ band. Unlike the break at 0.543 mag, this feature does not correspond to a known signal-to-noise threshold or other explicit selection applied during catalog construction.


Most of the DELVE-MC region is covered by multiple exposures, which creates 
a valuable time-series dataset (\autoref{fig:map}). The histogram of the number of exposures per brick is shown in \autoref{fig:nexp_bricks}.  While it peaks at 24 exposures, there are many bricks with hundreds of exposures and even some with over 1200 exposures. Roughly 150 million objects have 30 or more photometric measurements (\autoref{fig:nepochs}), which should be enough to detect and classify many types of variable stars reliably \citep[e.g.,][]{Stringer2021}. In addition, 513 million have temporal baselines spanning more than 1000 days, making them ideal for studying long-term variability in objects like Long Period Variables (LPVs) and Active Galactic Nuclei (AGN).
\autoref{fig:cmds} shows color–magnitude diagrams for all stellar sources (left panel) and for photometrically variable objects identified at the 10$\sigma$ level (right panel). Distinct variable-star populations, including RR Lyrae (clumps at $g$$\sim$19 mag), long-period variables (the red AGB stars around $g$$\sim$17--18), and eclipsing binaries (along the main-sequence, $-$0.5 $\lesssim$ $g-i$ $\lesssim$ 1.0), are visible, demonstrating the power of the DELVE-MC time-series dataset.

\begin{center}
\begin{figure*}[t]
\centering
\includegraphics[width=\hsize,angle=0]{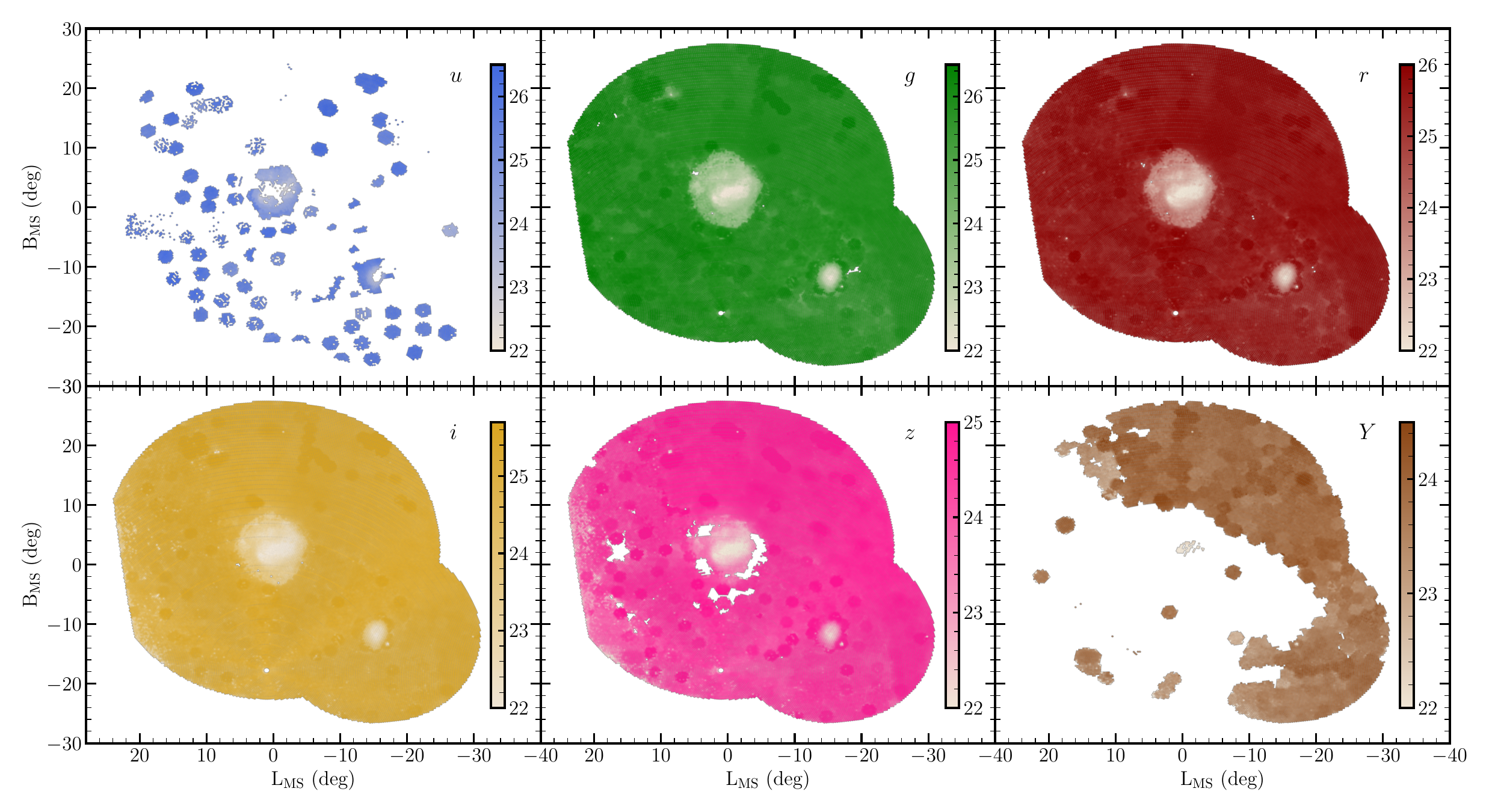}
\caption{DELVE-MC 50\% completeness magnitude. Estimated using the peak of the luminosity function for stellar sources in each band, and using exclusively the forced photometry step. SMASH AST results have been used to derive an empirical relation that depends on the crowding of the field.}
\label{fig:completeness}
\end{figure*}
\end{center}


The total stellar luminosity functions for the $ugrizY$ bands are shown in \autoref{fig:lumfunction}. The turnover at faint magnitudes reflects the declining completeness of the survey and provides the basis for the empirical completeness estimates discussed below.



Although dedicated DELVE-MC artificial-star tests, in which artificial stars are injected into the images and the photometry is reprocessed to characterize completeness and photometric uncertainties over a range of colors and magnitudes, were not performed for this data release and are not currently planned, we use existing SMASH artificial-star tests to estimate the DELVE-MC completeness. Specifically, we use the relationship between the peak of the point-source luminosity function and the completeness levels measured by the SMASH artificial-star tests. We apply this method exclusively to the forced-photometry catalog, which provides the deepest DELVE-MC photometry. \autoref{fig:completeness} shows the resulting estimated 50\% completeness maps in the six bands. For each processed area, the point-source luminosity function \citep[i.e., $dN/d{\rm mag}$;][]{Conn2007,Harris2016} was fit with a sigmoid function, and the SMASH artificial-star tests were used to relate the fitted luminosity-function turnover to the 50\% completeness limit.



DELVE-MC is being released through the NOIRLab's Astro Data Lab\footnote{\url{https://datalab.noirlab.edu}} \citep{Fitzpatrick2016,Nikutta2020}.  The database tables can be accessed via direct SQL queries using the Data Lab client software (Python) or via a TAP service.\footnote{\url{https://datalab.noirlab.edu/tap}} The column descriptions can be viewed using the Data Lab query interface page.\footnote{\url{https://datalab.noirlab.edu/query.php}}  Data analysis and exploration can be performed using the Astro Data Lab's Jupyter Hub Notebook server running next to the data which provides fast access.  A Simple Image Access (SIA) service for cutouts of exposure-level and CP-generated stacked images (not the custom \delvered coadded images) are available at \url{https://datalab.noirlab.edu/data-explorer}.

There are five DELVE-MC tables in the Astro Data Lab database:
\begin{itemize}
\item {\bf Brick}. Information for each unique brick (35,052 rows).
\item {\bf Chip}. Information for each unique chip image with zero-point, astrometric correction coefficients, and uncertainties ({\bf 2,038,678} rows).
\item {\bf Exposure}. Information for each exposure with zero-point values and uncertainties ({\bf 34,393} rows).
\item {\bf Object}. Information on each object with average photometric, astrometric and morphology values, and photometric variability indices (1,451,130,182 rows).
\item {\bf Measurement}. Information on each individual object measurement in the single-epoch images (25,944,585,949 rows).
\end{itemize}

\begin{center}
\begin{figure}[t]
\includegraphics[width=1.0\hsize,angle=0]{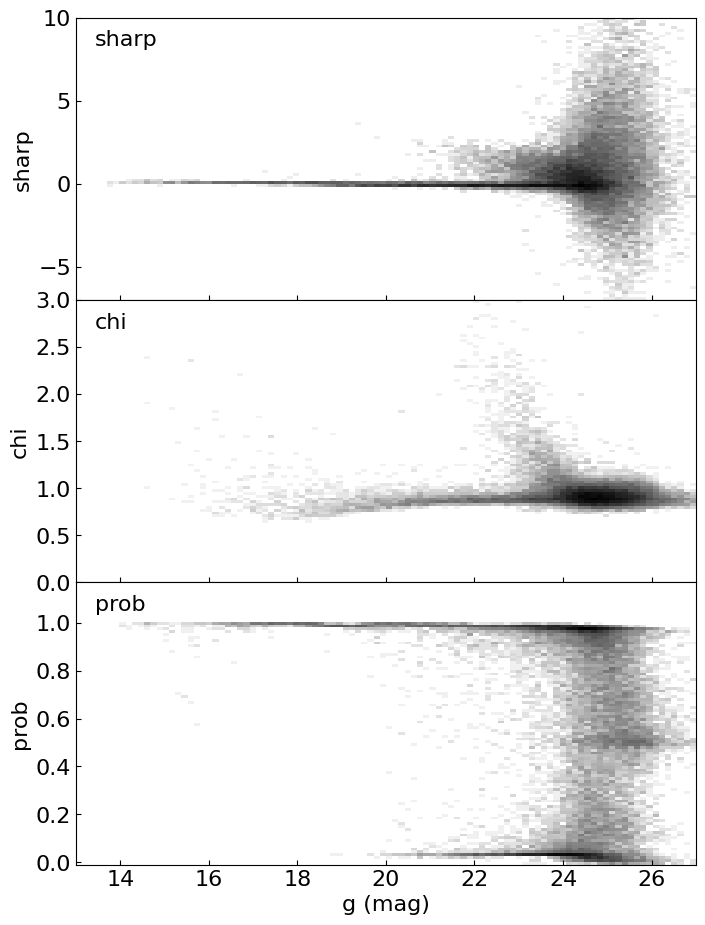}
\caption{Quality metrics for four bricks near ($\alpha$,$\delta$)=(101.8\degr,$-$52.25\degr) with \texttt{depthflag}$>$1 (sources with forced photometry). ({\em Top}) \texttt{sharp}, ({\em middle}) \texttt{chi}, and ({\em bottom}) \texttt{prob}.}
\label{fig:qaplots}
\end{figure}
\end{center}

\section{Constructing Reliable Star Samples}
\label{sec:stargalaxy}

To make the best scientific use of the DELVE-MC data to study
diffuse stellar populations, it is important to have a
reliable star catalog (including star-galaxy separation) especially in the MC periphery where the relative number of background galaxies to foreground stars is significant. The DELVE-MC catalog contains three morphology and quality metrics that can be used to separate stars from galaxies. The (a) \daophot \texttt{chi} which is a robust estimate of the scatter in the residual flux (data minus model PSF profile) divided by the expected scatter, the (b) \daophot sharpness (\texttt{sharp}) value which measures how much an object's light profile deviates from the PSF, and (c) the \SExtractor stellaricity metric (\texttt{prob}, also known as \texttt{CLASS\_STAR}) which gives the probability (from 0 to 1) of the source being a star using a neural network method.
\autoref{fig:qaplots} shows these distributions for four bricks in the LMC northeastern periphery.
In addition, the Source Extractor \texttt{FWHM} value can be used to identify extended objects such as galaxies.
Finally, the number of detections can be used to remove spurious sources.

\begin{figure*}[t!]
\centering
\includegraphics[width=0.40\textwidth]{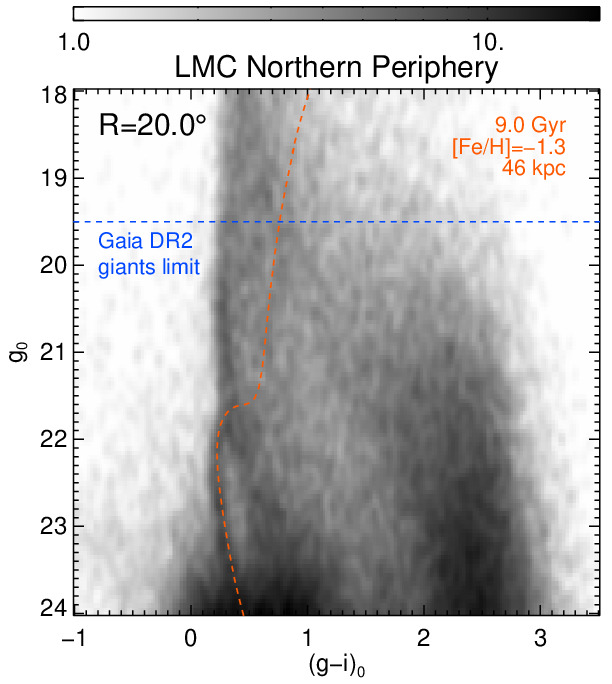}
\includegraphics[width=0.48\textwidth]{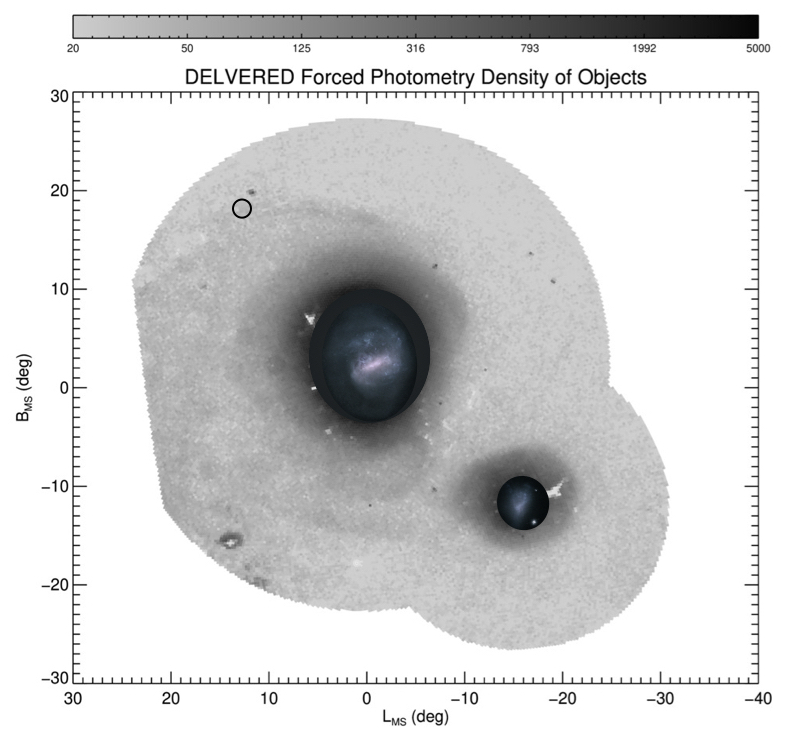}
\caption{
\textit{Left:} A DELVE-MC color-magnitude diagram in the LMC periphery (at $R$=20\dgr indicated by the black circle in the right panel) demonstrates that MSTO stars accessible to DELVE are $\roughly50$ times more numerous than the brighter, proper-motion-selected RGB stars observed by \Gaia.
\textit{Right:} DELVE-MC map of old main sequence turnoff starcounts per brick in the Magellanic periphery. 
}
\label{fig:mcmapcmd}
\end{figure*}

An example stellar selection is,


{\par\small\begin{align}\begin{split}
\var{stars} = ~
& (\var{chi} < 3) ~\& ~(|\var{sharp}|<1) \\
& \&~ (\var{prob} > 0.7) ~\& ~ (\var{ndet} > 3)
\end{split}\end{align}}

Note that \texttt{prob} is only calculated by \SExtractor for the forced photometry and is not available for exposure-level \allstar measurements.  We have found that there are bricks in the main bodies of the MCs (dominated by MC stars) where the \texttt{prob} parameter can be unreliable and remove many stars.  Therefore, we suggest caution in using \texttt{prob} in these areas.




\section{Example Science Cases}
\label{sec:science}

There are several science projects that the DELVE Collaboration is pursuing with the DELVE-MC catalog that tie in to high impact science topics with the Clouds.  Below we describe a few of them: extended stellar substructures in the periphery, SFHs and stellar variability.



\subsection{Stellar Substructures}
\label{sec:science_stellarsubstructures}

The stellar periphery of the Magellanic Clouds has been a topic of interest over the last couple of decades. Recent all-sky and wide-field surveys such as \Gaia and other ground-based surveys have enabled low-surface-brightness searches around the periphery. Both DECam and \Gaia are able to overcome the MW foreground contamination that has traditionally plagued low-surface-brightness searches around the MCs. \Gaia can use its excellent astrometric results to separate out stars based on their proper motion \citep{Belokurov2019, Luri2021}. On the other hand, deep DECam photometry can reach the oldest main sequence turn off (MSTO) in a color-magnitude diagram, which is an area with a much higher number of MC stellar populations and these stars are in much higher contrast compared to other MW halo populations, as can be seen in \autoref{fig:mcmapcmd}.

The left panel of \autoref{fig:mcmapcmd} shows the DELVE-MC CMD of a 3.8 deg$^2$ region in the northeastern LMC periphery ($\alpha=101.8$\degr, $\delta=-52.1$\degr) at a distance of 20\dgr from the LMC center.  The old LMC main-sequence is clearly visible along with a by-eye best-fit PARSEC isochrone \citep{Girardi2001} of 9.0 Gyr, [Fe/H]$=-$1.3 and 46 kpc. 
The \Gaia DR2 limit for reliable selection of giants is also shown at $g$=19.5.

The right panel of \autoref{fig:mcmapcmd} shows a preliminary map of the main sequence turn off (MSTO) stellar density in the MC periphery.  The extended structure of the LMC disk and the main body of the SMC are clearly visible.  Some of the known substructures such as the northern LMC arm \citep{Mackey2016} and the southern hooks \citep{Mackey2018} are also visible.  A more sophisticated analysis of the MC stellar periphery, including a more careful foreground decontamination process and detailed structure analysis will be presented in \citet{Massana2026}.



\subsection{Star Formation Histories}
\label{sec:science_SFH}


One of the principal scientific drivers of DELVE-MC is to re-derive the spatially resolved SFHs of the Magellanic Clouds across their full spatial extent. The exquisite depth and contiguous coverage of the DELVE-MC dataset enable the construction of precise CMDs down to the oldest main-sequence turnoff (oMSTO) over thousands of square degrees.

While HST was initially considered essential for reaching the oMSTO in the MCs with sufficient precision \citep[e.g.,][]{Olsen1999,Holtzman1999}, subsequent work demonstrated that ground-based observations from good-seeing sites could achieve comparable depth—not only in the outskirts \citep[e.g.,][]{Gallart2004,Nidever2011}, but also in the more crowded inner disk \citep{Noel2007,Gallart2008} and even within the bar \citep{Monteagudo2018}. The first ground-based SFHs for the MCs were derived for relatively wide fields \citep{Harris2004,Harris&Zaritsky2009,Noel2009,Meschin2014,Monteagudo2018}, but these still relied on carefully chosen pencil-beam regions. A major advance came with large-area surveys capable of reaching the oMSTO, such as VMC \citep[e.g., SFH studies by][]{Rubele2012,Rubele2015,Rubele2018,Mazzi2021} and SMASH \citep[e.g.,][]{RuizLara2020,Massana2022,Sakowska2026}, which extended precise SFH measurements over more contiguous coverage.

DELVE-MC now enables these techniques to be pushed further to the full periphery of the Clouds, allowing tidal features, warps, and halos to be examined with the same star formation age resolution as the central regions.


A detailed SFH analysis will be presented in \citet{Oden2026}, including the full derivation methodology and scientific interpretation. Here we briefly describe the main process and initial results. The SFH analysis compares the observed CMDs with synthetic CMDs generated using the BaSTI stellar-evolution models \citep{Hidalgo2018}, spanning ages from 0.02 to 13.9 Gyr and metallicities from $-2.27 \leq \mathrm{[Fe/H]} \leq 0.12$. Synthetic-CMD fitting is a well-established approach that has been implemented in several SFH-recovery packages, including \texttt{MATCH} \citep{Dolphin2002}, \texttt{StarFISH} \citep{Harris&Zaritsky2009}, and \texttt{TheStorm} \citep{Bernard2015a,Bernard2018}. For the DELVE-MC analysis presented here, we use only \texttt{TheStorm}, which determines the linear combination of simple stellar populations (SSPs) that best reproduces the observed CMD.


Following established methodology from previous large-area SFH studies \citep[e.g.,][]{Bernard2015a,Bernard2018,RuizLara2020,Sakowska2024}, we partition each CMD into diagnostic ``bundles'' that isolate evolutionary phases with different sensitivities to age and metallicity. To reduce contamination, we apply standard DAOPHOT \citep{Stetson1987,Stetson1990,Stetson1994} and Source Extractor \citep{Bertin1996} quality cuts, and we incorporate a foreground MW model constructed from five representative DELVE–MC fields in the MC periphery. Through these steps, we are able to obtain a best-fit solution CMD that encodes the full star-formation history and age–metallicity relation (AMR) of each field.

\begin{figure}
\includegraphics[width=0.48\textwidth]{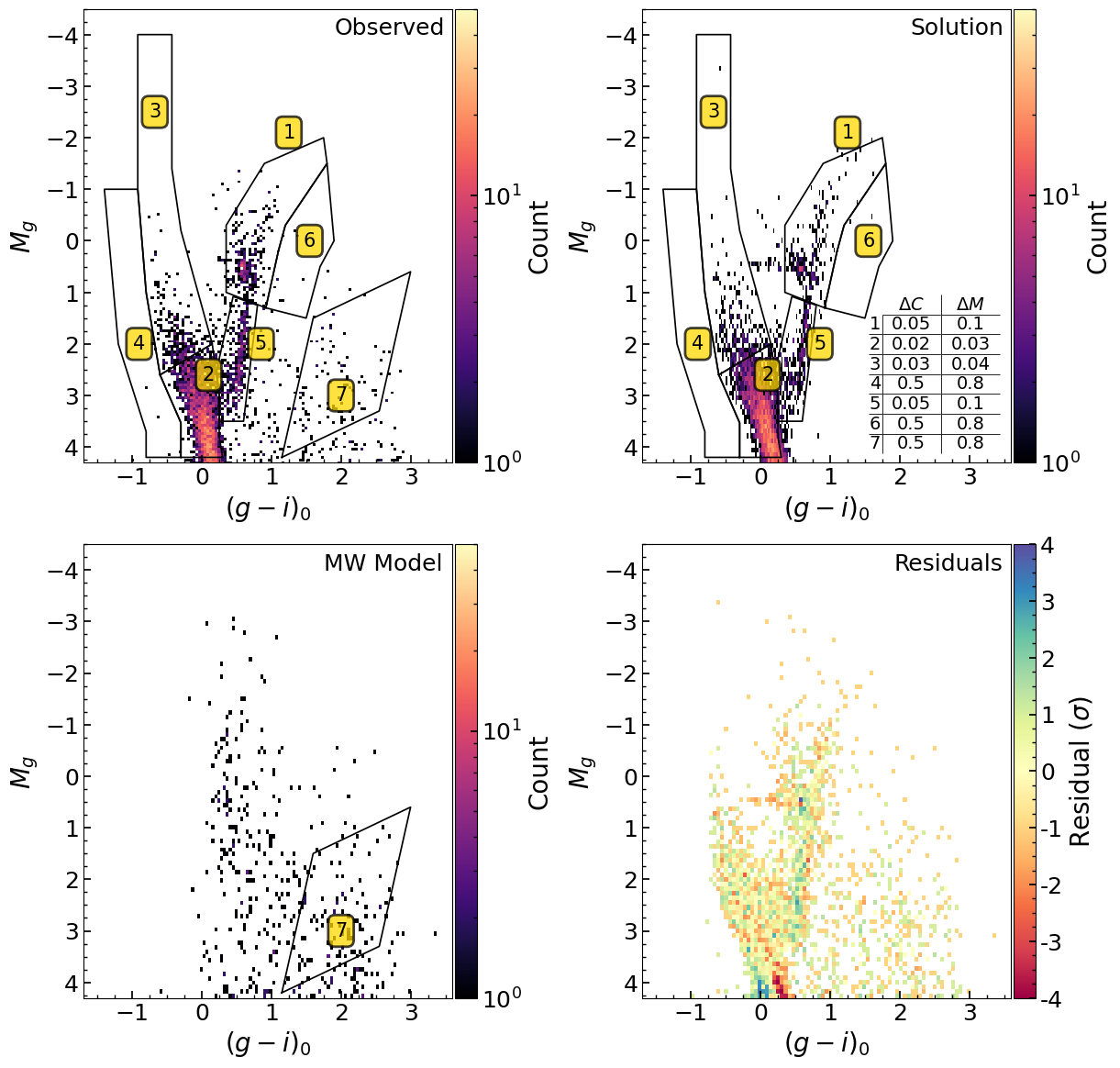}
\caption{Observed ({\em top left}), best-fit solution ({\em top right}), Milky Way field model ({\em bottom left}), and residual ({\em bottom right}) color–magnitude diagrams (CMDs) for an SMC DECam periphery field located at $(\alpha, \delta) = (0.29^{\circ}, -73.9^{\circ})$, corresponding to $R_{SMC} \sim 3.12^{\circ}$ and covering an area of $\sim$ 0.06 $\mathrm{deg}^2$. Black polygons delineate the CMD “bundle” regions used to construct the solution CMD, while the inset table summarizes the adopted bin sizes in color and magnitude for each bundle.} 
\label{fig:SFH_CMD}
\end{figure}

\begin{figure}
\includegraphics[width=0.47\textwidth]{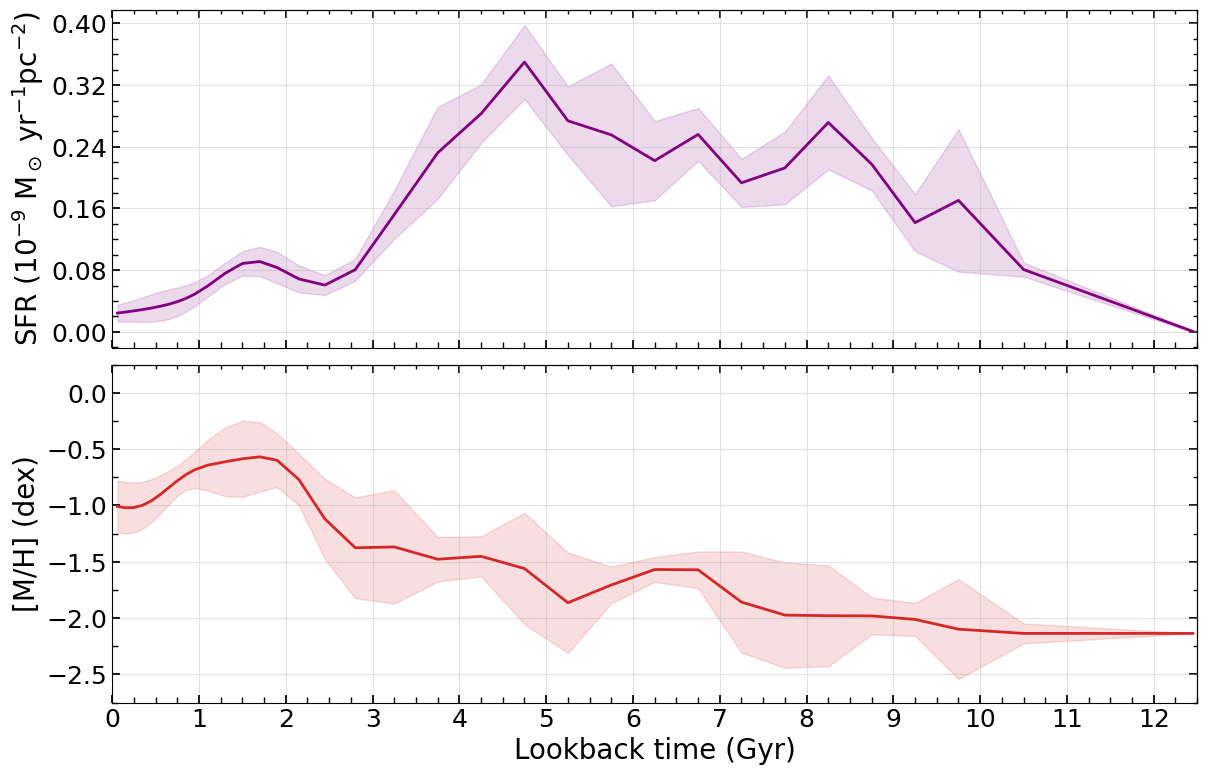}
\caption{Star formation history for the SMC periphery field shown in \autoref{fig:SFH_CMD}. {\em Top:} Star formation rate (SFR) as a function of lookback time, with the shaded region indicating the associated uncertainty. The field exhibits a prominent episode of enhanced star formation at intermediate ages ($\sim$5–8 Gyr), followed by a gradual decline toward older ages. {\em Bottom:} Age–metallicity relation (AMR) as a function of lookback time, showing a steady decrease in metallicity with increasing age, from $\mathrm{[M/H]} \sim -0.6$ at recent times to $\lesssim -2$ at early epochs. Shaded regions denote the uncertainties in the recovered metallicity.} 
\label{fig:SFH}
\end{figure}

As a demonstration, Figures~\ref{fig:SFH_CMD} and~\ref{fig:SFH} present the recovered SFH for a representative field in the southern SMC periphery. This field lies at a projected radius of $R_{SMC} \sim 3.12^{\circ}$, with equatorial coordinates $(\alpha,\delta) = (0.29^\circ, -73.9^\circ)$, and covers an area of approximately $0.06,\mathrm{deg}^2$.
\autoref{fig:SFH_CMD} shows a comparison of the observed,  solution, and residual CMDs for the best fit. \autoref{fig:SFH} shows the Star Formation Rate (SFR) and AMR for the field as a function of look-back time. The results reveal a prominent intermediate-aged population ($
\sim$5-8 Gyrs) with a negligible young population of stars, reminiscent of MC periphery fields where there exist a prominent age-gradient \citep{Harris2004,Gallart2008,Noel2009}. Systematic mapping of such variations across the full DELVE-MC footprint will enable tests of interaction-driven star formation models \citep[e.g.,][]{Besla2012,Erkal2019} and comparisons to cosmological
simulations of dwarf galaxy evolution \citep[e.g.,][]{Elbadry2018,Munshi2021,Tan2025}.


Ultimately, the SFHs derived from DELVE-MC will provide the most comprehensive temporal record of star formation across the entire Magellanic System to date, spanning from the crowded central regions to the diffuse periphery. These results will complement spectroscopic kinematical analyses of the Clouds and provide robust age–metallicity models for stellar populations. Such models will be invaluable not only for reconstructing the assembly history of the MCs, but also for applying population-dependent corrections to standard-candle distance indicators such as red clump stars --- thereby offering a holistic view of how the Clouds assembled and evolved in their current environment.


\subsection{Variable Stars}
\label{sec:variables}


The DELVE-MC dataset contains a wealth of temporal information that enables the construction of high-quality light curves for hundreds of millions of sources across the Magellanic System. This time-domain capability is particularly powerful for studying pulsating variable stars in the MCs, including RR Lyrae, Classical and Type II Cepheids, and $\delta$ Scuti stars, as well as long-period variables and eclipsing binaries. With typical bricks containing tens to thousands of epochs spanning baselines of up to $\sim$10 years, DELVE-MC provides one of the deepest wide-area optical time-series datasets ever assembled for the Magellanic Clouds. It complements higher-cadence but shallower surveys such as OGLE-IV by extending variability studies to fainter magnitudes and much larger spatial scales in the Magellanic periphery
\citep{Udalski2015,Soszynski2016}.

Roughly 150 million objects in the catalog have 30 or more photometric measurements, sufficient for robust variability detection and period determination for many classes of variables. The joint object table includes several photometric variability metrics (e.g., RMS scatter, $\chi^2$, and related indices; see \citealt{Nidever2021b}) that can be used to efficiently identify candidate variable stars, while the full measurement table enables detailed light-curve modeling and classification. \autoref{fig:cmds} (right panel) shows the color–magnitude diagram of photometrically variable sources, which highlights the loci of classical instability-strip populations as well as red long-period variables and RR Lyrae.

RR Lyrae and Cepheids are of particular importance, as they serve as precise standard candles for mapping the 3-D structure of the Clouds \citep[e.g.,][]{Haschke2012a,Haschke2012b,Skowron2016}. DELVE-MC’s depth allows RR Lyrae to be detected well into the periphery of both the LMC and SMC, extending distance mapping far beyond the central regions traditionally covered by surveys such as OGLE. The combination of wide spatial coverage and uniform photometry will enable systematic studies of line-of-sight depth, warps, and tidal distortions across the full Magellanic System, and will provide an independent check on structures inferred from red clump stars and other tracers.

For RR Lyrae and Cepheid variables, the DELVE-MC multi-epoch photometry should enable distance measurements with typical precisions of a few percent when combined with template fitting. At MC distances, this corresponds to line-of-sight distance uncertainties of order $\sim$1 kpc for individual stars, sufficient to probe the depth structure of the Clouds and measure relative distance variations across tidal features, stellar streams, and diffuse substructures. The large number of variable stars expected across the DELVE-MC footprint will additionally allow ensemble averaging to reveal coherent 3-D structures at even finer spatial scales.

Beyond distance measurements, the spatial distributions and population properties of different classes of variables encode valuable information about the formation history of the Clouds. RR Lyrae trace the oldest ($\gtrsim$10 Gyr) populations, while Cepheids and $\delta$ Scuti stars probe recent and intermediate-age star formation \citep[e.g.,][]{Catelan2015,Ripepi2017,Soszynski2015,Soszynski2022}. Mapping these populations over thousands of square degrees will allow direct comparisons between ancient halos, intermediate-age disks, and young star-forming structures, and will help constrain the timing and impact of past LMC–SMC interactions.

\begin{center}
\begin{figure}[t]
\includegraphics[width=1.0\hsize,angle=0]{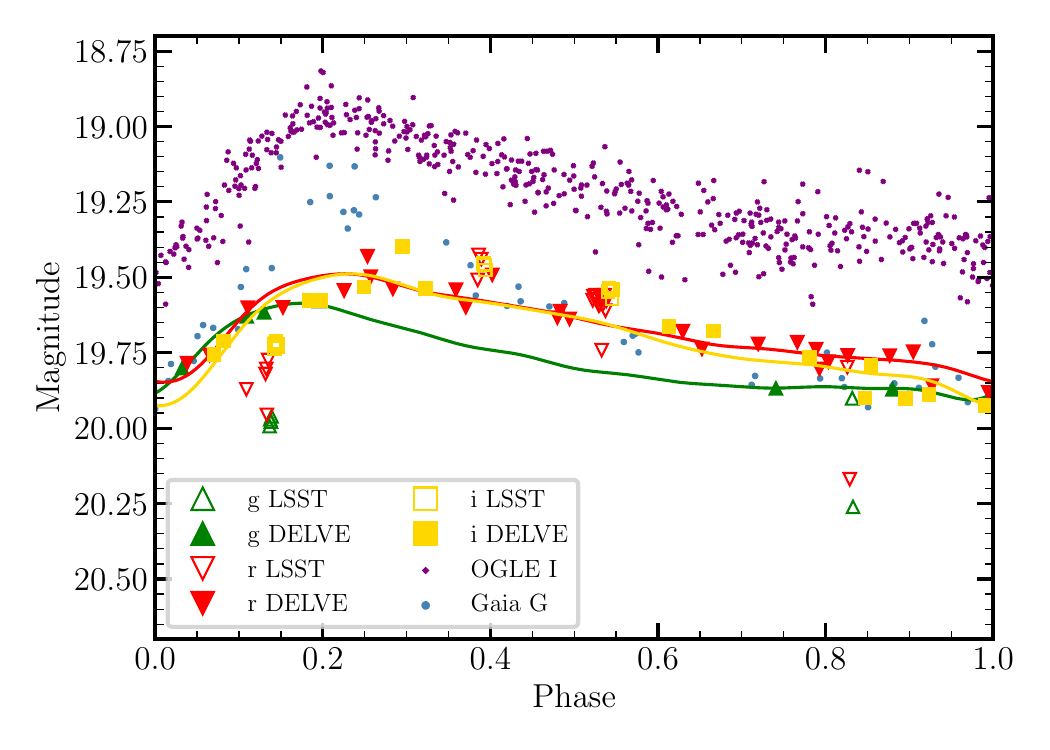}
\caption{Example lightcurve of a RRab variable star (see text for details on the star). Large filled symbols show DELVE $gri$ measurements, large open symbols show LSST DP1 $gri$ measurements, small filled purple circles show the well-sampled OGLE $I$ data, and small filled blue circles are Gaia DR3 $G$ data. Example templates extracted from \cite{Baeza-Villagra2025} for the $gri$ bands are also plotted as solid lines.}
\label{fig:lightcurve}
\end{figure}
\end{center}

The depth of DELVE-MC also opens a new window on faint and low-amplitude variables in crowded and low-surface-brightness regions, complementing shallower but higher-cadence surveys. When combined with {\em Gaia} proper motions and forthcoming spectroscopic surveys such as SDSS-V, DELVE-MC variable stars will provide powerful joint constraints on distances, kinematics, and stellar populations throughout the Magellanic System. In order to showcase the power of the dataset, we present in \autoref{fig:lightcurve} an example lightcurve of a known RRab star with DELVE-MC, LSST DP1\footnote{\url{https://dp1.lsst.io/index.html}} \citep{RubinDP1}, {\em Gaia} and OGLE data.
The star is identified in {\it Gaia} DR3 as \texttt{source\_id} 4689835147403404928 and in OGLE as \texttt{OGLE-SMC-RRLYR-0031}, with both classifying it as an RRab variable. {\it Gaia} DR3 reports a period of $0.53179052$ days, and OGLE-IV reports $0.5318053 \pm 0.0000015$ days. 
Applying the hybrid period search of \citet{Saha2017} with an additional verification step (Meerdink et al., in preparation) to the DELVE-MC photometry alone (objid \texttt{0059m717.2923}), we recover a period of $0.53180933 \pm 0.00000054$ days, which is 1.62 seconds longer than the {\it Gaia} DR3 period and 0.35 seconds longer than the OGLE-IV period.
This SMC star lies along the line of sight to the Milky Way globular cluster 47 Tuc, allowing us to include Rubin observations of that field. We also show RRab light-curve templates from \cite{Baeza-Villagra2025} for the native DECam filters.



A comprehensive catalog and analysis of variable stars in DELVE-MC, including period determination, classification, and 3-D mapping of the Clouds, will be presented in forthcoming work. These results will establish DELVE-MC as a cornerstone dataset for time-domain studies of nearby dwarf galaxies.

\section{Summary}
\label{sec:summary}

We present an overview of the DELVE-MC survey, its associated data reduction pipeline, and the first data release.  The data are based on over 33,120 new and archival images from CTIO-4m + DECam.  The catalog contains 19 billion individual measurements to depths of $\sim$25th magnitude of 1.3 billion unique objects over 2170 square degrees and baselines up to $\sim$10 years.  Due to the wealth of temporal information --- roughly 175 million objects have 30 measurements or more --- the DELVE-MC enables lightcurve analysis for variable stars and AGN.  The catalog enables a number of exciting science topics to be explored, including (1) searches for stellar substructure, stellar streams, and dwarf satellite galaxies, (2) spatially-resolved and precise star formation histories, and (3) cataloging various
types of variable stars.

DELVE-MC also provides an important foundation for future studies of the Magellanic System with the Vera C. Rubin Observatory Legacy Survey of Space and Time. In combination with Rubin time-domain observations, DELVE-MC will extend temporal baselines, improve studies of faint variable populations, and provide complementary deep imaging for investigations of low-surface-brightness structure throughout the Magellanic System.
Future facilities will further complement these datasets at infrared wavelengths: Euclid will image the SMC and its surroundings \citep{Euclid2025}, while the Nancy Grace Roman Space Telescope is capable of providing high-resolution infrared imaging of crowded stellar populations.

\begin{acknowledgments}

D.L.N. acknowledges support from NSF grants AST 1928856, 1908331, 2109196, and 2408159.
J.S. acknowledges financial support from project PID2022-138896NB-C53 and the Severo Ochoa grant CEX2021-001131-S funded by MCIN/AEI/ 10.13039/501100011033. 
S.J.O. acknowledges support from NSF grant AST 2408159.
D.J.S. acknowledges support from NSF grants AST-2205863 and 2508746. 
We thank Brett Meerdink for carrying out the independent period analysis of the example RRab star using DELVE-MC photometry.
We thank the anonymous referee for useful comments that improved the manuscript.

This project used data obtained with the Dark Energy Camera (DECam) at the Blanco 4m telescope at Cerro Tololo Inter-American Observatory. DECam was constructed by the Dark Energy Survey (DES) collaborating institutions: Argonne National Lab, University of California Santa Cruz, University of Cambridge, Centro de Investigaciones Energeticas, Medioambientales y Tecnologicas-Madrid, University of Chicago, University College London, DES-Brazil consortium, University of Edinburgh, ETH-Zurich, University of Illinois at Urbana-Champaign, Institut de Ciencies de l'Espai, Institut de Fisica d'Altes Energies, Lawrence Berkeley National Lab, Ludwig-Maximilians Universit\"at, University of Michigan, National Optical Astronomy Observatory, University of Nottingham, Ohio State University, University of Pennsylvania, University of Portsmouth, SLAC National Lab, Stanford University, University of Sussex, and Texas A\&M University. Funding for DES, including DECam, has been provided by the U.S. Department of Energy, National Science Foundation, Ministry of Education and Science (Spain), Science and Technology Facilities Council (UK), Higher Education Funding Council (England), National Center for Supercomputing Applications, Kavli Institute for Cosmological Physics, Financiadora de Estudos e Projetos, Funda\c{c}\~ao Carlos Chagas Filho de Amparo a Pesquisa, Conselho Nacional de Desenvolvimento Cientfico e Tecnol\'ogico and the Minist\'erio da Ci\^encia e Tecnologia (Brazil), the German Research Foundation-sponsored cluster of excellence "Origin and Structure of the Universe" and the DES collaborating institutions. The Cerro Tololo Inter-American Observatory, National Optical Astronomy Observatory is operated by the Association of Universities for Research in Astronomy (AURA) under a cooperative agreement with the National Science Foundation. 

The DELVE Collaboration gratefully acknowledges support from Fermilab LDRD L2019-011, the NASA Fermi Guest Investigator Program Cycle 9 No. 91201, and the NSF under Grant No. AST-2108168, AST-2108169, AST-2307126, and AST-2407526. This research is partially funded by a generous gift from Charles Simonyi to the NSF Division of Astronomical Sciences. The award is made in recognition of significant contributions to Rubin Observatory's Legacy Survey of Space and Time.

This research uses services or data provided by the Astro Data Lab at NSF's National Optical-Infrared Astronomy Research Laboratory. NSF's NOIRLab is operated by the Association of Universities for Research in Astronomy (AURA), Inc. under a cooperative agreement with the National Science Foundation.

NOIRLab IRAF is distributed by the Community Science and Data Center at NSF NOIRLab, which is managed by the Association of Universities for Research in Astronomy (AURA) under a cooperative agreement with the U.S. National Science Foundation.

This publication makes use of data from the Pan-STARRS1 Surveys (PS1) and the PS1 public science archive, which have been made possible through contributions by the Institute for Astronomy, the University of Hawaii, the Pan-STARRS Project Office, the Max-Planck Society and its participating institutes, the Max Planck Institute for Astronomy, Heidelberg and the Max Planck Institute for Extraterrestrial Physics, Garching, The Johns Hopkins University, Durham University, the University of Edinburgh, the Queen's University Belfast, the Harvard-Smithsonian Center for Astrophysics, the Las Cumbres Observatory Global Telescope Network Incorporated, the National Central University of Taiwan, the Space Telescope Science Institute, the National Aeronautics and Space Administration under Grant No. NNX08AR22G issued through the Planetary Science Division of the NASA Science Mission Directorate, the National Science Foundation Grant No. AST-1238877, the University of Maryland, Eotvos Lorand University (ELTE), the Los Alamos National Laboratory, and the Gordon and Betty Moore Foundation.

This work has made use of data from the European Space Agency (ESA)
mission {\it Gaia} (\url{https://www.cosmos.esa.int/gaia}), processed by
the {\it Gaia} Data Processing and Analysis Consortium (DPAC,
\url{https://www.cosmos.esa.int/web/gaia/dpac/consortium}). Funding
for the DPAC has been provided by national institutions, in particular
the institutions participating in the {\it Gaia} Multilateral Agreement.

This publication makes use of data products from the Two Micron All Sky Survey, which is a joint project of the University of Massachusetts and the Infrared Processing and Analysis Center/California Institute of Technology, funded by the National Aeronautics and Space Administration and the National Science Foundation.

Some of the results in this paper have been derived using the healpy and HEALPix package.

Portions of the writing and editing in this paper were assisted with the help of the ChatGPT large language model (OpenAI, 2025). The tool was used to improve clarity, phrasing, grammar, and structure of draft text, while all scientific content, interpretations, and conclusions remain entirely the responsibility of the authors.

\end{acknowledgments}


\software{
    \package{DAOPHOT} \citep{Stetson1987,Stetson1990,Stetson1994},
    \package{Astropy} \citep{astropy},
    \package{IPython} \citep{ipython},
    \package{matplotlib} \citep{mpl},
    \package{numpy} \citep{numpy},
    \package{scipy} \citep{scipy},
    \package{healpy} \citep{Zonca2019},
    \package{SExtractor} \citep{Bertin1996},
    \package{IRAF} \citep{Tody1986,Tody1993,Fitzpatrick2024},
    \package{scikit-learn} \citep{scikit-learn}
}

\facilities{CTIO:Blanco (DECam), Gaia, CTIO:2MASS, FLWO:2MASS, Skymapper, WISE, Spitzer, Astro Data Lab}

\appendix
\section{Additional Survey Performance Metrics}
\label{sec:appendix}

\begin{center}
\begin{figure*}[t]
\includegraphics[width=1.0\hsize,angle=0]{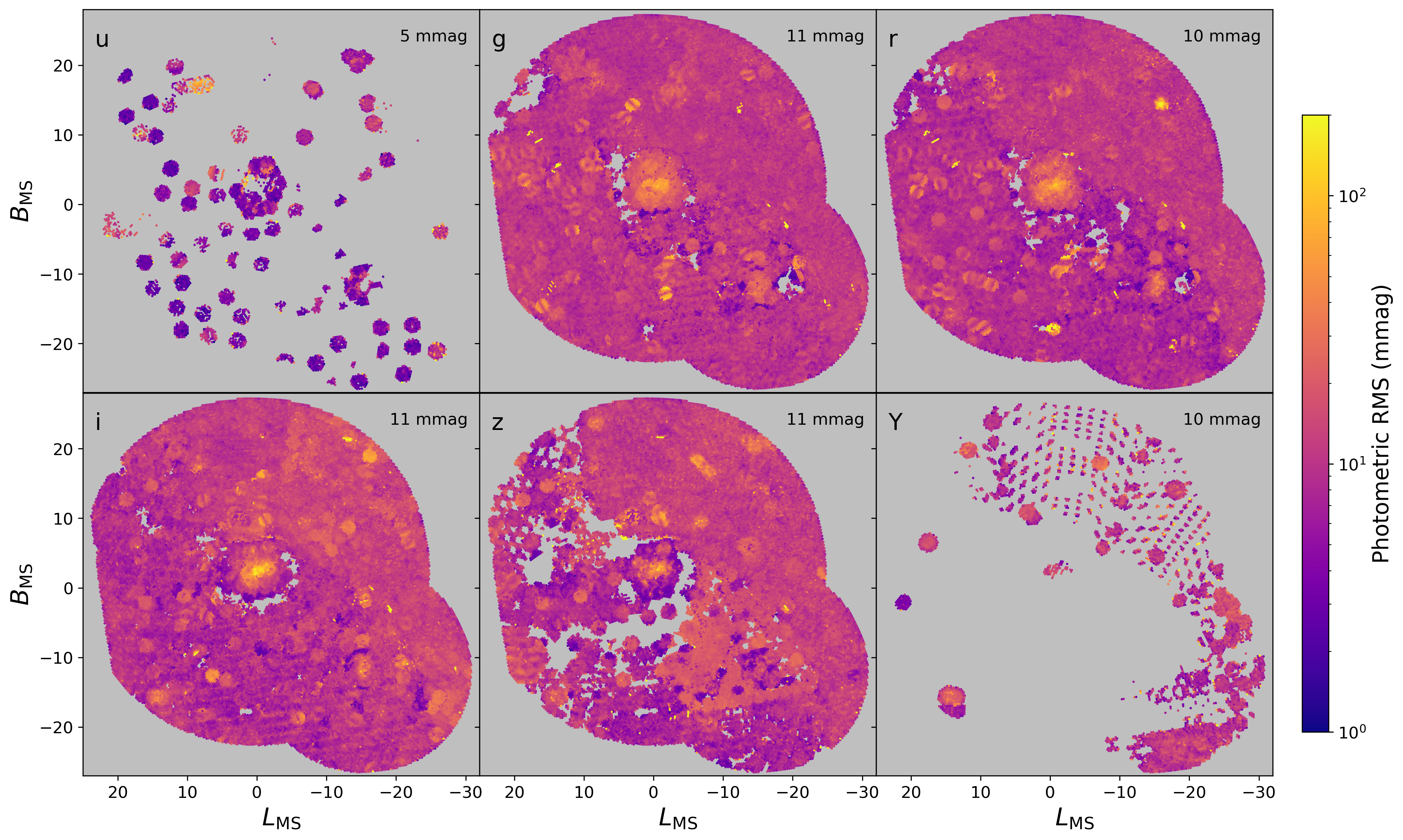}
\caption{
Photometric precision as determined by finding the lower threshold of RMS scatter as a function of magnitude for each brick (see \autoref{fig:photrms_example}). The median photometric scatter across all bricks is given in millimagnitudes in the upper-right corner of each panel. The patchier $Y$-band map results from the small number of bricks with more than two overlapping $Y$-band exposures.
}
\label{fig:photrms}
\end{figure*}
\end{center}

This appendix presents supplementary figures illustrating the photometric performance of the DELVE-MC survey across the survey footprint.
\autoref{fig:photrms} shows the spatial distribution of the photometric precision for each photometric band. The precision is estimated from the minimum median RMS scatter of bright stars in 0.5 magnitude bins for each brick (see \autoref{fig:photrms_example}). Variations across the footprint are driven primarily by differences in crowding, observing conditions, and the number of available exposures. The photometric scatter is generally quite low around $\sim$11 mmag for the $griz$ bands.

\bibliographystyle{aasjournalv7}
\bibliography{ref_og.bib}



\end{document}

%% file: shorthand.tex
\usepackage{xspace}

\newcommand{\dgr}{$^{\circ}~$}

\newcommand{\h}{$^{\rm h}$}
\newcommand{\m}{$^{\rm m}$}
\newcommand{\s}{$^{\rm s}~$}

\newcommand{\teff}{${\rm t}_{\rm eff}$}
\newcommand{\code}[1]{\texttt{#1}\xspace}
\newcommand{\SExtractor}{\code{SourceExtractor}}

\newcommand{\delvered}{\code{delvered}}
\newcommand{\photred}{\code{PHOTRED}}
\newcommand{\smashred}{\code{SMASHRED}}
\newcommand{\daophot}{\code{DAOPHOT}}
\newcommand{\daogrow}{\code{DAOGROW}}
\newcommand{\allstar}{\code{ALLSTAR}}
\newcommand{\allframe}{\code{ALLFRAME}}
\newcommand{\wcs}{\code{WCS}}
\newcommand{\match}{\code{MATCH}}
\newcommand{\apcor}{\code{APCOR}}
\newcommand{\zeropoint}{\code{ZEROPOINT}}
\newcommand{\astrom}{\code{ASTROM}}
\newcommand{\calib}{\code{CALIB}}
\newcommand{\combine}{\code{COMBINE}}
\newcommand{\deredden}{\code{DEREDDEN}}
\newcommand{\save}{\code{SAVE}}
\newcommand{\psf}{\code{PSF}}
\newcommand{\ccdnum}{\texttt{CCDNUM}}
\newcommand{\roughly}{\ensuremath{ {\sim}\,} }
\newcommand{\Gaia}{{\it Gaia}\xspace}
\newcommand{\dotdeg}{$\rlap{.}^\circ$}

%% file: authors.tex
\correspondingauthor{David L. Nidever}
\email{david.nidever@montana.edu}

\author[0000-0002-1793-3689]{D.~L.~Nidever}
\affiliation{Department of Physics, Montana State University, P.O. Box 173840, Bozeman, MT 59717-3840}
\affiliation{NSF's National Optical-Infrared Astronomy Research Laboratory, 950 N. Cherry Ave., Tucson, AZ 85719, USA}
\email{david.nidever@montana.edu}

\author[0000-0002-8093-7471]{P.~Massana}
\affiliation{Cerro Tololo Inter-American Observatory/NSF NOIRLab, Casilla 603, La Serena, Chile}
\affiliation{Department of Physics, Montana State University, P.O. Box 173840, Bozeman, MT 59717-3840}
\email{pol.massana@noirlab.edu}

\author[0000-0003-1680-1884]{Y.~Choi}
\affiliation{NSF's National Optical-Infrared Astronomy Research Laboratory, 950 N. Cherry Ave., Tucson, AZ 85719, USA}
\email{}

\author[0000-0002-1594-1466]{J.~Sakowska}
\affiliation{Instituto de Astrofísica de Andalucía, CSIC, Glorieta de la Astronom\'\i a,  E-18080 Granada, Spain}
\affiliation{Department of Physics, University of Surrey, Guildford, GU2 7XH, UK}
\email{}

\author[0000-0001-8251-933X]{A.~Drlica-Wagner}
\affiliation{Fermi National Accelerator Laboratory, P.O.\ Box 500, Batavia, IL 60510, USA}
\affiliation{Kavli Institute for Cosmological Physics, University of Chicago, Chicago, IL 60637, USA}
\affiliation{Department of Astronomy \& Astrophysics, University of Chicago, Chicago, IL 60637, USA}
\email{}

\author[0000-0002-7134-8296]{K.~A.~G.~Olsen}
\affiliation{NSF's National Optical-Infrared Astronomy Research Laboratory, 950 N. Cherry Ave., Tucson, AZ 85719, USA}
\email{}

\author[0009-0000-9037-1697]{S.~J.~Oden}
\affiliation{Department of Physics, Montana State University, P.O. Box 173840, Bozeman, MT 59717-3840}
\email{}

\author[0000-0001-7827-7825]{Roeland P. van der Marel}
\affiliation{Space Telescope Science Institute, 3700 San Martin Drive, Baltimore, MD 21218, USA}
\affiliation{Center for Astrophysical Sciences, The William H. Miller III Department of Physics \& Astronomy, Johns Hopkins University, Baltimore, MD 21218, USA}
\email{marel@stsci.edu}

\author[0000-0001-9438-5228]{M.~Navabi}
\affiliation{Department of Physics, University of Surrey, Guildford, GU2 7XH, UK}
\email{}

\author[0000-0001-6728-806X]{C.~Gallart}
\affiliation{Instituto de Astrof\'{i}sica de Canarias, La Laguna, Tenerife, Spain}
\affiliation{Departamento de Astrof\'{i}sica, Universidad de La Laguna, Tenerife, Spain}
\email{}

\author[0000-0002-8282-469X]{N.~E.~D.~No\"el}
\affiliation{Department of Physics, University of Surrey, Guildford, GU2 7XH, UK}
\email{}

\author[0000-0001-5143-1255]{A.~Chaturvedi}
\affiliation{Department of Physics, University of Surrey, Guildford, GU2 7XH, UK}
\email{}

\author[0000-0002-9269-8287]{G.~Limberg}
\affiliation{Department of Astronomy \& Astrophysics, University of Chicago, Chicago, IL 60637, USA}
\affiliation{Kavli Institute for Cosmological Physics, University of Chicago, Chicago, IL 60637, USA}
\email{}

\author[0000-0001-5805-5766]{A.~H.~Riley}
\affiliation{Lund Observatory, Division of Astrophysics, Department of Physics, Lund University, SE-221 00 Lund, Sweden}
\email{}

\author[0000-0002-1763-4128]{D.~Crnojevi\'c}
\affiliation{Department of Physics and Astronomy, University of Tampa, 401 West Kennedy Boulevard, Tampa, FL 33606, USA}
\email{}

\author[0000-0002-7123-8943]{A.~R.~Walker}
\affiliation{Cerro Tololo Inter-American Observatory/NSF NOIRLab, Casilla 603, La Serena, Chile}
\email{}

\author[0000-0002-6021-8760]{A.~B.~Pace}
\affiliation{Department of Astronomy, University of Virginia, 530 McCormick Road, Charlottesville, VA 22904, USA}
\email{}

\author[0000-0002-6797-696X]{M-R.~L.~Cioni}
\affiliation{Leibniz-Institut f\"{u}r Astrophysik Potsdam (AIP), An der Sternwarte 16, 14482 Potsdam Germany}
\email{}

\author[0000-0002-3204-1742]{N.~Kallivayalil}
\affiliation{Department of Astronomy, University of Virginia, 530 McCormick Road, Charlottesville, VA 22904, USA}
\email{}

\author[0000-0003-4102-380X]{D.~J.~Sand}
\affiliation{Steward Observatory, University of Arizona, 933 North Cherry Avenue, Tucson, AZ 85721-0065, USA}
\email{}

\author[0000-0003-2025-3147]{S.~R.~Majewski}
\affiliation{Department of Astronomy, University of Virginia, 530 McCormick Road, Charlottesville, VA 22904, USA}
\email{}


\author[0000-0002-3690-105X]{J.~A.~Carballo-Bello}
\affiliation{Instituto de Alta Investigaci\'on, Universidad de Tarapac\'a, Casilla 7D, Arica, Chile}
\email{}

\author[0000-0003-1697-7062]{W.~Cerny}
\affiliation{Department of Astronomy, Yale University, New Haven, CT 06520, USA}
\email{}

\author[0000-0001-6957-1627]{P.~S.~Ferguson}
\affiliation{DiRAC Institute, Department of Astronomy, University of Washington, 3910 15th Ave NE, Seattle, WA, 98195, USA}
\email{}

\author[0000-0002-9144-7726]{C.~E.~Mart\'inez-V\'azquez}
\affiliation{NSF NOIRLab, 670 N. A'ohoku Place, Hilo, Hawai'i, 96720, USA}
\email{}

\author[0000-0001-9649-4815]{B.~Mutlu-Pakdil}
\affiliation{Department of Physics and Astronomy, Dartmouth College, Hanover, NH 03755, USA}
\email{}

\author[0000-0003-3519-4004]{S.~Mau}
\affiliation{Department of Physics, Duke University Durham, NC 27708, USA}
\email{}

\author[0000-0003-1479-3059]{G.~S.~Stringfellow}
\affiliation{University of Colorado Boulder, Boulder, CO 80309 USA}
\email{}

\author[0000-0003-0478-0473]{C.~Y.~Tan}
\affiliation{Kavli Institute for Cosmological Physics, University of Chicago, Chicago, IL 60637, USA}
\email{}

\author[0000-0003-4341-6172]{A.~K.~Vivas}
\affiliation{Cerro Tololo Inter-American Observatory/NSF NOIRLab, Casilla 603, La Serena, Chile}
\email{}

%% file: ref_og.bib
@ARTICLE{Saha2017,
       author = {{Saha}, Abhijit and {Vivas}, A. Katherina},
        title = "{A Hybrid Algorithm for Period Analysis from Multiband Data with Sparse and Irregular Sampling for Arbitrary Light-curve Shapes}",
      journal = {\aj},
         year = 2017,
        month = dec,
       volume = {154},
       number = {6},
          eid = {231},
        pages = {231},
          doi = {10.3847/1538-3881/aa8fd3},
archivePrefix = {arXiv},
       eprint = {1709.10156},
 primaryClass = {astro-ph.IM},
       adsurl = {https://ui.adsabs.harvard.edu/abs/2017AJ....154..231S}
}

@INPROCEEDINGS{DaCosta1991,
       author = {{Da Costa}, G. S.},
        title = "{The Age-Abundance Relations and Age Distributions for the Star Clusters of the Magellanic Clouds}",
    booktitle = {The Magellanic Clouds},
         year = 1991,
       editor = {{Haynes}, R. and {Milne}, D.},
       series = {IAU Symposium},
       volume = {148},
        pages = {183--192},
          doi = {10.1017/S0074180900200296}
}

@ARTICLE{Euclid2025,
       author = {{Euclid Collaboration} and Mellier, Y. and others},
        title = "{Euclid. I. Overview of the Euclid mission}",
      journal = {\aap},
         year = 2025,
        month = may,
       volume = {697},
          eid = {A1},
        pages = {A1},
          doi = {10.1051/0004-6361/202450810},
archivePrefix = {arXiv},
       eprint = {2405.13491},
 primaryClass = {astro-ph.CO}
}

@ARTICLE{DrlicaWagner2022,
       author = {{Drlica-Wagner}, A. and {Ferguson}, P.~S. and {Adam{\'o}w}, M. and {Aguena}, M. and {Allam}, S. and {Andrade-Oliveira}, F. and {Bacon}, D. and {Bechtol}, K. and {Bell}, E.~F. and {Bertin}, E. and {Bilaji}, P. and {Bocquet}, S. and {Bom}, C.~R. and {Brooks}, D. and {Burke}, D.~L. and {Carballo-Bello}, J.~A. and {Carlin}, J.~L. and {Carnero Rosell}, A. and {Carrasco Kind}, M. and {Carretero}, J. and {Castander}, F.~J. and {Cerny}, W. and {Chang}, C. and {Choi}, Y. and {Conselice}, C. and {Costanzi}, M. and {Crnojevi{\'c}}, D. and {da Costa}, L.~N. and {de Vicente}, J. and {Desai}, S. and {Esteves}, J. and {Everett}, S. and {Ferrero}, I. and {Fitzpatrick}, M. and {Flaugher}, B. and {Friedel}, D. and {Frieman}, J. and {Garc{\'\i}a-Bellido}, J. and {Gatti}, M. and {Gaztanaga}, E. and {Gerdes}, D.~W. and {Gruen}, D. and {Gruendl}, R.~A. and {Gschwend}, J. and {Hartley}, W.~G. and {Hernandez-Lang}, D. and {Hinton}, S.~R. and {Hollowood}, D.~L. and {Honscheid}, K. and {Hughes}, A.~K. and {Jacques}, A. and {James}, D.~J. and {Johnson}, M.~D. and {Kuehn}, K. and {Kuropatkin}, N. and {Lahav}, O. and {Li}, T.~S. and {Lidman}, C. and {Lin}, H. and {March}, M. and {Marshall}, J.~L. and {Mart{\'\i}nez-Delgado}, D. and {Mart{\'\i}nez-V{\'a}zquez}, C.~E. and {Massana}, P. and {Mau}, S. and {McNanna}, M. and {Melchior}, P. and {Menanteau}, F. and {Miller}, A.~E. and {Miquel}, R. and {Mohr}, J.~J. and {Morgan}, R. and {Mutlu-Pakdil}, B. and {Mu{\~n}oz}, R.~R. and {Neilsen}, E.~H. and {Nidever}, D.~L. and {Nikutta}, R. and {Nilo Castellon}, J.~L. and {No{\"e}l}, N.~E.~D. and {Ogando}, R.~L.~C. and {Olsen}, K.~A.~G. and {Pace}, A.~B. and {Palmese}, A. and {Paz-Chinch{\'o}n}, F. and {Pereira}, M.~E.~S. and {Pieres}, A. and {Plazas Malag{\'o}n}, A.~A. and {Prat}, J. and {Riley}, A.~H. and {Rodriguez-Monroy}, M. and {Romer}, A.~K. and {Roodman}, A. and {Sako}, M. and {Sakowska}, J.~D. and {Sanchez}, E. and {S{\'a}nchez}, F.~J. and {Sand}, D.~J. and {Santana-Silva}, L. and {Santiago}, B. and {Schubnell}, M. and {Serrano}, S. and {Sevilla-Noarbe}, I. and {Simon}, J.~D. and {Smith}, M. and {Soares-Santos}, M. and {Stringfellow}, G.~S. and {Suchyta}, E. and {Suson}, D.~J. and {Tan}, C.~Y. and {Tarle}, G. and {Tavangar}, K. and {Thomas}, D. and {To}, C. and {Tollerud}, E.~J. and {Troxel}, M.~A. and {Tucker}, D.~L. and {Varga}, T.~N. and {Vivas}, A.~K. and {Walker}, A.~R. and {Weller}, J. and {Wilkinson}, R.~D. and {Wu}, J.~F. and {Yanny}, B. and {Zaborowski}, E. and {Zenteno}, A. and {Delve Collaboration} and {Des Collaboration} and {Astro Data Lab}},
        title = "{The DECam Local Volume Exploration Survey Data Release 2}",
      journal = {\apjs},
         year = 2022,
        month = aug,
       volume = {261},
       number = {2},
          eid = {38},
        pages = {38},
          doi = {10.3847/1538-4365/ac78eb},
archivePrefix = {arXiv},
       eprint = {2203.16565},
 primaryClass = {astro-ph.IM},
       adsurl = {https://ui.adsabs.harvard.edu/abs/2022ApJS..261...38D}
}

@article{Deason2022,
  author = {{Deason}, Alis J. and {Belokurov}, Vasily and {Kravtsov}, Andrey V. and others},
  title = {The edge of galaxy formation: tides and faint stellar haloes around dwarf galaxies},
  journal = {Monthly Notices of the Royal Astronomical Society},
  volume = {511},
  number = {3},
  pages = {4044--4060},
  year = {2022},
  doi = {10.1093/mnras/stac241},
  eprint = {2112.03949},
  archivePrefix = {arXiv},
  primaryClass = {astro-ph.GA}
}

@article{KadoFong2022,
  author = {{Kado-Fong}, Erin and {Greene}, Jenny E. and {Lancaster}, Lachlan and others},
  title = {Stellar Halos around Low-mass Galaxies in the SAGA Survey},
  journal = {The Astrophysical Journal},
  volume = {931},
  number = {2},
  pages = {152},
  year = {2022},
  doi = {10.3847/1538-4357/ac67a3},
  eprint = {2203.09781},
  archivePrefix = {arXiv},
  primaryClass = {astro-ph.GA}
}

@article{ElBadry2018,
  author = {{El-Badry}, Kareem and {Quataert}, Eliot and {Wetzel}, Andrew and others},
  title = {Gas kinematics in FIRE simulations of dwarf galaxies: bursty star formation and disequilibrium rotation support},
  journal = {Monthly Notices of the Royal Astronomical Society},
  volume = {473},
  number = {2},
  pages = {1930--1955},
  year = {2018},
  doi = {10.1093/mnras/stx2457},
  eprint = {1705.10321},
  archivePrefix = {arXiv},
  primaryClass = {astro-ph.GA}
}

@article{Munshi2021,
  author = {{Munshi}, Ferah and {Brooks}, Alyson and {Applebaum}, Elizabeth and others},
  title = {The DC Justice League: a comparison of cosmological hydrodynamical simulations of dwarf galaxies},
  journal = {The Astrophysical Journal},
  volume = {923},
  number = {1},
  pages = {35},
  year = {2021},
  doi = {10.3847/1538-4357/ac23b2},
  eprint = {2109.09709},
  archivePrefix = {arXiv},
  primaryClass = {astro-ph.GA}
}

@article{Bechtol2015,
  author = {{Bechtol}, Keith and others},
  title = {Eight New Milky Way Companions Discovered in First-Year Dark Energy Survey Data},
  journal = {The Astrophysical Journal},
  volume = {807},
  number = {1},
  pages = {50},
  year = {2015},
  doi = {10.1088/0004-637X/807/1/50},
  eprint = {1503.02584},
  archivePrefix = {arXiv},
  primaryClass = {astro-ph.GA}
}

@article{DrlicaWagner2015,
  author = {{Drlica-Wagner}, Alex and others},
  title = {Eight Ultra-faint Galaxy Candidates Discovered in Year Two of the Dark Energy Survey},
  journal = {The Astrophysical Journal},
  volume = {813},
  number = {2},
  pages = {109},
  year = {2015},
  doi = {10.1088/0004-637X/813/2/109},
  eprint = {1508.03622},
  archivePrefix = {arXiv},
  primaryClass = {astro-ph.GA}
}

@article{Koposov2015,
  author = {{Koposov}, Sergey E. and {Belokurov}, Vasily and {Torrealba}, Gabriel and {Evans}, N. Wyn},
  title = {Beasts of the Southern Wild: Discovery of Nine Ultra-Faint Satellites in the Vicinity of the Magellanic Clouds},
  journal = {The Astrophysical Journal},
  volume = {805},
  number = {2},
  pages = {130},
  year = {2015},
  doi = {10.1088/0004-637X/805/2/130},
  eprint = {1503.02079},
  archivePrefix = {arXiv},
  primaryClass = {astro-ph.GA}
}

@article{Torrealba2018,
  author = {{Torrealba}, G. and {Belokurov}, V. and {Koposov}, S. E. and 
            {Bechtol}, K. and {Drlica-Wagner}, A. and {Olsen}, K. A. G. and 
            {Vivas}, A. K. and {Yanny}, B. and {Jethwa}, P. and 
            {Walker}, A. R. and et al.},
  title = {Discovery of two neighboring satellites in the Carina constellation with MagLiteS},
  journal = {Monthly Notices of the Royal Astronomical Society},
  volume = {475},
  number = {4},
  pages = {5085--5101},
  year = {2018},
  doi = {10.1093/mnras/sty170},
  eprint = {1801.07279},
  archivePrefix = {arXiv},
  primaryClass = {astro-ph.GA}
}

@article{vdM2001,
  author = {{van der Marel}, R. P.},
  title = {Magellanic Cloud Structure from Near-Infrared Surveys. I. The Viewing Angles of the Large Magellanic Cloud},
  journal = {AJ},
  volume = {122},
  pages = {1827--1853},
  year = {2001},
  doi = {10.1086/323099}
}

@article{vdMarelCioni2001,
  author = {{van der Marel}, R. P. and {Cioni}, M.-R. L.},
  title = {Magellanic Cloud Structure from Near-Infrared Surveys. II. Star Count Maps and the Intrinsic Elongation of the Large Magellanic Cloud},
  journal = {AJ},
  volume = {122},
  pages = {1807--1826},
  year = {2001},
  doi = {10.1086/323097}
}

@article{Leavitt1908,
  author = {{Leavitt}, Henrietta S.},
  title = {1777 Variables in the Magellanic Clouds},
  journal = {Annals of Harvard College Observatory},
  volume = {60},
  pages = {87--108},
  year = {1908}
}

@article{Leavitt1912,
  author = {{Leavitt}, Henrietta S. and {Pickering}, Edward C.},
  title = {Periods of 25 Variable Stars in the Small Magellanic Cloud},
  journal = {Harvard College Observatory Circular},
  volume = {173},
  pages = {1--3},
  year = {1912}
}

@book{PayneGaposchkin1971,
  author = {{Payne-Gaposchkin}, Cecilia},
  title = {The Magellanic Clouds},
  publisher = {D. Reidel Publishing Company},
  address = {Dordrecht},
  year = {1971}
}

@BOOK{Catelan2015,
       author = {{Catelan}, M. and {Smith}, H.~A.},
        title = "{Pulsating Stars}",
         year = 2015,
       adsurl = {https://ui.adsabs.harvard.edu/abs/2015pust.book.....C}
}

@ARTICLE{Soszynski2015,
       author = {{Soszy{\'n}ski}, I. and {Udalski}, A. and {Szyma{\'n}ski}, M.~K. and {Skowron}, D. and {Pietrzy{\'n}ski}, G. and {Poleski}, R. and {Pietrukowicz}, P. and {Skowron}, J. and {Mr{\'o}z}, P. and {Koz{\l}owski}, S. and {Wyrzykowski}, {\L}. and {Ulaczyk}, K. and {Pawlak}, M.},
        title = "{The OGLE Collection of Variable Stars. Classical Cepheids in the Magellanic System}",
      journal = {\actaa},
         year = 2015,
        month = dec,
       volume = {65},
       number = {4},
        pages = {297-312},
          doi = {10.48550/arXiv.1601.01318},
archivePrefix = {arXiv},
       eprint = {1601.01318},
 primaryClass = {astro-ph.SR},
       adsurl = {https://ui.adsabs.harvard.edu/abs/2015AcA....65..297S}
}

@ARTICLE{Soszynski2022,
       author = {{Soszy{\'n}ski}, I. and {Udalski}, A. and {Skowron}, J. and {Pietrukowicz}, P. and {Szyma{\'n}ski}, M.~K. and {Poleski}, R. and {Skowron}, D.~M. and {Koz{\l}owski}, S. and {Mr{\'o}z}, P. and {Iwanek}, P. and {Wrona}, M. and {Ulaczyk}, K. and {Rybicki}, K. and {Gromadzki}, M.},
        title = "{The OGLE Collection of Variable Stars. Over 2600 {\ensuremath{\delta}} Scuti Stars in the Small Magellanic Cloud}",
      journal = {\actaa},
         year = 2022,
        month = jan,
       volume = {72},
       number = {4},
        pages = {245-256},
          doi = {10.32023/0001-5237/72.4.1},
archivePrefix = {arXiv},
       eprint = {2301.04663},
 primaryClass = {astro-ph.SR},
       adsurl = {https://ui.adsabs.harvard.edu/abs/2022AcA....72..245S}
}

@ARTICLE{Ripepi2017,
       author = {{Ripepi}, Vincenzo and {Cioni}, Maria-Rosa L. and {Moretti}, Maria Ida and {Marconi}, Marcella and {Bekki}, Kenji and {Clementini}, Gisella and {de Grijs}, Richard and {Emerson}, Jim and {Groenewegen}, Martin A.~T. and {Ivanov}, Valentin D. and {Molinaro}, Roberto and {Muraveva}, Tatiana and {Oliveira}, Joana M. and {Piatti}, Andr{\'e}s E. and {Subramanian}, Smitha and {van Loon}, Jacco Th.},
        title = "{The VMC survey - XXV. The 3D structure of the Small Magellanic Cloud from Classical Cepheids}",
      journal = {\mnras},
         year = 2017,
        month = nov,
       volume = {472},
       number = {1},
        pages = {808-827},
          doi = {10.1093/mnras/stx2096},
archivePrefix = {arXiv},
       eprint = {1707.04500},
 primaryClass = {astro-ph.GA},
       adsurl = {https://ui.adsabs.harvard.edu/abs/2017MNRAS.472..808R}
}

@ARTICLE{Haschke2012a,
       author = {{Haschke}, Raoul and {Grebel}, Eva K. and {Duffau}, Sonia},
        title = "{Three-dimensional Maps of the Magellanic Clouds using RR Lyrae Stars and Cepheids. I. The Large Magellanic Cloud}",
      journal = {\aj},
         year = 2012,
        month = oct,
       volume = {144},
       number = {4},
          eid = {106},
        pages = {106},
          doi = {10.1088/0004-6256/144/4/106},
archivePrefix = {arXiv},
       eprint = {1207.5791},
 primaryClass = {astro-ph.GA},
       adsurl = {https://ui.adsabs.harvard.edu/abs/2012AJ....144..106H}
}

@ARTICLE{Haschke2012b,
       author = {{Haschke}, Raoul and {Grebel}, Eva K. and {Duffau}, Sonia},
        title = "{Three-dimensional Maps of the Magellanic Clouds using RR Lyrae Stars and Cepheids. II. The Small Magellanic Cloud}",
      journal = {\aj},
         year = 2012,
        month = oct,
       volume = {144},
       number = {4},
          eid = {107},
        pages = {107},
          doi = {10.1088/0004-6256/144/4/107},
archivePrefix = {arXiv},
       eprint = {1209.3769},
 primaryClass = {astro-ph.GA},
       adsurl = {https://ui.adsabs.harvard.edu/abs/2012AJ....144..107H}
}

@ARTICLE{Skowron2016,
       author = {{Skowron}, D.~M. and {Soszy{\'n}ski}, I. and {Udalski}, A. and {Szyma{\'n}ski}, M.~K. and {Pietrukowicz}, P. and {Skowron}, J. and {Poleski}, R. and {Wyrzykowski}, {\L}. and {Ulaczyk}, K. and {Koz{\l}owski}, S. and {Mr{\'o}z}, P. and {Pawlak}, M.},
        title = "{OGLE-ing the Magellanic System: Photometric Metallicity from Fundamental Mode RR Lyrae Stars}",
      journal = {\actaa},
         year = 2016,
        month = sep,
       volume = {66},
       number = {3},
        pages = {269-292},
          doi = {10.48550/arXiv.1608.00013},
archivePrefix = {arXiv},
       eprint = {1608.00013},
 primaryClass = {astro-ph.SR},
       adsurl = {https://ui.adsabs.harvard.edu/abs/2016AcA....66..269S}
}

@misc{RubinDP1,
    author = "{NSF-DOE Vera C.\ Rubin Observatory}",
    doi = "10.71929/RUBIN/2570308",
    url = "https://www.osti.gov//servlets/purl/2570308",
    title = "{Legacy Survey of Space and Time Data Preview 1 [Data set]}",
    publisher = "NSF-DOE Vera C. Rubin Observatory",
    year = "2025"
    }

@ARTICLE{Kollmeier2026,
       author = {{Kollmeier}, Juna A. and {Rix}, Hans-Walter and {Aerts}, Conny and {Aird}, James and {Alfaro}, Pablo Vera and {Almeida}, Andr{\'e}s and {Anderson}, Scott F. and {Jim{\'e}nez Arranz}, {\'O}scar and {Arseneau}, Stefan M. and {Assef}, Roberto and {Aviram}, Shir and {Aydar}, Catarina and {Badenes}, Carles and {Bandyopadhyay}, Avrajit and {Barger}, Kat and {Barkhouser}, Robert H. and {Bauer}, Franz E. and {Bender}, Chad and {Besser}, Felipe and {Bhattarai}, Binod and {Bilgi}, Pavaman and {Bird}, Jonathan and {Bizyaev}, Dmitry and {Blanc}, Guillermo A. and {Blanton}, Michael R. and {Bochanski}, John and {Bovy}, Jo and {Brandon}, Christopher and {Brandt}, William Nielsen and {Brownstein}, Joel R. and {Buchner}, Johannes and {Burchett}, Joseph N. and {Carlberg}, Joleen and {Casey}, Andrew R. and {Castaneda-Carlos}, Lesly and {Chakraborty}, Priyanka and {Chanam{\'e}}, Julio and {Chandra}, Vedant and {Cherinka}, Brian and {Chilingarian}, Igor and {Comparat}, Johan and {Cosens}, Maren and {Covey}, Kevin and {Crane}, Jeffrey D. and {Crumpler}, Nicole R. and {Cunha}, Katia and {Cunningham}, Tim and {Dai}, Xinyu and {Darling}, Jeremy and {Davidson}, Jr., James W. and {Davis}, Megan C. and {De Lee}, Nathan and {Deacon}, Niall and {M{\'e}ndez Delgado}, Jos{\'e} Eduardo and {Demasi}, Sebastian and {Demianenko}, Mariia and {Derwent}, Mark and {D'Onghia}, Elena and {Di Mille}, Francesco and {Dias}, Bruno and {Donor}, John and {Drory}, Niv and {Dwelly}, Tom and {Egorov}, Oleg and {Egorova}, Evgeniya and {El-Badry}, Kareem and {Engelman}, Mike and {Eracleous}, Mike and {Fan}, Xiaohui and {Farr}, Emily and {Fries}, Logan and {Frinchaboy}, Peter and {Froning}, Cynthia S. and {G{\"a}nsicke}, Boris T. and {Garc{\'\i}a}, Pablo and {Gelfand}, Joseph and {Gentile Fusillo}, Nicola Pietro and {Glover}, Simon and {Grabowski}, Katie and {Grebel}, Eva K. and {Green}, Paul J and {Grier}, Catherine and {Gupta}, Pramod and {Gray}, Aidan C. and {H{\"a}berle}, Maximilian and {Hall}, Patrick B. and {Hammond}, Randolph P. and {Hawkins}, Keith and {Harding}, Albert C. and {Heged{\H{u}}s}, Viola and {Herbst}, Tom and {Hermes}, J.~J. and {Rodr{\'\i}guez Hidalgo}, Paola and {Hilder}, Thomas and {Hogg}, David W and {Holtzman}, Jon A. and {Horta}, Danny and {Huang}, Yang and {Hwang}, Hsiang-Chih and {Ibarra-Medel}, Hector Javier and {Imig}, Julie and {Inight}, Keith and {Jana}, Arghajit and {Ji}, Alexander P. and {Jofre}, Paula and {Johns}, Matt and {Johnson}, Jennifer and {Johnson}, James W. and {Johnston}, Evelyn J. and {Jones}, Amy M and {Katkov}, Ivan and {Koekemoer}, Anton M. and {Kounkel}, Marina and {Kreckel}, Kathryn and {Krishnarao}, Dhanesh and {Krumpe}, Mirko and {Kumari}, Nimisha and {Kupfer}, Thomas and {Lacerna}, Ivan and {Laporte}, Chervin and {Lepine}, Sebastien and {Li}, Jing and {Liu}, Xin and {Loebman}, Sarah and {Long}, Knox and {Roman-Lopes}, Alexandre and {Lu}, Yuxi and {Majewski}, Steven Raymond and {Maoz}, Dan and {McKinnon}, Kevin A. and {Medan}, Ilija and {Merloni}, Andrea and {Minniti}, Dante and {Morrison}, Sean and {Myers}, Natalie and {M{\'e}sz{\'a}ros}, Szabolcs and {Nandra}, Kirpal and {Nayak}, Prasanta K. and {Ness}, Melissa K and {Nidever}, David L. and {O'Brien}, Thomas and {Oeur}, Micah and {Oravetz}, Audrey and {Oravetz}, Daniel and {Otto}, Jonah and {Adamane Pallathadka}, Gautham and {Palunas}, Povilas and {Pan}, Kaike and {Pappalardo}, Daniel and {Pandey}, Rakesh and {Negrete Pe{\~n}aloza}, Castalia Alenka and {Pinsonneault}, Marc H. and {Pogge}, Richard W. and {Taghizadeh Popp}, Manuchehr and {Price-Whelan}, Adrian M. and {Pulatova}, Nadiia and {Qiu}, Dan and {Ramirez}, Solange and {Rankine}, Amy and {Ricci}, Claudio and {Runnoe}, Jessie C. and {Sanchez}, Sebastian and {Salvato}, Mara and {Sattler}, Natascha and {Saydjari}, Andrew K. and {Sayres}, Conor and {Schlaufman}, Kevin C. and {Schneider}, Donald P. and {Schreiber}, Matthias R. and {Schwope}, Axel and {Serna}, Javier and {Shen}, Yue and {Sif{\'o}n}, Crist{\'o}bal and {Singh}, Amrita and {Sinha}, Amaya and {Smee}, Stephen and {Song}, Ying-Yi and {Souto}, Diogo and {Stassun}, Keivan G. and {Steinmetz}, Matthias and {Stone-Martinez}, Alexander and {Stringfellow}, Guy and {Stutz}, Amelia and {Jos{\'e}} and {S{\'a}} and {nchez-Gallego} and {Tan}, Jonathan C. and {Tayar}, Jamie and {Thai}, Riley and {Thakar}, Ani and {Ting}, Yuan-Sen and {Tkachenko}, Andrew and {Tovmasian}, Gagik and {Trakhtenbrot}, Benny and {Fern{\'a}ndez-Trincado}, Jos{\'e} G. and {Troup}, Nicholas and {Trump}, Jonathan and {Tuttle}, Sarah and {van der Marel}, Roeland P. and {Villanova}, Sandro},
        title = "{Sloan Digital Sky Survey-V: Pioneering Panoptic Spectroscopy}",
      journal = {AJ, in press},
         year = {2026},
 primaryClass = {astro-ph.IM},
       adsurl = {https://ui.adsabs.harvard.edu/abs/2025arXiv250706989K}
}

@ARTICLE{Dennefeld2020,
       author = {{Dennefeld}, M.},
        title = "{A History of the Magellanic Clouds and the European Exploration of the Southern Hemisphere}",
      journal = {The Messenger},
         year = 2020,
        month = sep,
       volume = {181},
        pages = {37-42},
          doi = {10.18727/0722-6691/5210},
archivePrefix = {arXiv},
       eprint = {2009.04973},
 primaryClass = {physics.hist-ph},
       adsurl = {https://ui.adsabs.harvard.edu/abs/2020Msngr.181...37D}
}

@ARTICLE{DrlicaWagner2016,
       author = {{Drlica-Wagner}, A. and {Bechtol}, K. and {Allam}, S. and {Tucker}, D.~L. and {Gruendl}, R.~A. and {Johnson}, M.~D. and {Walker}, A.~R. and {James}, D.~J. and {Nidever}, D.~L. and {Olsen}, K.~A.~G. and {Wechsler}, R.~H. and {Cioni}, M.~R.~L. and {Conn}, B.~C. and {Kuehn}, K. and {Li}, T.~S. and {Mao}, Y.-Y. and {Martin}, N.~F. and {Neilsen}, E. and {Noel}, N.~E.~D. and {Pieres}, A. and {Simon}, J.~D. and {Stringfellow}, G.~S. and {van der Marel}, R.~P. and {Yanny}, B.},
        title = "{An Ultra-faint Galaxy Candidate Discovered in Early Data from the Magellanic Satellites Survey}",
      journal = {\apjl},
         year = 2016,
        month = dec,
       volume = {833},
       number = {1},
          eid = {L5},
        pages = {L5},
          doi = {10.3847/2041-8205/833/1/L5},
archivePrefix = {arXiv},
       eprint = {1609.02148},
 primaryClass = {astro-ph.GA},
       adsurl = {https://ui.adsabs.harvard.edu/abs/2016ApJ...833L...5D}
}

@ARTICLE{Harris2016,
       author = {{Harris}, William E. and {Blakeslee}, John P. and {Whitmore}, Bradley C. and {Gnedin}, Oleg Y. and {Geisler}, Douglas and {Rothberg}, Barry},
        title = "{Globular Cluster Systems in Brightest Cluster Galaxies. II. NGC 6166}",
      journal = {\apj},
         year = 2016,
        month = jan,
       volume = {817},
       number = {1},
          eid = {58},
        pages = {58},
          doi = {10.3847/0004-637X/817/1/58},
archivePrefix = {arXiv},
       eprint = {1511.08493},
 primaryClass = {astro-ph.GA},
       adsurl = {https://ui.adsabs.harvard.edu/abs/2016ApJ...817...58H}
}

@ARTICLE{Conn2007,
       author = {{Conn}, Blair C. and {Lane}, Richard R. and {Lewis}, Geraint F. and {Gil-Merino}, Rodrigo and {Irwin}, Mike J. and {Ibata}, Rodrigo A. and {Martin}, Nicolas F. and {Bellazzini}, Michele and {Sharp}, Robert and {Tuntsov}, Artem V. and {Ferguson}, Annette M.~N.},
        title = "{The AAT/WFI survey of the Monoceros Ring and Canis Major dwarf galaxy - I. From l = (193-276){\textdegree}}",
      journal = {\mnras},
         year = 2007,
        month = apr,
       volume = {376},
       number = {3},
        pages = {939-959},
          doi = {10.1111/j.1365-2966.2007.11503.x},
archivePrefix = {arXiv},
       eprint = {astro-ph/0701664},
 primaryClass = {astro-ph},
       adsurl = {https://ui.adsabs.harvard.edu/abs/2007MNRAS.376..939C}
}

@ARTICLE{Abbott2021,
       author = {{Abbott}, T.~M.~C. and {Adam{\'o}w}, M. and {Aguena}, M. and {Allam}, S. and {Amon}, A. and {Annis}, J. and {Avila}, S. and {Bacon}, D. and {Banerji}, M. and {Bechtol}, K. and {Becker}, M.~R. and {Bernstein}, G.~M. and {Bertin}, E. and {Bhargava}, S. and {Bridle}, S.~L. and {Brooks}, D. and {Burke}, D.~L. and {Carnero Rosell}, A. and {Carrasco Kind}, M. and {Carretero}, J. and {Castander}, F.~J. and {Cawthon}, R. and {Chang}, C. and {Choi}, A. and {Conselice}, C. and {Costanzi}, M. and {Crocce}, M. and {da Costa}, L.~N. and {Davis}, T.~M. and {De Vicente}, J. and {DeRose}, J. and {Desai}, S. and {Diehl}, H.~T. and {Dietrich}, J.~P. and {Drlica-Wagner}, A. and {Eckert}, K. and {Elvin-Poole}, J. and {Everett}, S. and {Evrard}, A.~E. and {Ferrero}, I. and {Fert{\'e}}, A. and {Flaugher}, B. and {Fosalba}, P. and {Friedel}, D. and {Frieman}, J. and {Garc{\'\i}a-Bellido}, J. and {Gaztanaga}, E. and {Gelman}, L. and {Gerdes}, D.~W. and {Giannantonio}, T. and {Gill}, M.~S.~S. and {Gruen}, D. and {Gruendl}, R.~A. and {Gschwend}, J. and {Gutierrez}, G. and {Hartley}, W.~G. and {Hinton}, S.~R. and {Hollowood}, D.~L. and {Honscheid}, K. and {Huterer}, D. and {James}, D.~J. and {Jeltema}, T. and {Johnson}, M.~D. and {Kent}, S. and {Kron}, R. and {Kuehn}, K. and {Kuropatkin}, N. and {Lahav}, O. and {Li}, T.~S. and {Lidman}, C. and {Lin}, H. and {MacCrann}, N. and {Maia}, M.~A.~G. and {Manning}, T.~A. and {Maloney}, J.~D. and {March}, M. and {Marshall}, J.~L. and {Martini}, P. and {Melchior}, P. and {Menanteau}, F. and {Miquel}, R. and {Morgan}, R. and {Myles}, J. and {Neilsen}, E. and {Ogando}, R.~L.~C. and {Palmese}, A. and {Paz-Chinch{\'o}n}, F. and {Petravick}, D. and {Pieres}, A. and {Plazas}, A.~A. and {Pond}, C. and {Rodriguez-Monroy}, M. and {Romer}, A.~K. and {Roodman}, A. and {Rykoff}, E.~S. and {Sako}, M. and {Sanchez}, E. and {Santiago}, B. and {Scarpine}, V. and {Serrano}, S. and {Sevilla-Noarbe}, I. and {Smith}, J. Allyn and {Smith}, M. and {Soares-Santos}, M. and {Suchyta}, E. and {Swanson}, M.~E.~C. and {Tarle}, G. and {Thomas}, D. and {To}, C. and {Tremblay}, P.~E. and {Troxel}, M.~A. and {Tucker}, D.~L. and {Turner}, D.~J. and {Varga}, T.~N. and {Walker}, A.~R. and {Wechsler}, R.~H. and {Weller}, J. and {Wester}, W. and {Wilkinson}, R.~D. and {Yanny}, B. and {Zhang}, Y. and {Nikutta}, R. and {Fitzpatrick}, M. and {Jacques}, A. and {Scott}, A. and {Olsen}, K. and {Huang}, L. and {Herrera}, D. and {Juneau}, S. and {Nidever}, D. and {Weaver}, B.~A. and {Adean}, C. and {Correia}, V. and {de Freitas}, M. and {Freitas}, F.~N. and {Singulani}, C. and {Vila-Verde}, G. and {Linea Science Server}},
        title = "{The Dark Energy Survey Data Release 2}",
      journal = {\apjs},
         year = 2021,
        month = aug,
       volume = {255},
       number = {2},
          eid = {20},
        pages = {20},
          doi = {10.3847/1538-4365/ac00b3},
archivePrefix = {arXiv},
       eprint = {2101.05765},
 primaryClass = {astro-ph.IM},
       adsurl = {https://ui.adsabs.harvard.edu/abs/2021ApJS..255...20A}
}

@ARTICLE{Stringer2021,
       author = {{Stringer}, K.~M. and {Drlica-Wagner}, A. and {Macri}, L. and {Mart{\'\i}nez-V{\'a}zquez}, C.~E. and {Vivas}, A.~K. and {Ferguson}, P. and {Pace}, A.~B. and {Walker}, A.~R. and {Neilsen}, E. and {Tavangar}, K. and {Wester}, W. and {Abbott}, T.~M.~C. and {Aguena}, M. and {Allam}, S. and {Bacon}, D. and {Bechtol}, K. and {Bertin}, E. and {Brooks}, D. and {Burke}, D.~L. and {Carnero Rosell}, A. and {Carrasco Kind}, M. and {Carretero}, J. and {Costanzi}, M. and {Crocce}, M. and {da Costa}, L.~N. and {Pereira}, M.~E.~S. and {De Vicente}, J. and {Desai}, S. and {Diehl}, H.~T. and {Doel}, P. and {Ferrero}, I. and {Garc{\'\i}a-Bellido}, J. and {Gaztanaga}, E. and {Gerdes}, D.~W. and {Gruen}, D. and {Gruendl}, R.~A. and {Gschwend}, J. and {Gutierrez}, G. and {Hinton}, S.~R. and {Hollowood}, D.~L. and {Honscheid}, K. and {Hoyle}, B. and {James}, D.~J. and {Kuehn}, K. and {Kuropatkin}, N. and {Li}, T.~S. and {Maia}, M.~A.~G. and {Marshall}, J.~L. and {Menanteau}, F. and {Miquel}, R. and {Morgan}, R. and {Ogando}, R.~L.~C. and {Palmese}, A. and {Paz-Chinch{\'o}n}, F. and {Plazas}, A.~A. and {Roodman}, A. and {Sanchez}, E. and {Schubnell}, M. and {Serrano}, S. and {Sevilla-Noarbe}, I. and {Smith}, M. and {Soares-Santos}, M. and {Suchyta}, E. and {Tarle}, G. and {Thomas}, D. and {To}, C. and {Varga}, T.~N. and {Wilkinson}, R.~D. and {Zhang}, Y. and {DES Collaboration}},
        title = "{Identifying RR Lyrae Variable Stars in Six Years of the Dark Energy Survey}",
      journal = {\apj},
         year = 2021,
        month = apr,
       volume = {911},
       number = {2},
          eid = {109},
        pages = {109},
          doi = {10.3847/1538-4357/abe873},
archivePrefix = {arXiv},
       eprint = {2011.13930},
 primaryClass = {astro-ph.GA},
       adsurl = {https://ui.adsabs.harvard.edu/abs/2021ApJ...911..109S}
}

@ARTICLE{Freedman2010,
       author = {{Freedman}, Wendy L. and {Madore}, Barry F.},
        title = "{The Hubble Constant}",
      journal = {\araa},
         year = 2010,
        month = sep,
       volume = {48},
        pages = {673-710},
          doi = {10.1146/annurev-astro-082708-101829},
archivePrefix = {arXiv},
       eprint = {1004.1856},
 primaryClass = {astro-ph.CO},
       adsurl = {https://ui.adsabs.harvard.edu/abs/2010ARA&A..48..673F}
}

@ARTICLE{Tonry2018,
       author = {{Tonry}, J.~L. and {Denneau}, L. and {Flewelling}, H. and
         {Heinze}, A.~N. and {Onken}, C.~A. and {Smartt}, S.~J. and
         {Stalder}, B. and {Weiland}, H.~J. and {Wolf}, C.},
        title = "{The ATLAS All-Sky Stellar Reference Catalog}",
      journal = {\apj},
         year = 2018,
        month = nov,
       volume = {867},
       number = {2},
          eid = {105},
        pages = {105},
          doi = {10.3847/1538-4357/aae386},
archivePrefix = {arXiv},
       eprint = {1809.09157},
 primaryClass = {astro-ph.IM},
       adsurl = {https://ui.adsabs.harvard.edu/abs/2018ApJ...867..105T}
}

@ARTICLE{Oden2025,
       author = {{Oden}, Slater J. and {Nidever}, David L. and {Povick}, Joshua and {Massana}, Pol and {Choi}, Yumi and {van der Marel}, Roeland P. and {Cioni}, Maria-Rosa L. and {Sakowska}, Joanna and {Olsen}, Knut A.~G. and {Cullinane}, Lara and {Carballo-Bello}, J.~A. and {Crnojevi{\'c}}, D. and {Ferguson}, P.~S. and {Mart{\'\i}nez-V{\'a}zquez}, C.~E. and {Medina}, G.~E. and {Mutlu-Pakdil}, B. and {Navabi}, M. and {Pace}, A.~B. and {Riley}, A.~H. and {Stringfellow}, Guy S. and {Vivas}, A.~K.},
        title = "{Warped \& Hooked: Mapping the Magellanic Clouds in 3D using Red Clump stars}",
      journal = {arXiv e-prints},
         year = 2025,
        month = dec,
          eid = {arXiv:2512.04200},
        pages = {arXiv:2512.04200},
          doi = {10.48550/arXiv.2512.04200},
archivePrefix = {arXiv},
       eprint = {2512.04200},
 primaryClass = {astro-ph.GA},
       adsurl = {https://ui.adsabs.harvard.edu/abs/2025arXiv251204200O}
}

@article{Nidever2026,
  author = {{Nidever}, D. L. and {Horta}, D. and {Majewski}, S. R. and et al.},
  title = {Overview of the SDSS-V Magellanic Genesis Survey},
  journal = {AJ, submitted},
  year = {2026},
  eprint = {2602.02693},
  archivePrefix = {arXiv},
  primaryClass = {astro-ph.GA}
}

@ARTICLE{Oden2026,
       author = {{Oden}, S. and {Nidever}, D. L. and {Massana}, P. and others},
        title = "{DELVE-MC Star Formation Histories}",
	Year = {2026, in prep.}
}

@ARTICLE{Erkal2019,
       author = {{Erkal}, D. and {Belokurov}, V. and {Laporte}, C.~F.~P. and {Koposov}, S.~E. and {Li}, T.~S. and {Grillmair}, C.~J. and {Kallivayalil}, N. and {Price-Whelan}, A.~M. and {Evans}, N.~W. and {Hawkins}, K. and {Hendel}, D. and {Mateu}, C. and {Navarro}, J.~F. and {del Pino}, A. and {Slater}, C.~T. and {Sohn}, S.~T. and {Orphan Aspen Treasury Collaboration}},
        title = "{The total mass of the Large Magellanic Cloud from its perturbation on the Orphan stream}",
      journal = {\mnras},
         year = 2019,
        month = aug,
       volume = {487},
       number = {2},
        pages = {2685-2700},
          doi = {10.1093/mnras/stz1371},
archivePrefix = {arXiv},
       eprint = {1812.08192},
 primaryClass = {astro-ph.GA},
       adsurl = {https://ui.adsabs.harvard.edu/abs/2019MNRAS.487.2685E}
}

@ARTICLE{Nidever2008,
       author = {{Nidever}, David L. and {Majewski}, Steven R. and {Butler Burton}, W.},
        title = "{The Origin of the Magellanic Stream and Its Leading Arm}",
      journal = {\apj},
         year = 2008,
        month = may,
       volume = {679},
       number = {1},
        pages = {432-459},
          doi = {10.1086/587042},
       adsurl = {https://ui.adsabs.harvard.edu/abs/2008ApJ...679..432N}
}

@ARTICLE{Onken2024,
       author = {{Onken}, Christopher A. and {Wolf}, Christian and {Bessell}, Michael S. and {Chang}, Seo-Won and {Luvaul}, Lance C. and {Tonry}, John L. and {White}, Marc C. and {Da Costa}, Gary S.},
        title = "{SkyMapper Southern Survey: Data release 4}",
      journal = {\pasa},
         year = 2024,
        month = oct,
       volume = {41},
          eid = {e061},
        pages = {e061},
          doi = {10.1017/pasa.2024.53},
archivePrefix = {arXiv},
       eprint = {2402.02015},
 primaryClass = {astro-ph.CO},
       adsurl = {https://ui.adsabs.harvard.edu/abs/2024PASA...41...61O}
}

@ARTICLE{Wolf2018,
       author = {{Wolf}, Christian and {Onken}, Christopher A. and {Luvaul}, Lance C. and
         {Schmidt}, Brian P. and {Bessell}, Michael S. and {Chang}, Seo-Won and
         {Da Costa}, Gary S. and {Mackey}, Dougal and {Martin-Jones}, Tony and
         {Murphy}, Simon J. and {Preston}, Tim and {Scalzo}, Richard A. and
         {Shao}, Li and {Smillie}, Jon and {Tisserand}, Patrick and
         {White}, Marc C. and {Yuan}, Fang},
        title = "{SkyMapper Southern Survey: First Data Release (DR1)}",
      journal = {\pasa},
         year = 2018,
        month = feb,
       volume = {35},
          eid = {e010},
        pages = {e010},
          doi = {10.1017/pasa.2018.5},
archivePrefix = {arXiv},
       eprint = {1801.07834},
 primaryClass = {astro-ph.IM},
       adsurl = {https://ui.adsabs.harvard.edu/abs/2018PASA...35...10W}
}

@ARTICLE{Bernard2015a,
       author = {{Bernard}, Edouard J. and {Ferguson}, Annette M.~N. and {Chapman}, Scott C. and {Ibata}, Rodrigo A. and {Irwin}, Mike J. and {Lewis}, Geraint F. and {McConnachie}, Alan W.},
        title = "{The spatially-resolved star formation history of the M31 outer disc}",
      journal = {\mnras},
         year = 2015,
        month = oct,
       volume = {453},
       number = {1},
        pages = {L113-L117},
          doi = {10.1093/mnrasl/slv116},
archivePrefix = {arXiv},
       eprint = {1508.01559},
 primaryClass = {astro-ph.GA},
       adsurl = {https://ui.adsabs.harvard.edu/abs/2015MNRAS.453L.113B}
}

@ARTICLE{Bernard2018,
       author = {{Bernard}, Edouard J. and {Schultheis}, Mathias and {Di Matteo}, Paola and {Hill}, Vanessa and {Haywood}, Misha and {Calamida}, Annalisa},
        title = "{Star formation history of the Galactic bulge from deep HST imaging of low reddening windows}",
      journal = {\mnras},
         year = 2018,
        month = jul,
       volume = {477},
       number = {3},
        pages = {3507-3519},
          doi = {10.1093/mnras/sty902},
archivePrefix = {arXiv},
       eprint = {1801.01426},
 primaryClass = {astro-ph.GA},
       adsurl = {https://ui.adsabs.harvard.edu/abs/2018MNRAS.477.3507B}
}

@ARTICLE{Nidever2013,
       author = {{Nidever}, David L. and {Monachesi}, Antonela and {Bell}, Eric F. and {Majewski}, Steven R. and {Mu{\~n}oz}, Ricardo R. and {Beaton}, Rachael L.},
        title = "{A Tidally Stripped Stellar Component of the Magellanic Bridge}",
      journal = {\apj},
         year = 2013,
        month = dec,
       volume = {779},
       number = {2},
          eid = {145},
        pages = {145},
          doi = {10.1088/0004-637X/779/2/145},
archivePrefix = {arXiv},
       eprint = {1310.4824},
 primaryClass = {astro-ph.GA},
       adsurl = {https://ui.adsabs.harvard.edu/abs/2013ApJ...779..145N}
}

@ARTICLE{Massana2026,
       author = {{Massana}, P. and {Nidever}, D. L. and others},
        title = "{The Clouds extended stellar periphery: a DELVE-MC view}",
	Year = {2026, in prep.}
}

@ARTICLE{Mackey2018,
       author = {{Mackey}, Dougal and {Koposov}, Sergey and {Da Costa}, Gary and {Belokurov}, Vasily and {Erkal}, Denis and {Kuzma}, Pete},
        title = "{Substructures and Tidal Distortions in the Magellanic Stellar Periphery}",
      journal = {\apjl},
         year = 2018,
        month = may,
       volume = {858},
       number = {2},
          eid = {L21},
        pages = {L21},
          doi = {10.3847/2041-8213/aac175},
archivePrefix = {arXiv},
       eprint = {1804.06431},
 primaryClass = {astro-ph.GA},
       adsurl = {https://ui.adsabs.harvard.edu/abs/2018ApJ...858L..21M}
}

@ARTICLE{Mackey2016,
       author = {{Mackey}, A.~D. and {Koposov}, S.~E. and {Erkal}, D. and {Belokurov}, V. and {Da Costa}, G.~S. and {G{\'o}mez}, F.~A.},
        title = "{A 10 kpc stellar substructure at the edge of the Large Magellanic Cloud: perturbed outer disc or evidence for tidal stripping?}",
      journal = {\mnras},
         year = 2016,
        month = jun,
       volume = {459},
       number = {1},
        pages = {239-255},
          doi = {10.1093/mnras/stw497},
archivePrefix = {arXiv},
       eprint = {1508.01356},
 primaryClass = {astro-ph.GA},
       adsurl = {https://ui.adsabs.harvard.edu/abs/2016MNRAS.459..239M}
}

@ARTICLE{Nikutta2020,
       author = {{Nikutta}, R. and {Fitzpatrick}, M. and {Scott}, A. and {Weaver}, B.~A.},
        title = "{Data Lab-A community science platform}",
      journal = {Astronomy and Computing},
         year = 2020,
        month = oct,
       volume = {33},
          eid = {100411},
        pages = {100411},
          doi = {10.1016/j.ascom.2020.100411},
       adsurl = {https://ui.adsabs.harvard.edu/abs/2020A&C....3300411N}
}

@INPROCEEDINGS{Fitzpatrick2016,
       author = {{Fitzpatrick}, Michael J. and {Graham}, Matthew J. and {Mighell}, Kenneth J. and {Olsen}, Knut and {Norris}, Patrick and {Ridgway}, Stephen T. and {Stobie}, Elizabeth B. and {Bolton}, Adam S. and {Saha}, Abhijit and {Huang}, Lijuan W.},
        title = "{The NOAO data lab: science-driven development}",
    booktitle = {Software and Cyberinfrastructure for Astronomy IV},
         year = 2016,
       editor = {{Chiozzi}, Gianluca and {Guzman}, Juan C.},
       series = {Society of Photo-Optical Instrumentation Engineers (SPIE) Conference Series},
       volume = {9913},
        month = aug,
          eid = {99130L},
        pages = {99130L},
          doi = {10.1117/12.2233791},
       adsurl = {https://ui.adsabs.harvard.edu/abs/2016SPIE.9913E..0LF}
}

@techreport{Neilsen2015,
       author = {{Neilsen}, E. and {Bernstein}, G. and {Gruendl}, R. and {Kent}, S.}, 
        title = "{$\tau$, $T_{\rm eff}$, and Image Quality in DES Year 1}",
  institution = {FERMILAB-TM-2610-AE-CD},
         year = 2015,
}

@ARTICLE{Fitzpatrick2024,
       author = {{Fitzpatrick}, Michael and {Placco}, Vinicius and {Bolton}, Adam and {Merino}, Brian and {Ridgway}, Susan and {Stanghellini}, Letizia},
        title = "{Modernizing IRAF to Support Gemini Data Reduction}",
      journal = {arXiv e-prints},
         year = 2024,
        month = jan,
          eid = {arXiv:2401.01982},
        pages = {arXiv:2401.01982},
          doi = {10.48550/arXiv.2401.01982},
archivePrefix = {arXiv},
       eprint = {2401.01982},
 primaryClass = {astro-ph.IM},
       adsurl = {https://ui.adsabs.harvard.edu/abs/2024arXiv240101982F}
}

@INPROCEEDINGS{Tody1993,
       author = {{Tody}, Doug},
        title = "{IRAF in the Nineties}",
    booktitle = {Astronomical Data Analysis Software and Systems II},
         year = 1993,
       editor = {{Hanisch}, R.~J. and {Brissenden}, R.~J.~V. and {Barnes}, J.},
       series = {Astronomical Society of the Pacific Conference Series},
       volume = {52},
        month = jan,
        pages = {173},
       adsurl = {https://ui.adsabs.harvard.edu/abs/1993ASPC...52..173T}
}

@INPROCEEDINGS{Tody1986,
       author = {{Tody}, Doug},
        title = "{The IRAF Data Reduction and Analysis System}",
    booktitle = {Instrumentation in astronomy VI},
         year = 1986,
       editor = {{Crawford}, David L.},
       series = {Society of Photo-Optical Instrumentation Engineers (SPIE) Conference Series},
       volume = {627},
        month = jan,
        pages = {733},
          doi = {10.1117/12.968154},
       adsurl = {https://ui.adsabs.harvard.edu/abs/1986SPIE..627..733T}
}

@INPROCEEDINGS{Valdes2014,
       author = {{Valdes}, F. and {Gruendl}, R. and {DES Project}},
        title = "{The DECam Community Pipeline}",
    booktitle = {Astronomical Data Analysis Software and Systems XXIII},
         year = 2014,
       editor = {{Manset}, N. and {Forshay}, P.},
       series = {Astronomical Society of the Pacific Conference Series},
       volume = {485},
        month = may,
        pages = {379},
       adsurl = {https://ui.adsabs.harvard.edu/abs/2014ASPC..485..379V}
}

@ARTICLE{Nidever2021b,
       author = {{Nidever}, David L. and {Dey}, Arjun and {Fasbender}, Katie and {Juneau}, St{\'e}phanie and {Meisner}, Aaron M. and {Wishart}, Joseph and {Scott}, Adam and {Matt}, Kyle and {Nikutta}, Robert and {Pucha}, Ragadeepika},
        title = "{Second Data Release of the All-sky NOIRLab Source Catalog}",
      journal = {\aj},
         year = {2021b},
        month = apr,
       volume = {161},
       number = {4},
          eid = {192},
        pages = {192},
          doi = {10.3847/1538-3881/abd6e1},
archivePrefix = {arXiv},
       eprint = {2011.08868},
 primaryClass = {astro-ph.GA},
       adsurl = {https://ui.adsabs.harvard.edu/abs/2021AJ....161..192N}
}

@article{scikit-learn,
 title={Scikit-learn: Machine Learning in {P}ython},
 author={Pedregosa, F. and Varoquaux, G. and Gramfort, A. and Michel, V.
         and Thirion, B. and Grisel, O. and Blondel, M. and Prettenhofer, P.
         and Weiss, R. and Dubourg, V. and Vanderplas, J. and Passos, A. and
         Cournapeau, D. and Brucher, M. and Perrot, M. and Duchesnay, E.},
 journal={Journal of Machine Learning Research},
 volume={12},
 pages={2825--2830},
 year={2011}
}

@ARTICLE{Bertin1996,
       author = {{Bertin}, E. and {Arnouts}, S.},
        title = "{SExtractor: Software for source extraction.}",
      journal = {\aaps},
         year = 1996,
        month = jun,
       volume = {117},
        pages = {393-404},
          doi = {10.1051/aas:1996164},
       adsurl = {https://ui.adsabs.harvard.edu/abs/1996A&AS..117..393B}
}

@article{Zonca2019,
  doi = {10.21105/joss.01298},
  url = {https://doi.org/10.21105/joss.01298},
  year = {2019},
  month = mar,
  publisher = {The Open Journal},
  volume = {4},
  number = {35},
  pages = {1298},
  author = {Andrea Zonca and Leo Singer and Daniel Lenz and Martin Reinecke and Cyrille Rosset and Eric Hivon and Krzysztof Gorski},
  title = {healpy: equal area pixelization and spherical harmonics transforms for data on the sphere in Python},
  journal = {Journal of Open Source Software}
}

@article{ipython,
        Author = {P\'erez, Fernando and Granger, Brian E.},
        Doi = {10.1109/MCSE.2007.53},
        Issn = {1521-9615},
        Journal = {Computing in Science and Engineering},
        Month = may,
        Number = {3},
        Pages = {21--29},
        Publisher = {IEEE Computer Society},
        Title = {{IP}ython: a System for Interactive Scientific Computing},
        Url = {http://ipython.org},
        Volume = {9},
        Year = 2007}

@misc{scipy,
        Author = {Eric Jones and Travis Oliphant and Pearu Peterson and others},
        Title = {{SciPy}: Open source scientific tools for {Python}},
        Url = {http://www.scipy.org/},
        Year = {2001--}}

@article{numpy,
        Acmid = {1957466},
        Address = {Piscataway, NJ, USA},
        Author = {Walt, Stefan van der and Colbert, S. Chris and Varoquaux, Gael},
        Doi = {10.1109/MCSE.2011.37},
        Issn = {1521-9615},
        Issue_Date = {March 2011},
        Journal = {Computing in Science and Engg.},
        Month = mar,
        Number = {2},
        Numpages = {9},
        Pages = {22--30},
        Publisher = {IEEE Educational Activities Department},
        Title = {The NumPy Array: A Structure for Efficient Numerical Computation},
        Url = {http://dx.doi.org/10.1109/MCSE.2011.37},
        Volume = {13},
        Year = {2011}}

@article{mpl,
        Adsurl = {http://adsabs.harvard.edu/abs/2007CSE.....9...90H},
        Author = {{Hunter}, J.~D.},
        Doi = {10.1109/MCSE.2007.55},
        Journal = {Computing in Science and Engineering},
        Month = may,
        Pages = {90-95},
        Title = {{Matplotlib: A 2D Graphics Environment}},
        Volume = 9,
        Year = 2007}

@ARTICLE{Stetson1994,
       author = {{Stetson}, Peter B.},
        title = "{The Center of the Core-Cusp Globular Cluster M15: CFHT and HST Observations, ALLFRAME Reductions}",
      journal = {\pasp},
         year = 1994,
        month = mar,
       volume = {106},
        pages = {250},
          doi = {10.1086/133378},
       adsurl = {https://ui.adsabs.harvard.edu/abs/1994PASP..106..250S}
}

@ARTICLE{Stetson1990,
       author = {{Stetson}, Peter B.},
        title = "{On the Growth-Curve Method for Calibrating Stellar Photometry with CCDs}",
      journal = {\pasp},
         year = 1990,
        month = aug,
       volume = {102},
        pages = {932},
          doi = {10.1086/132719},
       adsurl = {https://ui.adsabs.harvard.edu/abs/1990PASP..102..932S}
}

@ARTICLE{Stetson1987,
       author = {{Stetson}, Peter B.},
        title = "{DAOPHOT: A Computer Program for Crowded-Field Stellar Photometry}",
      journal = {\pasp},
         year = 1987,
        month = mar,
       volume = {99},
        pages = {191},
          doi = {10.1086/131977},
       adsurl = {https://ui.adsabs.harvard.edu/abs/1987PASP...99..191S}
}

@software{photred:2025,
       author = {{Nidever}, David},
        title = "{dnidever/PHOTRED: First DELVE-MC Data Release}",
         year = 2025,
        month = jun,
          eid = {10.5281/zenodo.16812729},
          doi = {10.5281/zenodo.16812729},
      version = {v1.2.0},
    publisher = {Zenodo},
       adsurl = {https://ui.adsabs.harvard.edu/abs/2025zndo...XXXXX}
}

@ARTICLE{Dey2019,
       author = {{Dey}, Arjun and {Schlegel}, David J. and {Lang}, Dustin and {Blum}, Robert and {Burleigh}, Kaylan and {Fan}, Xiaohui and {Findlay}, Joseph R. and {Finkbeiner}, Doug and {Herrera}, David and {Juneau}, St{\'e}phanie and {Landriau}, Martin and {Levi}, Michael and {McGreer}, Ian and {Meisner}, Aaron and {Myers}, Adam D. and {Moustakas}, John and {Nugent}, Peter and {Patej}, Anna and {Schlafly}, Edward F. and {Walker}, Alistair R. and {Valdes}, Francisco and {Weaver}, Benjamin A. and {Y{\`e}che}, Christophe and {Zou}, Hu and {Zhou}, Xu and {Abareshi}, Behzad and {Abbott}, T.~M.~C. and {Abolfathi}, Bela and {Aguilera}, C. and {Alam}, Shadab and {Allen}, Lori and {Alvarez}, A. and {Annis}, James and {Ansarinejad}, Behzad and {Aubert}, Marie and {Beechert}, Jacqueline and {Bell}, Eric F. and {BenZvi}, Segev Y. and {Beutler}, Florian and {Bielby}, Richard M. and {Bolton}, Adam S. and {Brice{\~n}o}, C{\'e}sar and {Buckley-Geer}, Elizabeth J. and {Butler}, Karen and {Calamida}, Annalisa and {Carlberg}, Raymond G. and {Carter}, Paul and {Casas}, Ricard and {Castander}, Francisco J. and {Choi}, Yumi and {Comparat}, Johan and {Cukanovaite}, Elena and {Delubac}, Timoth{\'e}e and {DeVries}, Kaitlin and {Dey}, Sharmila and {Dhungana}, Govinda and {Dickinson}, Mark and {Ding}, Zhejie and {Donaldson}, John B. and {Duan}, Yutong and {Duckworth}, Christopher J. and {Eftekharzadeh}, Sarah and {Eisenstein}, Daniel J. and {Etourneau}, Thomas and {Fagrelius}, Parker A. and {Farihi}, Jay and {Fitzpatrick}, Mike and {Font-Ribera}, Andreu and {Fulmer}, Leah and {G{\"a}nsicke}, Boris T. and {Gaztanaga}, Enrique and {George}, Koshy and {Gerdes}, David W. and {Gontcho}, Satya Gontcho A. and {Gorgoni}, Claudio and {Green}, Gregory and {Guy}, Julien and {Harmer}, Diane and {Hernandez}, M. and {Honscheid}, Klaus and {Huang}, Lijuan Wendy and {James}, David J. and {Jannuzi}, Buell T. and {Jiang}, Linhua and {Joyce}, Richard and {Karcher}, Armin and {Karkar}, Sonia and {Kehoe}, Robert and {Kneib}, Jean-Paul and {Kueter-Young}, Andrea and {Lan}, Ting-Wen and {Lauer}, Tod R. and {Le Guillou}, Laurent and {Le Van Suu}, Auguste and {Lee}, Jae Hyeon and {Lesser}, Michael and {Perreault Levasseur}, Laurence and {Li}, Ting S. and {Mann}, Justin L. and {Marshall}, Robert and {Mart{\'\i}nez-V{\'a}zquez}, C.~E. and {Martini}, Paul and {du Mas des Bourboux}, H{\'e}lion and {McManus}, Sean and {Meier}, Tobias Gabriel and {M{\'e}nard}, Brice and {Metcalfe}, Nigel and {Mu{\~n}oz-Guti{\'e}rrez}, Andrea and {Najita}, Joan and {Napier}, Kevin and {Narayan}, Gautham and {Newman}, Jeffrey A. and {Nie}, Jundan and {Nord}, Brian and {Norman}, Dara J. and {Olsen}, Knut A.~G. and {Paat}, Anthony and {Palanque-Delabrouille}, Nathalie and {Peng}, Xiyan and {Poppett}, Claire L. and {Poremba}, Megan R. and {Prakash}, Abhishek and {Rabinowitz}, David and {Raichoor}, Anand and {Rezaie}, Mehdi and {Robertson}, A.~N. and {Roe}, Natalie A. and {Ross}, Ashley J. and {Ross}, Nicholas P. and {Rudnick}, Gregory and {Safonova}, Sasha and {Saha}, Abhijit and {S{\'a}nchez}, F. Javier and {Savary}, Elodie and {Schweiker}, Heidi and {Scott}, Adam and {Seo}, Hee-Jong and {Shan}, Huanyuan and {Silva}, David R. and {Slepian}, Zachary and {Soto}, Christian and {Sprayberry}, David and {Staten}, Ryan and {Stillman}, Coley M. and {Stupak}, Robert J. and {Summers}, David L. and {Sien Tie}, Suk and {Tirado}, H. and {Vargas-Maga{\~n}a}, Mariana and {Vivas}, A. Katherina and {Wechsler}, Risa H. and {Williams}, Doug and {Yang}, Jinyi and {Yang}, Qian and {Yapici}, Tolga and {Zaritsky}, Dennis and {Zenteno}, A. and {Zhang}, Kai and {Zhang}, Tianmeng and {Zhou}, Rongpu and {Zhou}, Zhimin},
        title = "{Overview of the DESI Legacy Imaging Surveys}",
      journal = {\aj},
         year = 2019,
        month = may,
       volume = {157},
       number = {5},
          eid = {168},
        pages = {168},
          doi = {10.3847/1538-3881/ab089d},
archivePrefix = {arXiv},
       eprint = {1804.08657},
 primaryClass = {astro-ph.IM},
       adsurl = {https://ui.adsabs.harvard.edu/abs/2019AJ....157..168D}
}

@ARTICLE{Miller2022,
       author = {{Miller}, Amy E. and {Cioni}, Maria-Rosa L. and {de Grijs}, Richard and {Sun}, Ning-Chen and {Bell}, Cameron P.~M. and {Choudhury}, Samyaday and {Ivanov}, Valentin D. and {Marconi}, Marcella and {Oliveira}, Joana M. and {Petr-Gotzens}, Monika and {Ripepi}, Vincenzo and {van Loon}, Jacco Th},
        title = "{The VMC survey - XLVII. Turbulence-controlled hierarchical star formation in the Large Magellanic Cloud}",
      journal = {\mnras},
         year = 2022,
        month = may,
       volume = {512},
       number = {1},
        pages = {1196-1213},
          doi = {10.1093/mnras/stac508},
archivePrefix = {arXiv},
       eprint = {2202.09267},
 primaryClass = {astro-ph.GA},
       adsurl = {https://ui.adsabs.harvard.edu/abs/2022MNRAS.512.1196M}
}

@ARTICLE{Piatti2015,
       author = {{Piatti}, Andr{\'e}s E. and {de Grijs}, Richard and {Ripepi}, Vincenzo and {Ivanov}, Valentin D. and {Cioni}, Maria-Rosa L. and {Marconi}, Marcella and {Rubele}, Stefano and {Bekki}, Kenji and {For}, Bi-Qing},
        title = "{The VMC survey - XVI. Spatial variation of the cluster formation activity in the innermost regions of the Large Magellanic Cloud}",
      journal = {\mnras},
         year = 2015,
        month = nov,
       volume = {454},
       number = {1},
        pages = {839-848},
          doi = {10.1093/mnras/stv2054},
archivePrefix = {arXiv},
       eprint = {1509.00827},
 primaryClass = {astro-ph.GA},
       adsurl = {https://ui.adsabs.harvard.edu/abs/2015MNRAS.454..839P}
}

@ARTICLE{Choudhury2020,
       author = {{Choudhury}, Samyaday and {de Grijs}, Richard and {Rubele}, Stefano and {Bekki}, Kenji and {Cioni}, Maria-Rosa L. and {Ivanov}, Valentin D. and {van Loon}, Jacco Th and {Niederhofer}, Florian and {Oliveira}, Joana M. and {Ripepi}, Vincenzo},
        title = "{The VMC survey - XXXIX. Mapping metallicity trends in the Small Magellanic Cloud using near-infrared passbands}",
      journal = {\mnras},
         year = 2020,
        month = sep,
       volume = {497},
       number = {3},
        pages = {3746-3760},
          doi = {10.1093/mnras/staa2140},
archivePrefix = {arXiv},
       eprint = {2007.08753},
 primaryClass = {astro-ph.GA},
       adsurl = {https://ui.adsabs.harvard.edu/abs/2020MNRAS.497.3746C}
}

@ARTICLE{ElYoussoufi2019,
       author = {{El Youssoufi}, Dalal and {Cioni}, Maria-Rosa L. and {Bell}, Cameron P.~M. and {Rubele}, Stefano and {Bekki}, Kenji and {de Grijs}, Richard and {Girardi}, L{\'e}o and {Ivanov}, Valentin D. and {Matijevic}, Gal and {Niederhofer}, Florian and {Oliveira}, Joana M. and {Ripepi}, Vincenzo and {Subramanian}, Smitha and {van Loon}, Jacco Th},
        title = "{The VMC survey - XXXIV. Morphology of stellar populations in the Magellanic Clouds}",
      journal = {\mnras},
         year = 2019,
        month = nov,
       volume = {490},
       number = {1},
        pages = {1076-1093},
          doi = {10.1093/mnras/stz2400},
archivePrefix = {arXiv},
       eprint = {1908.08545},
 primaryClass = {astro-ph.GA},
       adsurl = {https://ui.adsabs.harvard.edu/abs/2019MNRAS.490.1076E}
}

@ARTICLE{Rubele2015,
       author = {{Rubele}, Stefano and {Girardi}, L{\'e}o and {Kerber}, Leandro and {Cioni}, Maria-Rosa L. and {Piatti}, Andr{\'e}s E. and {Zaggia}, Simone and {Bekki}, Kenji and {Bressan}, Alessandro and {Clementini}, Gisella and {de Grijs}, Richard and {Emerson}, Jim P. and {Groenewegen}, Martin A.~T. and {Ivanov}, Valentin D. and {Marconi}, Marcella and {Marigo}, Paola and {Moretti}, Maria-Ida and {Ripepi}, Vincenzo and {Subramanian}, Smitha and {Tatton}, Benjamin L. and {van Loon}, Jacco Th.},
        title = "{The VMC survey - XIV. First results on the look-back time star formation rate tomography of the Small Magellanic Cloud}",
      journal = {\mnras},
         year = 2015,
        month = may,
       volume = {449},
       number = {1},
        pages = {639-661},
          doi = {10.1093/mnras/stv141},
archivePrefix = {arXiv},
       eprint = {1501.05347},
 primaryClass = {astro-ph.SR},
       adsurl = {https://ui.adsabs.harvard.edu/abs/2015MNRAS.449..639R}
}

@ARTICLE{Massana2024,
       author = {{Massana}, Pol and {Nidever}, David L. and {Olsen}, Knut},
        title = "{The Magellanic Puzzle: origin of the periphery}",
      journal = {\mnras},
         year = 2024,
        month = jan,
       volume = {527},
       number = {3},
        pages = {8706-8717},
          doi = {10.1093/mnras/stad3788},
archivePrefix = {arXiv},
       eprint = {2307.14342},
 primaryClass = {astro-ph.GA},
       adsurl = {https://ui.adsabs.harvard.edu/abs/2024MNRAS.527.8706M}
}

@INPROCEEDINGS{Irwin1991,
       author = {{Irwin}, M.~J.},
        title = "{Optical Surveys of the Magellanic Bridge, Stream and Outer Halos}",
    booktitle = {The Magellanic Clouds},
         year = 1991,
       editor = {{Haynes}, Raymond and {Milne}, Douglas},
       series = {IAU Symposium},
       volume = {148},
        month = jan,
        pages = {453},
       adsurl = {https://ui.adsabs.harvard.edu/abs/1991IAUS..148..453I}
}

@ARTICLE{Gardiner1992,
       author = {{Gardiner}, L.~T. and {Hatzidimitriou}, D.},
        title = "{Stellar populations and large-scale structure of the Small Magellanic Cloud. IV. Age distribution studies of the outer regions.}",
      journal = {\mnras},
         year = 1992,
        month = jul,
       volume = {257},
        pages = {195-224},
          doi = {10.1093/mnras/257.2.195},
       adsurl = {https://ui.adsabs.harvard.edu/abs/1992MNRAS.257..195G}
}

@ARTICLE{Westerlund1970,
       author = {{Westerlund}, Bengt E.},
        title = "{The stellar content of the Magellanic Clouds}",
      journal = {Vistas in Astronomy},
         year = 1970,
        month = jan,
       volume = {12},
       number = {1},
        pages = {335-355},
          doi = {10.1016/0083-6656(70)90047-4},
       adsurl = {https://ui.adsabs.harvard.edu/abs/1970VA.....12..335W}
}

@ARTICLE{Zaritsky2002,
       author = {{Zaritsky}, Dennis and {Harris}, Jason and {Thompson}, Ian B. and {Grebel}, Eva K. and {Massey}, Philip},
        title = "{The Magellanic Clouds Photometric Survey: The Small Magellanic Cloud Stellar Catalog and Extinction Map}",
      journal = {\aj},
         year = 2002,
        month = feb,
       volume = {123},
       number = {2},
        pages = {855-872},
          doi = {10.1086/338437},
archivePrefix = {arXiv},
       eprint = {astro-ph/0110665},
 primaryClass = {astro-ph},
       adsurl = {https://ui.adsabs.harvard.edu/abs/2002AJ....123..855Z}
}

@ARTICLE{Zaritsky2004,
       author = {{Zaritsky}, Dennis and {Harris}, Jason and {Thompson}, Ian B. and {Grebel}, Eva K.},
        title = "{The Magellanic Clouds Photometric Survey: The Large Magellanic Cloud Stellar Catalog and Extinction Map}",
      journal = {\aj},
         year = 2004,
        month = oct,
       volume = {128},
       number = {4},
        pages = {1606-1614},
          doi = {10.1086/423910},
archivePrefix = {arXiv},
       eprint = {astro-ph/0407006},
 primaryClass = {astro-ph},
       adsurl = {https://ui.adsabs.harvard.edu/abs/2004AJ....128.1606Z}
}

@ARTICLE{J-D2016,
       author = {{Jacyszyn-Dobrzeniecka}, A.~M. and {Skowron}, D.~M. and {Mr{\'o}z}, P. and {Skowron}, J. and {Soszy{\'n}ski}, I. and {Udalski}, A. and {Pietrukowicz}, P. and {Koz{\l}owski}, S. and {Wyrzykowski}, {\L}. and {Poleski}, R. and {Pawlak}, M. and {Szyma{\'n}ski}, M.~K. and {Ulaczyk}, K.},
        title = "{OGLE-ing the Magellanic System: Three-Dimensional Structure of the Clouds and the Bridge Using Classical Cepheids}",
      journal = {\actaa},
         year = 2016,
        month = jun,
       volume = {66},
       number = {2},
        pages = {149-196},
          doi = {10.48550/arXiv.1602.09141},
archivePrefix = {arXiv},
       eprint = {1602.09141},
 primaryClass = {astro-ph.GA},
       adsurl = {https://ui.adsabs.harvard.edu/abs/2016AcA....66..149J}
}

@ARTICLE{J-D2017,
       author = {{Jacyszyn-Dobrzeniecka}, A.~M. and {Skowron}, D.~M. and {Mr{\'o}z}, P. and {Soszy{\'n}ski}, I. and {Udalski}, A. and {Pietrukowicz}, P. and {Skowron}, J. and {Poleski}, R. and {Koz{\l}owski}, S. and {Wyrzykowski}, {\L}. and {Pawlak}, M. and {Szyma{\'n}ski}, M.~K. and {Ulaczyk}, K.},
        title = "{OGLE-ing the Magellanic System: Three-Dimensional Structure of the Clouds and the Bridge using RR Lyrae Stars}",
      journal = {\actaa},
         year = 2017,
        month = mar,
       volume = {67},
       number = {1},
        pages = {1-35},
          doi = {10.32023/0001-5237/67.1.1},
archivePrefix = {arXiv},
       eprint = {1611.02709},
 primaryClass = {astro-ph.GA},
       adsurl = {https://ui.adsabs.harvard.edu/abs/2017AcA....67....1J}
}

@ARTICLE{Skowron2014,
       author = {{Skowron}, D.~M. and {Jacyszyn}, A.~M. and {Udalski}, A. and {Szyma{\'n}ski}, M.~K. and {Skowron}, J. and {Poleski}, R. and {Koz{\l}owski}, S. and {Kubiak}, M. and {Pietrzy{\'n}ski}, G. and {Soszy{\'n}ski}, I. and {Mr{\'o}z}, P. and {Pietrukowicz}, P. and {Ulaczyk}, K. and {Wyrzykowski}, {\L}.},
        title = "{OGLE-ING the Magellanic System: Stellar Populations in the Magellanic Bridge}",
      journal = {\apj},
         year = 2014,
        month = nov,
       volume = {795},
       number = {2},
          eid = {108},
        pages = {108},
          doi = {10.1088/0004-637X/795/2/108},
archivePrefix = {arXiv},
       eprint = {1405.7364},
 primaryClass = {astro-ph.SR},
       adsurl = {https://ui.adsabs.harvard.edu/abs/2014ApJ...795..108S}
}

@ARTICLE{Soszynski2016,
       author = {{Soszy{\'n}ski}, I. and {Udalski}, A. and {Szyma{\'n}ski}, M.~K. and {Wyrzykowski}, {\L}. and {Ulaczyk}, K. and {Poleski}, R. and {Pietrukowicz}, P. and {Koz{\l}owski}, S. and {Skowron}, D.~M. and {Skowron}, J. and {Mr{\'o}z}, P. and {Pawlak}, M.},
        title = "{The OGLE Collection of Variable Stars. Over 45 000 RR Lyrae Stars in the Magellanic System}",
      journal = {\actaa},
         year = 2016,
        month = jun,
       volume = {66},
       number = {2},
        pages = {131-147},
          doi = {10.48550/arXiv.1606.02727},
archivePrefix = {arXiv},
       eprint = {1606.02727},
 primaryClass = {astro-ph.SR},
       adsurl = {https://ui.adsabs.harvard.edu/abs/2016AcA....66..131S}
}

@ARTICLE{Udalski2015,
       author = {{Udalski}, A. and {Szyma{\'n}ski}, M.~K. and {Szyma{\'n}ski}, G.},
        title = "{OGLE-IV: Fourth Phase of the Optical Gravitational Lensing Experiment}",
      journal = {\actaa},
         year = 2015,
        month = mar,
       volume = {65},
       number = {1},
        pages = {1-38},
          doi = {10.48550/arXiv.1504.05966},
archivePrefix = {arXiv},
       eprint = {1504.05966},
 primaryClass = {astro-ph.SR},
       adsurl = {https://ui.adsabs.harvard.edu/abs/2015AcA....65....1U}
}

@ARTICLE{Gatto2024,
       author = {{Gatto}, M. and {Ripepi}, V. and {Bellazzini}, M. and {Tosi}, M. and {Cignoni}, M. and {Tortora}, C. and {Marconi}, M. and {Dall'Ora}, M. and {Cioni}, M. -R.~L. and {Musella}, I. and {Schipani}, P. and {Spavone}, M.},
        title = "{The YMCA (Yes, Magellanic Clouds Again) survey: Probing the outer regions of the Magellanic system with VST}",
      journal = {\aap},
         year = 2024,
        month = oct,
       volume = {690},
          eid = {A164},
        pages = {A164},
          doi = {10.1051/0004-6361/202450642},
archivePrefix = {arXiv},
       eprint = {2407.07163},
 primaryClass = {astro-ph.GA},
       adsurl = {https://ui.adsabs.harvard.edu/abs/2024A&A...690A.164G}
}

@ARTICLE{Luri2021,
       author = {{Gaia Collaboration} and {Luri}, X. and {Chemin}, L. and {Clementini}, G. and {Delgado}, H.~E. and {McMillan}, P.~J. and {Romero-G{\'o}mez}, M. and {Balbinot}, E. and {Castro-Ginard}, A. and {Mor}, R. and {Ripepi}, V. and {Sarro}, L.~M. and {Cioni}, M. -R.~L. and {Fabricius}, C. and {Garofalo}, A. and {Helmi}, A. and {Muraveva}, T. and {Brown}, A.~G.~A. and {Vallenari}, A. and {Prusti}, T. and {de Bruijne}, J.~H.~J. and {Babusiaux}, C. and {Biermann}, M. and {Creevey}, O.~L. and {Evans}, D.~W. and {Eyer}, L. and {Hutton}, A. and {Jansen}, F. and {Jordi}, C. and {Klioner}, S.~A. and {Lammers}, U. and {Lindegren}, L. and {Mignard}, F. and {Panem}, C. and {Pourbaix}, D. and {Randich}, S. and {Sartoretti}, P. and {Soubiran}, C. and {Walton}, N.~A. and {Arenou}, F. and {Bailer-Jones}, C.~A.~L. and {Bastian}, U. and {Cropper}, M. and {Drimmel}, R. and {Katz}, D. and {Lattanzi}, M.~G. and {van Leeuwen}, F. and {Bakker}, J. and {Casta{\~n}eda}, J. and {De Angeli}, F. and {Ducourant}, C. and {Fouesneau}, M. and {Fr{\'e}mat}, Y. and {Guerra}, R. and {Guerrier}, A. and {Guiraud}, J. and {Jean-Antoine Piccolo}, A. and {Masana}, E. and {Messineo}, R. and {Mowlavi}, N. and {Nicolas}, C. and {Nienartowicz}, K. and {Pailler}, F. and {Panuzzo}, P. and {Riclet}, F. and {Roux}, W. and {Seabroke}, G.~M. and {Sordo}, R. and {Tanga}, P. and {Th{\'e}venin}, F. and {Gracia-Abril}, G. and {Portell}, J. and {Teyssier}, D. and {Altmann}, M. and {Andrae}, R. and {Bellas-Velidis}, I. and {Benson}, K. and {Berthier}, J. and {Blomme}, R. and {Brugaletta}, E. and {Burgess}, P.~W. and {Busso}, G. and {Carry}, B. and {Cellino}, A. and {Cheek}, N. and {Damerdji}, Y. and {Davidson}, M. and {Delchambre}, L. and {Dell'Oro}, A. and {Fern{\'a}ndez-Hern{\'a}ndez}, J. and {Galluccio}, L. and {Garc{\'\i}a-Lario}, P. and {Garcia-Reinaldos}, M. and {Gonz{\'a}lez-N{\'u}{\~n}ez}, J. and {Gosset}, E. and {Haigron}, R. and {Halbwachs}, J. -L. and {Hambly}, N.~C. and {Harrison}, D.~L. and {Hatzidimitriou}, D. and {Heiter}, U. and {Hern{\'a}ndez}, J. and {Hestroffer}, D. and {Hodgkin}, S.~T. and {Holl}, B. and {Jan{\ss}en}, K. and {Jevardat de Fombelle}, G. and {Jordan}, S. and {Krone-Martins}, A. and {Lanzafame}, A.~C. and {L{\"o}ffler}, W. and {Lorca}, A. and {Manteiga}, M. and {Marchal}, O. and {Marrese}, P.~M. and {Moitinho}, A. and {Mora}, A. and {Muinonen}, K. and {Osborne}, P. and {Pancino}, E. and {Pauwels}, T. and {Recio-Blanco}, A. and {Richards}, P.~J. and {Riello}, M. and {Rimoldini}, L. and {Robin}, A.~C. and {Roegiers}, T. and {Rybizki}, J. and {Siopis}, C. and {Smith}, M. and {Sozzetti}, A. and {Ulla}, A. and {Utrilla}, E. and {van Leeuwen}, M. and {van Reeven}, W. and {Abbas}, U. and {Abreu Aramburu}, A. and {Accart}, S. and {Aerts}, C. and {Aguado}, J.~J. and {Ajaj}, M. and {Altavilla}, G. and {{\'A}lvarez}, M.~A. and {{\'A}lvarez Cid-Fuentes}, J. and {Alves}, J. and {Anderson}, R.~I. and {Anglada Varela}, E. and {Antoja}, T. and {Audard}, M. and {Baines}, D. and {Baker}, S.~G. and {Balaguer-N{\'u}{\~n}ez}, L. and {Balog}, Z. and {Barache}, C. and {Barbato}, D. and {Barros}, M. and {Barstow}, M.~A. and {Bartolom{\'e}}, S. and {Bassilana}, J. -L. and {Bauchet}, N. and {Baudesson-Stella}, A. and {Becciani}, U. and {Bellazzini}, M. and {Bernet}, M. and {Bertone}, S. and {Bianchi}, L. and {Blanco-Cuaresma}, S. and {Boch}, T. and {Bombrun}, A. and {Bossini}, D. and {Bouquillon}, S. and {Bragaglia}, A. and {Bramante}, L. and {Breedt}, E. and {Bressan}, A. and {Brouillet}, N. and {Bucciarelli}, B. and {Burlacu}, A. and {Busonero}, D. and {Butkevich}, A.~G. and {Buzzi}, R. and {Caffau}, E. and {Cancelliere}, R. and {C{\'a}novas}, H. and {Cantat-Gaudin}, T. and {Carballo}, R. and {Carlucci}, T. and {Carnerero}, M.~I. and {Carrasco}, J.~M. and {Casamiquela}, L. and {Castellani}, M. and {Castro Sampol}, P. and {Chaoul}, L. and {Charlot}, P. and {Chiavassa}, A. and {Comoretto}, G. and {Cooper}, W.~J. and {Cornez}, T. and {Cowell}, S. and {Crifo}, F.},
        title = "{Gaia Early Data Release 3. Structure and properties of the Magellanic Clouds}",
      journal = {\aap},
         year = 2021,
        month = may,
       volume = {649},
          eid = {A7},
        pages = {A7},
          doi = {10.1051/0004-6361/202039588},
archivePrefix = {arXiv},
       eprint = {2012.01771},
 primaryClass = {astro-ph.GA},
       adsurl = {https://ui.adsabs.harvard.edu/abs/2021A&A...649A...7G}
}

@ARTICLE{GaiaDR3,
       author = {{Gaia Collaboration} and {Vallenari}, A. and {Brown}, A.~G.~A. and {Prusti}, T. and {de Bruijne}, J.~H.~J. and {Arenou}, F. and {Babusiaux}, C. and {Biermann}, M. and {Creevey}, O.~L. and {Ducourant}, C. and {Evans}, D.~W. and {Eyer}, L. and {Guerra}, R. and {Hutton}, A. and {Jordi}, C. and {Klioner}, S.~A. and {Lammers}, U.~L. and {Lindegren}, L. and {Luri}, X. and {Mignard}, F. and {Panem}, C. and {Pourbaix}, D. and {Randich}, S. and {Sartoretti}, P. and {Soubiran}, C. and {Tanga}, P. and {Walton}, N.~A. and {Bailer-Jones}, C.~A.~L. and {Bastian}, U. and {Drimmel}, R. and {Jansen}, F. and {Katz}, D. and {Lattanzi}, M.~G. and {van Leeuwen}, F. and {Bakker}, J. and {Cacciari}, C. and {Casta{\~n}eda}, J. and {De Angeli}, F. and {Fabricius}, C. and {Fouesneau}, M. and {Fr{\'e}mat}, Y. and {Galluccio}, L. and {Guerrier}, A. and {Heiter}, U. and {Masana}, E. and {Messineo}, R. and {Mowlavi}, N. and {Nicolas}, C. and {Nienartowicz}, K. and {Pailler}, F. and {Panuzzo}, P. and {Riclet}, F. and {Roux}, W. and {Seabroke}, G.~M. and {Sordo}, R. and {Th{\'e}venin}, F. and {Gracia-Abril}, G. and {Portell}, J. and {Teyssier}, D. and {Altmann}, M. and {Andrae}, R. and {Audard}, M. and {Bellas-Velidis}, I. and {Benson}, K. and {Berthier}, J. and {Blomme}, R. and {Burgess}, P.~W. and {Busonero}, D. and {Busso}, G. and {C{\'a}novas}, H. and {Carry}, B. and {Cellino}, A. and {Cheek}, N. and {Clementini}, G. and {Damerdji}, Y. and {Davidson}, M. and {de Teodoro}, P. and {Nu{\~n}ez Campos}, M. and {Delchambre}, L. and {Dell'Oro}, A. and {Esquej}, P. and {Fern{\'a}ndez-Hern{\'a}ndez}, J. and {Fraile}, E. and {Garabato}, D. and {Garc{\'\i}a-Lario}, P. and {Gosset}, E. and {Haigron}, R. and {Halbwachs}, J. -L. and {Hambly}, N.~C. and {Harrison}, D.~L. and {Hern{\'a}ndez}, J. and {Hestroffer}, D. and {Hodgkin}, S.~T. and {Holl}, B. and {Jan{\ss}en}, K. and {Jevardat de Fombelle}, G. and {Jordan}, S. and {Krone-Martins}, A. and {Lanzafame}, A.~C. and {L{\"o}ffler}, W. and {Marchal}, O. and {Marrese}, P.~M. and {Moitinho}, A. and {Muinonen}, K. and {Osborne}, P. and {Pancino}, E. and {Pauwels}, T. and {Recio-Blanco}, A. and {Reyl{\'e}}, C. and {Riello}, M. and {Rimoldini}, L. and {Roegiers}, T. and {Rybizki}, J. and {Sarro}, L.~M. and {Siopis}, C. and {Smith}, M. and {Sozzetti}, A. and {Utrilla}, E. and {van Leeuwen}, M. and {Abbas}, U. and {{\'A}brah{\'a}m}, P. and {Abreu Aramburu}, A. and {Aerts}, C. and {Aguado}, J.~J. and {Ajaj}, M. and {Aldea-Montero}, F. and {Altavilla}, G. and {{\'A}lvarez}, M.~A. and {Alves}, J. and {Anders}, F. and {Anderson}, R.~I. and {Anglada Varela}, E. and {Antoja}, T. and {Baines}, D. and {Baker}, S.~G. and {Balaguer-N{\'u}{\~n}ez}, L. and {Balbinot}, E. and {Balog}, Z. and {Barache}, C. and {Barbato}, D. and {Barros}, M. and {Barstow}, M.~A. and {Bartolom{\'e}}, S. and {Bassilana}, J. -L. and {Bauchet}, N. and {Becciani}, U. and {Bellazzini}, M. and {Berihuete}, A. and {Bernet}, M. and {Bertone}, S. and {Bianchi}, L. and {Binnenfeld}, A. and {Blanco-Cuaresma}, S. and {Blazere}, A. and {Boch}, T. and {Bombrun}, A. and {Bossini}, D. and {Bouquillon}, S. and {Bragaglia}, A. and {Bramante}, L. and {Breedt}, E. and {Bressan}, A. and {Brouillet}, N. and {Brugaletta}, E. and {Bucciarelli}, B. and {Burlacu}, A. and {Butkevich}, A.~G. and {Buzzi}, R. and {Caffau}, E. and {Cancelliere}, R. and {Cantat-Gaudin}, T. and {Carballo}, R. and {Carlucci}, T. and {Carnerero}, M.~I. and {Carrasco}, J.~M. and {Casamiquela}, L. and {Castellani}, M. and {Castro-Ginard}, A. and {Chaoul}, L. and {Charlot}, P. and {Chemin}, L. and {Chiaramida}, V. and {Chiavassa}, A. and {Chornay}, N. and {Comoretto}, G. and {Contursi}, G. and {Cooper}, W.~J. and {Cornez}, T. and {Cowell}, S. and {Crifo}, F. and {Cropper}, M. and {Crosta}, M. and {Crowley}, C. and {Dafonte}, C. and {Dapergolas}, A. and {David}, M. and {David}, P. and {de Laverny}, P. and {De Luise}, F. and {De March}, R.},
        title = "{Gaia Data Release 3. Summary of the content and survey properties}",
      journal = {\aap},
         year = 2023,
        month = jun,
       volume = {674},
          eid = {A1},
        pages = {A1},
          doi = {10.1051/0004-6361/202243940},
archivePrefix = {arXiv},
       eprint = {2208.00211},
 primaryClass = {astro-ph.GA},
       adsurl = {https://ui.adsabs.harvard.edu/abs/2023A&A...674A...1G}
}

@ARTICLE{Ivezic2019,
       author = {{Ivezi{\'c}}, {\v{Z}}eljko and {Kahn}, Steven M. and {Tyson}, J. Anthony and {Abel}, Bob and {Acosta}, Emily and {Allsman}, Robyn and {Alonso}, David and {AlSayyad}, Yusra and {Anderson}, Scott F. and {Andrew}, John and {Angel}, James Roger P. and {Angeli}, George Z. and {Ansari}, Reza and {Antilogus}, Pierre and {Araujo}, Constanza and {Armstrong}, Robert and {Arndt}, Kirk T. and {Astier}, Pierre and {Aubourg}, {\'E}ric and {Auza}, Nicole and {Axelrod}, Tim S. and {Bard}, Deborah J. and {Barr}, Jeff D. and {Barrau}, Aurelian and {Bartlett}, James G. and {Bauer}, Amanda E. and {Bauman}, Brian J. and {Baumont}, Sylvain and {Bechtol}, Ellen and {Bechtol}, Keith and {Becker}, Andrew C. and {Becla}, Jacek and {Beldica}, Cristina and {Bellavia}, Steve and {Bianco}, Federica B. and {Biswas}, Rahul and {Blanc}, Guillaume and {Blazek}, Jonathan and {Blandford}, Roger D. and {Bloom}, Josh S. and {Bogart}, Joanne and {Bond}, Tim W. and {Booth}, Michael T. and {Borgland}, Anders W. and {Borne}, Kirk and {Bosch}, James F. and {Boutigny}, Dominique and {Brackett}, Craig A. and {Bradshaw}, Andrew and {Brandt}, William Nielsen and {Brown}, Michael E. and {Bullock}, James S. and {Burchat}, Patricia and {Burke}, David L. and {Cagnoli}, Gianpietro and {Calabrese}, Daniel and {Callahan}, Shawn and {Callen}, Alice L. and {Carlin}, Jeffrey L. and {Carlson}, Erin L. and {Chandrasekharan}, Srinivasan and {Charles-Emerson}, Glenaver and {Chesley}, Steve and {Cheu}, Elliott C. and {Chiang}, Hsin-Fang and {Chiang}, James and {Chirino}, Carol and {Chow}, Derek and {Ciardi}, David R. and {Claver}, Charles F. and {Cohen-Tanugi}, Johann and {Cockrum}, Joseph J. and {Coles}, Rebecca and {Connolly}, Andrew J. and {Cook}, Kem H. and {Cooray}, Asantha and {Covey}, Kevin R. and {Cribbs}, Chris and {Cui}, Wei and {Cutri}, Roc and {Daly}, Philip N. and {Daniel}, Scott F. and {Daruich}, Felipe and {Daubard}, Guillaume and {Daues}, Greg and {Dawson}, William and {Delgado}, Francisco and {Dellapenna}, Alfred and {de Peyster}, Robert and {de Val-Borro}, Miguel and {Digel}, Seth W. and {Doherty}, Peter and {Dubois}, Richard and {Dubois-Felsmann}, Gregory P. and {Durech}, Josef and {Economou}, Frossie and {Eifler}, Tim and {Eracleous}, Michael and {Emmons}, Benjamin L. and {Fausti Neto}, Angelo and {Ferguson}, Henry and {Figueroa}, Enrique and {Fisher-Levine}, Merlin and {Focke}, Warren and {Foss}, Michael D. and {Frank}, James and {Freemon}, Michael D. and {Gangler}, Emmanuel and {Gawiser}, Eric and {Geary}, John C. and {Gee}, Perry and {Geha}, Marla and {Gessner}, Charles J.~B. and {Gibson}, Robert R. and {Gilmore}, D. Kirk and {Glanzman}, Thomas and {Glick}, William and {Goldina}, Tatiana and {Goldstein}, Daniel A. and {Goodenow}, Iain and {Graham}, Melissa L. and {Gressler}, William J. and {Gris}, Philippe and {Guy}, Leanne P. and {Guyonnet}, Augustin and {Haller}, Gunther and {Harris}, Ron and {Hascall}, Patrick A. and {Haupt}, Justine and {Hernandez}, Fabio and {Herrmann}, Sven and {Hileman}, Edward and {Hoblitt}, Joshua and {Hodgson}, John A. and {Hogan}, Craig and {Howard}, James D. and {Huang}, Dajun and {Huffer}, Michael E. and {Ingraham}, Patrick and {Innes}, Walter R. and {Jacoby}, Suzanne H. and {Jain}, Bhuvnesh and {Jammes}, Fabrice and {Jee}, M. James and {Jenness}, Tim and {Jernigan}, Garrett and {Jevremovi{\'c}}, Darko and {Johns}, Kenneth and {Johnson}, Anthony S. and {Johnson}, Margaret W.~G. and {Jones}, R. Lynne and {Juramy-Gilles}, Claire and {Juri{\'c}}, Mario and {Kalirai}, Jason S. and {Kallivayalil}, Nitya J. and {Kalmbach}, Bryce and {Kantor}, Jeffrey P. and {Karst}, Pierre and {Kasliwal}, Mansi M. and {Kelly}, Heather and {Kessler}, Richard and {Kinnison}, Veronica and {Kirkby}, David and {Knox}, Lloyd and {Kotov}, Ivan V. and {Krabbendam}, Victor L. and {Krughoff}, K. Simon and {Kub{\'a}nek}, Petr and {Kuczewski}, John and {Kulkarni}, Shri and {Ku}, John and {Kurita}, Nadine R. and {Lage}, Craig S. and {Lambert}, Ron and {Lange}, Travis and {Langton}, J. Brian and {Le Guillou}, Laurent and {Levine}, Deborah and {Liang}, Ming and {Lim}, Kian-Tat and {Lintott}, Chris J. and {Long}, Kevin E. and {Lopez}, Margaux and {Lotz}, Paul J. and {Lupton}, Robert H. and {Lust}, Nate B. and {MacArthur}, Lauren A. and {Mahabal}, Ashish and {Mandelbaum}, Rachel and {Markiewicz}, Thomas W. and {Marsh}, Darren S. and {Marshall}, Philip J. and {Marshall}, Stuart and {May}, Morgan and {McKercher}, Robert and {McQueen}, Michelle and {Meyers}, Joshua and {Migliore}, Myriam and {Miller}, Michelle and {Mills}, David J.},
        title = "{LSST: From Science Drivers to Reference Design and Anticipated Data Products}",
      journal = {\apj},
         year = 2019,
        month = mar,
       volume = {873},
       number = {2},
          eid = {111},
        pages = {111},
          doi = {10.3847/1538-4357/ab042c},
archivePrefix = {arXiv},
       eprint = {0805.2366},
 primaryClass = {astro-ph},
       adsurl = {https://ui.adsabs.harvard.edu/abs/2019ApJ...873..111I}
}

@ARTICLE{Cioni2011,
       author = {{Cioni}, M. -R.~L. and {Clementini}, G. and {Girardi}, L. and {Guandalini}, R. and {Gullieuszik}, M. and {Miszalski}, B. and {Moretti}, M. -I. and {Ripepi}, V. and {Rubele}, S. and {Bagheri}, G. and {Bekki}, K. and {Cross}, N. and {de Blok}, W.~J.~G. and {de Grijs}, R. and {Emerson}, J.~P. and {Evans}, C.~J. and {Gibson}, B. and {Gonzales-Solares}, E. and {Groenewegen}, M.~A.~T. and {Irwin}, M. and {Ivanov}, V.~D. and {Lewis}, J. and {Marconi}, M. and {Marquette}, J. -B. and {Mastropietro}, C. and {Moore}, B. and {Napiwotzki}, R. and {Naylor}, T. and {Oliveira}, J.~M. and {Read}, M. and {Sutorius}, E. and {van Loon}, J. Th. and {Wilkinson}, M.~I. and {Wood}, P.~R.},
        title = "{The VMC survey. I. Strategy and first data}",
      journal = {\aap},
         year = 2011,
        month = mar,
       volume = {527},
          eid = {A116},
        pages = {A116},
          doi = {10.1051/0004-6361/201016137},
archivePrefix = {arXiv},
       eprint = {1012.5193},
 primaryClass = {astro-ph.CO},
       adsurl = {https://ui.adsabs.harvard.edu/abs/2011A&A...527A.116C}
}

@ARTICLE{Martin2016,
       author = {{Martin}, Nicolas F. and {Jungbluth}, Valentin and {Nidever}, David L. and {Bell}, Eric F. and {Besla}, Gurtina and {Blum}, Robert D. and {Cioni}, Maria-Rosa L. and {Conn}, Blair C. and {Kaleida}, Catherine C. and {Gallart}, Carme and {Jin}, Shoko and {Majewski}, Steven R. and {Martinez-Delgado}, David and {Monachesi}, Antonela and {Mu{\~n}oz}, Ricardo R. and {No{\"e}l}, Noelia E.~D. and {Olsen}, Knut and {Stringfellow}, Guy S. and {van der Marel}, Roeland P. and {Vivas}, A. Katherina and {Walker}, Alistair R. and {Zaritsky}, Dennis},
        title = "{SMASH 1: A Very Faint Globular Cluster Disrupting in the Outer Reaches of the LMC?}",
      journal = {\apjl},
         year = 2016,
        month = oct,
       volume = {830},
       number = {1},
          eid = {L10},
        pages = {L10},
          doi = {10.3847/2041-8205/830/1/L10},
archivePrefix = {arXiv},
       eprint = {1609.05918},
 primaryClass = {astro-ph.GA},
       adsurl = {https://ui.adsabs.harvard.edu/abs/2016ApJ...830L..10M}
}

@ARTICLE{Martin2015,
       author = {{Martin}, Nicolas F. and {Nidever}, David L. and {Besla}, Gurtina and {Olsen}, Knut and {Walker}, Alistair R. and {Vivas}, A. Katherina and {Gruendl}, Robert A. and {Kaleida}, Catherine C. and {Mu{\~n}oz}, Ricardo R. and {Blum}, Robert D. and {Saha}, Abhijit and {Conn}, Blair C. and {Bell}, Eric F. and {Chu}, You-Hua and {Cioni}, Maria-Rosa L. and {de Boer}, Thomas J.~L. and {Gallart}, Carme and {Jin}, Shoko and {Kunder}, Andrea and {Majewski}, Steven R. and {Martinez-Delgado}, David and {Monachesi}, Antonela and {Monelli}, Matteo and {Monteagudo}, Lara and {No{\"e}l}, Noelia E.~D. and {Olszewski}, Edward W. and {Stringfellow}, Guy S. and {van der Marel}, Roeland P. and {Zaritsky}, Dennis},
        title = "{Hydra II: A Faint and Compact Milky Way Dwarf Galaxy Found in the Survey of the Magellanic Stellar History}",
      journal = {\apjl},
         year = 2015,
        month = may,
       volume = {804},
       number = {1},
          eid = {L5},
        pages = {L5},
          doi = {10.1088/2041-8205/804/1/L5},
archivePrefix = {arXiv},
       eprint = {1503.06216},
 primaryClass = {astro-ph.GA},
       adsurl = {https://ui.adsabs.harvard.edu/abs/2015ApJ...804L...5M}
}

@ARTICLE{Flaugher2015,
       author = {{Flaugher}, B. and {Diehl}, H.~T. and {Honscheid}, K. and
         {Abbott}, T.~M.~C. and {Alvarez}, O. and {Angstadt}, R. and
         {Annis}, J.~T. and {Antonik}, M. and {Ballester}, O. and
         {Beaufore}, L. and {Bernstein}, G.~M. and {Bernstein}, R.~A. and
         {Bigelow}, B. and {Bonati}, M. and {Boprie}, D. and {Brooks}, D. and
         {Buckley-Geer}, E.~J. and {Campa}, J. and {Cardiel-Sas}, L. and {Castand
        er}, F.~J. and {Castilla}, J. and {Cease}, H. and {Cela-Ruiz}, J.~M. and
         {Chappa}, S. and {Chi}, E. and {Cooper}, C. and {da Costa}, L.~N. and
         {Dede}, E. and {Derylo}, G. and {DePoy}, D.~L. and {de Vicente}, J. and
         {Doel}, P. and {Drlica-Wagner}, A. and {Eiting}, J. and
         {Elliott}, A.~E. and {Emes}, J. and {Estrada}, J. and
         {Fausti Neto}, A. and {Finley}, D.~A. and {Flores}, R. and
         {Frieman}, J. and {Gerdes}, D. and {Gladders}, M.~D. and {Gregory}, B. and
         {Gutierrez}, G.~R. and {Hao}, J. and {Holland}, S.~E. and {Holm}, S. and
         {Huffman}, D. and {Jackson}, C. and {James}, D.~J. and {Jonas}, M. and
         {Karcher}, A. and {Karliner}, I. and {Kent}, S. and {Kessler}, R. and
         {Kozlovsky}, M. and {Kron}, R.~G. and {Kubik}, D. and {Kuehn}, K. and
         {Kuhlmann}, S. and {Kuk}, K. and {Lahav}, O. and {Lathrop}, A. and
         {Lee}, J. and {Levi}, M.~E. and {Lewis}, P. and {Li}, T.~S. and {Mand
        richenko}, I. and {Marshall}, J.~L. and {Martinez}, G. and
         {Merritt}, K.~W. and {Miquel}, R. and {Mu{\~n}oz}, F. and
         {Neilsen}, E.~H. and {Nichol}, R.~C. and {Nord}, B. and {Ogando}, R. and
         {Olsen}, J. and {Palaio}, N. and {Patton}, K. and {Peoples}, J. and
         {Plazas}, A.~A. and {Rauch}, J. and {Reil}, K. and {Rheault}, J. -P. and
         {Roe}, N.~A. and {Rogers}, H. and {Roodman}, A. and {Sanchez}, E. and
         {Scarpine}, V. and {Schindler}, R.~H. and {Schmidt}, R. and
         {Schmitt}, R. and {Schubnell}, M. and {Schultz}, K. and {Schurter}, P. and
         {Scott}, L. and {Serrano}, S. and {Shaw}, T.~M. and {Smith}, R.~C. and
         {Soares-Santos}, M. and {Stefanik}, A. and {Stuermer}, W. and
         {Suchyta}, E. and {Sypniewski}, A. and {Tarle}, G. and {Thaler}, J. and
         {Tighe}, R. and {Tran}, C. and {Tucker}, D. and {Walker}, A.~R. and
         {Wang}, G. and {Watson}, M. and {Weaverdyck}, C. and {Wester}, W. and
         {Woods}, R. and {Yanny}, B. and {DES Collaboration}},
        title = "{The Dark Energy Camera}",
      journal = {\aj},
         year = "2015",
        month = "Nov",
       volume = {150},
       number = {5},
          eid = {150},
        pages = {150},
          doi = {10.1088/0004-6256/150/5/150},
archivePrefix = {arXiv},
       eprint = {1504.02900},
 primaryClass = {astro-ph.IM},
       adsurl = {https://ui.adsabs.harvard.edu/abs/2015AJ....150..150F}
}

@ARTICLE{Massana2022,
       author = {{Massana}, P. and {Ruiz-Lara}, T. and {No{\"e}l}, N.~E.~D. and {Gallart}, C. and {Nidever}, D.~L. and {Choi}, Y. and {Sakowska}, J.~D. and {Besla}, G. and {Olsen}, K.~A.~G. and {Monelli}, M. and {Dorta}, A. and {Stringfellow}, G.~S. and {Cassisi}, S. and {Bernard}, E.~J. and {Zaritsky}, D. and {Cioni}, M. -R.~L. and {Monachesi}, A. and {van der Marel}, R.~P. and {de Boer}, T.~J.~L. and {Walker}, A.~R.},
        title = "{The synchronized dance of the magellanic clouds' star formation history}",
      journal = {\mnras},
         year = 2022,
        month = jun,
       volume = {513},
       number = {1},
        pages = {L40-L45},
          doi = {10.1093/mnrasl/slac030},
archivePrefix = {arXiv},
       eprint = {2203.09523},
 primaryClass = {astro-ph.GA},
       adsurl = {https://ui.adsabs.harvard.edu/abs/2022MNRAS.513L..40M}
}

@software{delvered:2025,
       author = {{Nidever}, David},
        title = "{dnidever/delvered: first delvered release}",
         year = 2025,
        month = may,
          eid = {10.5281/zenodo.15548994},
          doi = {10.5281/zenodo.15548994},
      version = {1.0.0},
    publisher = {Zenodo},
       adsurl = {https://ui.adsabs.harvard.edu/abs/2025zndo...15548994}
}

@ARTICLE{Drlica-Wagner:2021,
       author = {{Drlica-Wagner}, A. and {Carlin}, J.~L. and {Nidever}, D.~L. and {Ferguson}, P.~S. and {Kuropatkin}, N. and {Adam{\'o}w}, M. and {Cerny}, W. and {Choi}, Y. and {Esteves}, J.~H. and {Mart{\'\i}nez-V{\'a}zquez}, C.~E. and {Mau}, S. and {Miller}, A.~E. and {Mutlu-Pakdil}, B. and {Neilsen}, E.~H. and {Olsen}, K.~A.~G. and {Pace}, A.~B. and {Riley}, A.~H. and {Sakowska}, J.~D. and {Sand}, D.~J. and {Santana-Silva}, L. and {Tollerud}, E.~J. and {Tucker}, D.~L. and {Vivas}, A.~K. and {Zaborowski}, E. and {Zenteno}, A. and {Abbott}, T.~M.~C. and {Allam}, S. and {Bechtol}, K. and {Bell}, C.~P.~M. and {Bell}, E.~F. and {Bilaji}, P. and {Bom}, C.~R. and {Carballo-Bello}, J.~A. and {Crnojevi{\'c}}, D. and {Cioni}, M. -R.~L. and {Diaz-Ocampo}, A. and {de Boer}, T.~J.~L. and {Erkal}, D. and {Gruendl}, R.~A. and {Hernandez-Lang}, D. and {Hughes}, A.~K. and {James}, D.~J. and {Johnson}, L.~C. and {Li}, T.~S. and {Mao}, Y. -Y. and {Mart{\'\i}nez-Delgado}, D. and {Massana}, P. and {McNanna}, M. and {Morgan}, R. and {Nadler}, E.~O. and {No{\"e}l}, N.~E.~D. and {Palmese}, A. and {Peter}, A.~H.~G. and {Rykoff}, E.~S. and {S{\'a}nchez}, J. and {Shipp}, N. and {Simon}, J.~D. and {Smercina}, A. and {Soares-Santos}, M. and {Stringfellow}, G.~S. and {Tavangar}, K. and {van der Marel}, R.~P. and {Walker}, A.~R. and {Wechsler}, R.~H. and {Wu}, J.~F. and {Yanny}, B. and {Fitzpatrick}, M. and {Huang}, L. and {Jacques}, A. and {Nikutta}, R. and {Scott}, A. and {Astro Data Lab}},
        title = "{The DECam Local Volume Exploration Survey: Overview and First Data Release}",
      journal = {\apjs},
         year = 2021,
        month = sep,
       volume = {256},
       number = {1},
          eid = {2},
        pages = {2},
          doi = {10.3847/1538-4365/ac079d},
archivePrefix = {arXiv},
       eprint = {2103.07476},
 primaryClass = {astro-ph.GA},
       adsurl = {https://ui.adsabs.harvard.edu/abs/2021ApJS..256....2D}
}

@ARTICLE{Cullinane2022a,
       author = {{Cullinane}, L.~R. and {Mackey}, A.~D. and {Da Costa}, G.~S. and {Erkal}, D. and {Koposov}, S.~E. and {Belokurov}, V.},
        title = "{The Magellanic Edges Survey - III. Kinematics of the disturbed LMC outskirts}",
      journal = {\mnras},
         year = 2022,
        month = jun,
       volume = {512},
       number = {4},
        pages = {4798-4818},
          doi = {10.1093/mnras/stac733},
archivePrefix = {arXiv},
       eprint = {2203.05450},
 primaryClass = {astro-ph.GA},
       adsurl = {https://ui.adsabs.harvard.edu/abs/2022MNRAS.512.4798C}
}

@ARTICLE{Nidever2019a,
       author = {{Nidever}, David L. and {Olsen}, Knut and {Choi}, Yumi and {de Boer}, Thomas J.~L. and {Blum}, Robert D. and {Bell}, Eric F. and {Zaritsky}, Dennis and {Martin}, Nicolas F. and {Saha}, Abhijit and {Conn}, Blair C. and {Besla}, Gurtina and {van der Marel}, Roeland P. and {No{\"e}l}, Noelia E.~D. and {Monachesi}, Antonela and {Stringfellow}, Guy S. and {Massana}, Pol and {Cioni}, Maria-Rosa L. and {Gallart}, Carme and {Monelli}, Matteo and {Martinez-Delgado}, David and {Mu{\~n}oz}, Ricardo R. and {Majewski}, Steven R. and {Vivas}, A. Katherina and {Walker}, Alistair R. and {Kaleida}, Catherine and {Chu}, You-Hua},
        title = "{Exploring the Very Extended Low-surface-brightness Stellar Populations of the Large Magellanic Cloud with SMASH}",
      journal = {\apj},
         year = 2019,
        month = apr,
       volume = {874},
       number = {2},
          eid = {118},
        pages = {118},
          doi = {10.3847/1538-4357/aafaf7},
archivePrefix = {arXiv},
       eprint = {1805.02671},
 primaryClass = {astro-ph.GA},
       adsurl = {https://ui.adsabs.harvard.edu/abs/2019ApJ...874..118N}
}

@ARTICLE{Sakowska2024,
       author = {{Sakowska}, Joanna D. and {No{\"e}l}, Noelia E.~D. and {Ruiz-Lara}, Tom{\'a}s and {Gallart}, Carme and {Massana}, Pol and {Nidever}, David L. and {Cassisi}, Santi and {Correa-Amaro}, Patricio and {Choi}, Yumi and {Besla}, Gurtina and {Erkal}, Denis and {Mart{\'\i}nez-Delgado}, David and {Monelli}, Matteo and {Olsen}, Knut A.~G. and {Stringfellow}, Guy S.},
        title = "{Unveiling the purely young star formation history of the SMC's northeastern shell from colour-magnitude diagram fitting}",
      journal = {\mnras},
         year = 2024,
        month = aug,
       volume = {532},
       number = {4},
        pages = {4272-4288},
          doi = {10.1093/mnras/stae1766},
archivePrefix = {arXiv},
       eprint = {2407.13876},
 primaryClass = {astro-ph.GA},
       adsurl = {https://ui.adsabs.harvard.edu/abs/2024MNRAS.532.4272S}
}

@article{Sakowska2026,
	Author = {{Sakowska}, J. and {Massana}, P. and {Noel}, N. and {Ruiz-Lara}, T. and {Gallart}, C. and {Nidever}, D. and {SMASH team} },
	Year = {2026, in prep.}
    }

@ARTICLE{RuizLara2020,
       author = {{Ruiz-Lara}, T. and {Gallart}, C. and {Monelli}, M. and {Nidever}, D. and
         {Dorta}, A. and {Choi}, Y. and {Olsen}, K. and {Besla}, G. and
         {Bernard}, E.~J. and {Cassisi}, S. and {Massana}, P. and
         {No{\"e}l}, N.~E.~D. and {P{\'e}rez}, I. and {Rusakov}, V. and
         {Cioni}, M. -R.~L. and {Majewski}, S.~R. and {van der Marel}, R.~P. and
         {Mart{\'\i}nez-Delgado}, D. and {Monachesi}, A. and {Monteagudo}, L. and
         {Mu{\~n}oz}, R.~R. and {Stringfellow}, G.~S. and {Surot}, F. and
         {Vivas}, A.~K. and {Walker}, A.~R. and {Zaritsky}, D.},
        title = "{The Large Magellanic Cloud stellar content with SMASH. I. Assessing the stability of the Magellanic spiral arms}",
      journal = {\aap},
         year = 2020,
        month = jul,
       volume = {639},
          eid = {L3},
        pages = {L3},
          doi = {10.1051/0004-6361/202038392},
archivePrefix = {arXiv},
       eprint = {2006.10759},
 primaryClass = {astro-ph.GA},
       adsurl = {https://ui.adsabs.harvard.edu/abs/2020A&A...639L...3R}
}

@ARTICLE{Nidever2021a,
       author = {{Nidever}, David L. and {Olsen}, Knut and {Choi}, Yumi and {Ruiz-Lara}, Tomas and {Miller}, Amy E. and {Johnson}, L. Clifton and {Bell}, Cameron P.~M. and {Blum}, Robert D. and {Cioni}, Maria-Rosa L. and {Gallart}, Carme and {Majewski}, Steven R. and {Martin}, Nicolas F. and {Massana}, Pol and {Monachesi}, Antonela and {No{\"e}l}, Noelia E.~D. and {Sakowska}, Joanna D. and {van der Marel}, Roeland P. and {Walker}, Alistair R. and {Zaritsky}, Dennis and {Bell}, Eric F. and {Conn}, Blair C. and {de Boer}, Thomas J.~L. and {Gruendl}, Robert A. and {Monelli}, Matteo and {Mu{\~n}oz}, Ricardo R. and {Saha}, Abhijit and {Vivas}, A. Katherina and {Bernard}, Edouard and {Besla}, Gurtina and {Carballo-Bello}, Julio A. and {Dorta}, Antonio and {Martinez-Delgado}, David and {Goater}, Alex and {Rusakov}, Vadim and {Stringfellow}, Guy S.},
        title = "{The Second Data Release of the Survey of the MAgellanic Stellar History (SMASH)}",
      journal = {\aj},
         year = {2021a},
        month = feb,
       volume = {161},
       number = {2},
          eid = {74},
        pages = {74},
          doi = {10.3847/1538-3881/abceb7},
archivePrefix = {arXiv},
       eprint = {2011.13943},
 primaryClass = {astro-ph.GA},
       adsurl = {https://ui.adsabs.harvard.edu/abs/2021AJ....161...74N}
}

@ARTICLE{Massana:2020,
       author = {{Massana}, Pol and {No{\"e}l}, Noelia E.~D. and {Nidever}, David L. and
         {Erkal}, Denis and {de Boer}, Thomas J.~L. and {Choi}, Yumi and
         {Majewski}, Steven R. and {Olsen}, Knut and {Monachesi}, Antonela and
         {Gallart}, Carme and {van der Marel}, Roeland P. and
         {Ruiz-Lara}, Tom{\'a}s and {Zaritsky}, Dennis and {Martin}, Nicolas F. and
         {Mu{\~n}oz}, Ricardo R. and {Cioni}, Maria-Rosa L. and
         {Bell}, Cameron P.~M. and {Bell}, Eric F. and {Stringfellow}, Guy S. and
         {Belokurov}, Vasily and {Monelli}, Matteo and {Walker}, Alistair R. and
         {Mart{\'\i}nez-Delgado}, David and {Vivas}, A. Katherina and
         {Conn}, Blair C.},
        title = "{SMASHing the low surface brightness SMC}",
      journal = {\mnras},
         year = 2020,
        month = aug,
       volume = {498},
       number = {1},
        pages = {1034-1049},
          doi = {10.1093/mnras/staa2451},
archivePrefix = {arXiv},
       eprint = {2008.00012},
 primaryClass = {astro-ph.GA},
       adsurl = {https://ui.adsabs.harvard.edu/abs/2020MNRAS.498.1034M}
}

@ARTICLE{Ruiz-Lara:2020,
       author = {{Ruiz-Lara}, T. and {Gallart}, C. and {Monelli}, M. and {Nidever}, D. and
         {Dorta}, A. and {Choi}, Y. and {Olsen}, K. and {Besla}, G. and
         {Bernard}, E.~J. and {Cassisi}, S. and {Massana}, P. and
         {No{\"e}l}, N.~E.~D. and {P{\'e}rez}, I. and {Rusakov}, V. and
         {Cioni}, M. -R.~L. and {Majewski}, S.~R. and {van der Marel}, R.~P. and
         {Mart{\'\i}nez-Delgado}, D. and {Monachesi}, A. and {Monteagudo}, L. and
         {Mu{\~n}oz}, R.~R. and {Stringfellow}, G.~S. and {Surot}, F. and
         {Vivas}, A.~K. and {Walker}, A.~R. and {Zaritsky}, D.},
        title = "{The Large Magellanic Cloud stellar content with SMASH. I. Assessing the stability of the Magellanic spiral arms}",
      journal = {\aap},
         year = 2020,
        month = jul,
       volume = {639},
          eid = {L3},
        pages = {L3},
          doi = {10.1051/0004-6361/202038392},
archivePrefix = {arXiv},
       eprint = {2006.10759},
 primaryClass = {astro-ph.GA},
       adsurl = {https://ui.adsabs.harvard.edu/abs/2020A&A...639L...3R}
}

@ARTICLE{Gallart2004,
       author = {{Gallart}, Carme and {Stetson}, Peter B. and {Hardy}, Eduardo and {Pont}, Frederic and {Zinn}, Robert},
        title = "{Surface Brightness and Stellar Populations at the Outer Edge of the Large Magellanic Cloud: No Stellar Halo Yet}",
      journal = {\apjl},
         year = 2004,
        month = oct,
       volume = {614},
       number = {2},
        pages = {L109-L112},
          doi = {10.1086/425866},
archivePrefix = {arXiv},
       eprint = {astro-ph/0409023},
 primaryClass = {astro-ph},
       adsurl = {https://ui.adsabs.harvard.edu/abs/2004ApJ...614L.109G}
}

@ARTICLE{Gallart2008,
       author = {{Gallart}, Carme and {Stetson}, Peter B. and {Meschin}, Ingrid P. and {Pont}, Frederic and {Hardy}, Eduardo},
        title = "{Outside-In Disk Evolution in the Large Magellanic Cloud}",
      journal = {\apjl},
         year = 2008,
        month = aug,
       volume = {682},
       number = {2},
        pages = {L89},
          doi = {10.1086/590552},
archivePrefix = {arXiv},
       eprint = {0806.3669},
 primaryClass = {astro-ph},
       adsurl = {https://ui.adsabs.harvard.edu/abs/2008ApJ...682L..89G}
}

@article{Harris&Zaritsky2009,
	Adsurl = {http://adsabs.harvard.edu/abs/2009AJ....138.1243H},
	Archiveprefix = {arXiv},
	Author = {{Harris}, J. and {Zaritsky}, D.},
	Doi = {10.1088/0004-6256/138/5/1243},
	Eprint = {0908.1422},
	Journal = {\aj},
	Month = nov,
	Pages = {1243-1260},
	Title = {{The Star Formation History of the Large Magellanic Cloud}},
	Volume = 138,
	Year = 2009}
